\documentclass[a4paper, amsfonts, amssymb, amsmath, reprint, showkeys, nofootinbib, twoside]{revtex4-1}
\usepackage[english]{babel}
\usepackage[utf8]{inputenc}
\usepackage[colorinlistoftodos, color=green!40, prependcaption]{todonotes}
\usepackage[pdftex, pdftitle={Article}, pdfauthor={Author}]{hyperref} % For hyperlinks in the PDF
\usepackage{xr}

\begin{document}

\title{Topology mediated organization of {\em E.coli} chromosome in fast growth conditions}
%as a consequence of
\author{Shreerang Pande, Debarshi Mitra and Apratim Chatterji}
  %  \email[Correspondence email address: ]{debarshi.mitra@students.iiserpune.ac.in}% Your name
    \affiliation{$^1$ Dept of Physics, IISER-Pune, Pune, India-411008.
    }

\date{\today}% It is always \today, today,

\begin{abstract}
The mechanism  underlying the spatio-temporal chromosome organization in {\em  Escherichia coli} cells  
remains an open question, though experiments have been able to visually see the evolving chromosome organization 
in fast and slow growing cells. 
We had proposed [D. Mitra et al., Soft Matter, 18, 5615-5631(2022)] that 
the DNA ring polymer adopts a specific polymer topology as it goes through its cell cycle, which in turn self-organizes the chromosome by entropic forces during 
slow growth. The fast growing {\em E.coli} cells have four (or more) copies of the replicating DNA, 
with overlapping rounds of replication going on simultaneously. This makes the spatial segregation and the subsequent
organization of the multiple generations of DNA a complex task. Here, we establish that the same simple principles of entropic repulsion between polymer segments
%entropy maximizatio
%of topologically modified confined ring DNA-polymers,  
which provided an understanding of self-organization of DNA
in slow-growth conditions, also explains the organization of chromosomes in the much more complex scenario of fast growth conditions. Repulsion between DNA-polymer segments through entropic mechanisms is harnessed by modifying polymer topology.
The ring-polymer topology is modified by introducing cross-links (emulating the effects of linker-proteins)
% {\color{blue}such as MukBef})
between specific segments. 
Our simulation reproduces the emergent evolution of the organization of chromosomes as seen {\em in-vivo} in FISH experiments. 
%We employ computer simulations of a replicating bead spring model of a polymer confined in a cylinder to investigate the problem. 
%Our simulation   also reproduces the evolution of the spatial organization of the chromosomes as observed in experiments.
Furthermore, we reconcile the mechanism of longitudinal organization of the chromosomes arms in fast growth conditions by a suitable adaptation of the model. Thus, polymer physics principles,  previously used to understand chromosome organization in slow growing
{\em  E.coli} cells also resolve DNA-organization in more complex scenarios with multiple rounds of replication occurring in parallel.

\end{abstract}

%%\dates{This manuscript was compiled on \today}
%\doi{\url{www.pnas.org/cgi/doi/10.1073/pnas.XXXXXXXXXX}}

%\begin{document}

\maketitle
%\thispagestyle{firststyle}
%\ifthenelse{\boolean{shortarticle}}%%%{\ifthenelse{\boolean{singlecolumn}}{\abscontentformatted}{\abscontent}}{}

\section{Introduction}

It is vital for the living cell to make a copy of its DNA and segregate it into two halves of the cell, before the cell division can occur
%can divide 
\cite{Phillips2012,Kuzminov2013}. These essential processes have been extensively studied for  
 one of the simplest single-celled organisms, the {\em Escherichia coli (E.coli)} bacteria. 
As the chromosomes replicate and segregate thereafter,
the mechanism of spatio-temporal organization of the chromosomes remains controversial \cite{KLECKNER2013,Dewachter2018,Japaridze2020,Sherratt2020,Wiggins2018,Badrinarayanan2015,WOLDRINGH2006273}. Unlike in higher organisms,
the bacterial cell does not have dedicated protein machinery to transfer its two daughter chromosomes to two halves of the cell \cite{BenYehuda2003}.
{\em E.coli} is a rod-shaped bacterium whose chromosome occupies the central region named the nucleoid. The bacterial cell does not have a nucleus.
The segregation of the daughter chromosomes occurs simultaneously as replication is in progress \cite{Kuzminov2013}. 
In contrast, in eukaryotes the mitotic spindle helps segregate the daughter chromosomes after the
replication is complete.  Most bacterial cells have just one chromosome, and  
is a ring polymer \cite{gogou2021mechanisms}. The chromosome of the bacteria  {\em E.coli} and 
{\em C.crescentus} consist of a single ring polymer with  $4.6$ million and $4$ million base-pairs (BPs), respectively\cite{Jun2010,crescentus_nature,caul_loci}.

% The living cell has to make a copy of its DNA, and segregate 
% it to two halves of the cell, before the cell can divide. As the chromosomes replicate and segregate there-after,
% the mechanism of spatio-temporal organization of the chromosomes
% of one of the simplest living single celled organism i.e. the
% {\em E.coli} bacteria, remains controversial \cite{Kleckner2014}. Unlike in higher organisms,
% the bacterial cell does not have a dedicated machinery to transfer its two daughter 
% chromosomes to two halves of the cell.  Two other major differences of bacterial cells 
% from the cells  of  eukaryotes are: 
% (a)  the nuclear membrane 
% which encapsulates the chromosomes are absent for bacterial cells, and 
% (b) segregation of chromosomes in bacteria proceeds even as replication is in progress.
% In
% eukaryotes, segregation of daughter chromosomes through the mitotic spindle occurs after
% replication is complete. Most bacterial cells have just one  chromosome, and each 
% chromosome is a ring polymer \cite{Jun2006,Jun2007,Jun2010,Jun2012}. The chromosome of the bacteria  {\em E. coli} and 
% {\em C.crescentus} consist of a single ring polymer
% with  $4.6$ million and $4$ million base-pairs (BPs), respectively \cite{Jun2010,crescentus_nature,caul_loci}. 

 In  {\em E.coli} and other bacteria, replication begins at a site called  {\em oriC} to
end at the  {\em dif-locus}  of the {\em ter} macrodomain and proceeds along the two arms of 
the ring DNA-polymer simultaneously \cite{Japaridze2020,Wiggins2018,Japaridze2020,wigginsrf}.
Approximately $1000$ base pairs (BPs) are replicated per second by the 
repliosome at the replication forks (RFs) \cite{nielsen2006progressive,marko2009linking}.
% Approximately  $1000$ base pairs (BPs) are replicated per second on an average by the 
% {\color{blue}replisome located at} replication forks (RFs) \cite{nielsen2006progressive,marko2009linking}.
By controlling the growth medium, the doubling time $\tau$ of the {\em E.coli} bacterial cells can be varied to have values from $20$ minutes to $3$ hours or more \cite{Helmstetter1968,zaritsky2007changes,zaritsky2011instructive}. The doubling time is time taken for one newly born cell to divide into two.  
The cell cycle typically consists of three periods. 
The `B period' refers to the time period between the birth of the cell and the start of replication. 
Once replication starts, the cell enters the `C period' and lasts till the next  $\tau_C$ minutes, i.e. till the time it 
% Once replication starts, the cell enters the so-called C period and lasts till the time it 
takes for the replication to be completed. Thereafter, the cell remains in the `D period' lasting $\tau_D$
minutes, i.e., till 
cell division occurs \cite{Helmstetter1968,Zaritsky2019}. 
The bacterial cells are said to be in fast growth if $\tau < \tau_C + \tau_D$.

\begin{figure*}[ht]
     \centering
     \includegraphics[width=2\columnwidth]{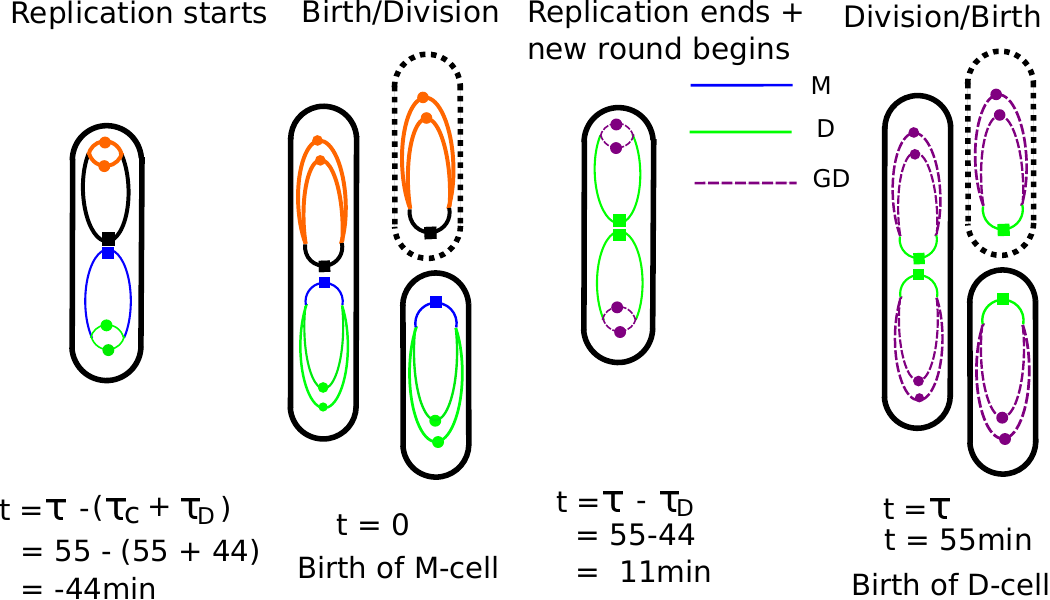}
     \caption{\textbf{Schematic of the cell cycle:}
      Given specific growth conditions\cite{Youngren2014}, the {\em E. coli} cells double every $\tau = 55$ minutes (min), the C-period is, $\tau_C=55$ min, and D-period is $\tau_D=44$ min.  
Since doubling time of $\tau = 55$min is less than $\tau_C + \tau_D = 99$ min, we can infer that the cells are undergoing fast growth. 
In the schematic, cell division takes place at time $t=0$, here the Mother cell (M-cell) is born. After $\tau = 55$ mins i.e. at $t=55$ mins,  another cell division takes place to form Daughter cells (D-cell), as shown. 
However, for the pair of daughter-chromosomes 
(green(light gray)) which divide into the two D-cells at $t=55$ min; their replication started
$\tau_C + \tau_D =99$ minutes  earlier i.e., at $t=-44$ mins.
%This is $44$ mins before  the M-cell  was born. 
We follow the ``green(light gray)" chromosomes from the start of replication. 
The {\em oriC} of blue(dark gray thin) chromosome in the grandmother-cell (GM-cell) starts a new 
round of replication to form two green(light gray) {\em oriCs}, and the replication forks proceed towards {\em dif-ter}. 
% The two RFs will start moving 
% along the arms of the blue chromosome to form two complete green chromosomes after $\tau_C$ mins,
% i.e., at $t=11$ mins. 
Meanwhile, the GM cell has divided to form two M-cells at $t=0$, 
and we follow the cell that has the green(light gray) chromosome. Since $44$ min is $80\%$ of the C-period, the mother cell is born with $80\%$ of $2$ daughter DNA's and $20\%$ of Mother DNA. There are two complete copies of the D-chromosome at $t=11$ mins
The other M-cell (shown in dashed outline) 
have the orange(light gray thick line) (\& black thick line ) chromosomes, which are fully equivalent to that of blue(dark gray thin) (\& green(light gray)) chromosomes. But 
we color it differently to distinguish it from the green(light gray) chain. From $t = 11$ to $t=55$ mins (D-period), we have two green(light gray) {\em dif-ters} connected 
to each other before cell division.
As explained before, a round of replication starts $44$ min before cell division, the purple(dashed) {\em oriCs} have formed from the existing green {\em oriCs} at $t = 11$ min, which is $44$ min before cell division at $t = 55$ min.
The D-cells are born with only $20\%$ green(light gray) D-chromosome and a pair of $80\%$ formed purple(dashed) grand-daughter 
chromosomes. In the legend, we show the colors of the M (blue(dark gray)), D (green(light gray)), GD (purple(dashed)) chromosomes. 
To visualize the different stages of the chromosomes in a cell-cycle the reader is referred to the section  titled `Movies' 
in the SI-1\cite{Suppli} (videos `Vid-1' and `Vid-2').
}
\label{fig:cellcycle}
\end{figure*}

In fast growing cells, the B period is absent, implying that the cells are continuously replicating  
and segregating. 
The conundrum of  cells doubling every $20$ minutes, even though the time taken for a chromosome to make a copy and divide is $\tau_C + \tau_D \approx 100$ minutes, was resolved by Helmstetter and Cooper and others \cite{Helmstetter1968,bremer1977examination,skarstad1985escherichia}
% The conundrum of  cells doubling every 20 minutes, even though the 
% replication time ($\tau_C$) is 40 minutes, was resolved by Helmstetter and Cooper and others \cite{Helmstetter1968,bremer1977examination,skarstad1985escherichia}
who showed that a second round of replication begins even before the first round is complete. 
Refer Figure \ref{fig:cellcycle} for a schematic of how the multiple rounds 
of replication proceed with overlapping cell cycles. Thus, the chromosome in fast-growth conditions undergoes multi-fork replication, with the replication process occurring simultaneously at two or more pairs of RFs \cite{Youngren2014,Helmstetter1968}.

It is accepted that for {\em E.coli} chromosomes, entropic forces between the ring polymers 
play a significant role  in the segregation of daughter chromosomes \cite{Jun2006,Jun2007,Jun2010,Jun2012,ha1,hamain,hamain2,hareview}, though proteins like 
MukBEF also plays a critical role in the process \cite{Sherratt2020,Boccard2021}.
%However, it has also been demonstrated that the Par-ABS system plays a role only in the organization of the two arms of the chromosome 
%rather than segregation {\bf REFERENCE FOR THIS}.
Moreover, researchers have used Fluorescent In Situ Hybridization (FISH) experiments to track 
the position of multiple DNA-loci at different points in the cell cycle, i.e., while the replication 
and segregation of the bacterial chromosome is in progress both in fast and slow growth conditions \cite{Youngren2014, Cass2016, woldy}. 
For slow growth, it is observed that the {\em oriC}  is initially found in the mid-cell position,
and after about 20 minutes into the C-period, the two {\em oriCs}  move to the quarter and three-quarter 
positions along the cell-long axis \cite{Cass2016,dna1}. The position of the {\em oriCs} is measured from one of the pole 
positions. In contrast, the {\em dif-ter} locus remains delocalized within the cell at the start 
of the C period but eventually moves to the mid-cell position before the end of the replication process. Other
loci also move to their respective `home positions' as segregation proceeds \cite{Cass2016}. The mechanism by which the different 
genomic loci identify their cellular addresses within the cell and then occupy the position at the appropriate stage
of the cell-cycle had remained an open question. 

In our previous work in slow growth conditions, we established that this DNA-organization can be 
obtained by adopting a suitably modified polymer topology by having long range contacts on the chain 
contour, which are likely mediated by MukBEF  or other linker molecules \cite{dna1}. We refer to these
contacts along the DNA-polymer as cross-links (CLs). 

\begin{figure}[!hbt]
    \centering
    \includegraphics[width=0.8\columnwidth]{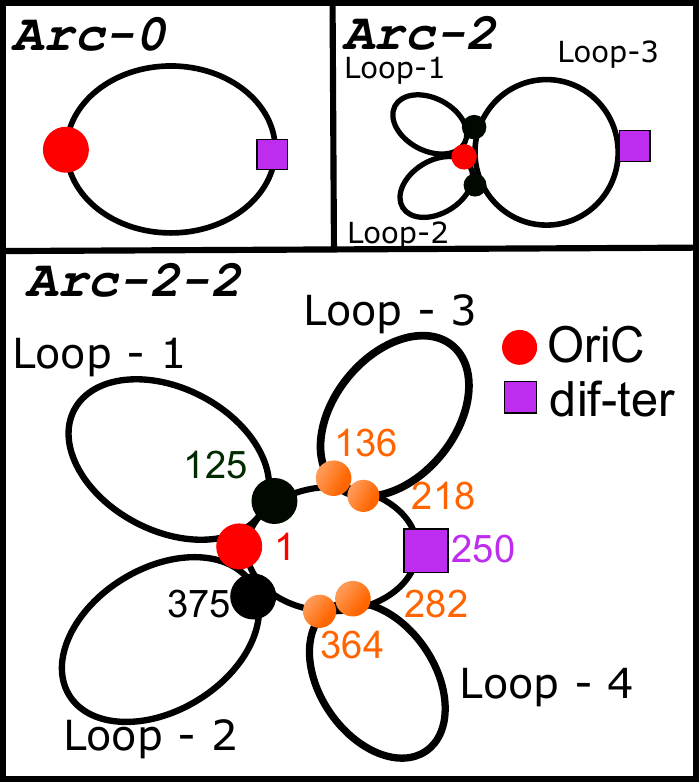}
    \caption{ \textbf{Schematic of the different polymer architectures:} The schematic  shows the Arc-2-2 topology of 
    the DNA-polymer with $500$ monomers. We start out with a ring polymer (Arc-0); thus, monomer $1$ is joined to $500$. We label monomer-1 as {\rm oriC} and $250$ as {\em dif-ter}.
    In addition, in our model, the monomer $125$ \& $375$ is cross-linked to monomer $1$ by harmonic springs modelling bridging proteins to create the Arc-2 
    architecture of the polymer.
    For the Arc2-2 we have additionally cross-linked the monomers
    $136$ \& $218$, as well as $282$ \& $364$. Also refer Fig.\ref{fig:Loci}.
    %We have studied these architectures in the case of slow growth in \cite{dna1}. 
    % This manuscript uses the same Arc-2-2 architecture 
    % to model mother and daughter chromosomes in cells with overlapping life cycles. Organization of 
    % tagged loci as seen in FISH experiments spontaneously emerges from our simulations. In simulations
    % presented in this paper, the daughter chromosomes are in the Arc-2-2 topology 
    % at time $t=0$ (refer Fig.\ref{fig:cellcycle}), whereas the grand-daughters adopt the Arc-2-2 topology after the relevant monomers are replicated 
    % and become available for cross-linking by springs.
    }
    \label{fig:Arc2_2}
\end{figure}

 We refer the reader to 
Fig.\ref{fig:Arc2_2} for a schematic of the modified polymer architectures we used for our 
study of chromosome organization in slow growth conditions. We use the 
same architectures for the current study of DNA organization in fast growth conditions.
We also showed that the sites of the CLs that we used in our DNA-polymer simulations, show  
high contact probabilities in the Hi-C map of 
{\em  E.coli} chromosomes \cite{dna1}.  As a consequence of the introduction of CLs, 
internal loops of the polymer segments are formed within the ring-polymers. Thus, the chromosome adopts 
a more complex polymer-topology than a simple ring polymer.  
 Though this choice of architecture 
might appear {\em ad-hoc} to the reader at first, the architecture was arrived at using intuition 
developed by systematically studying the emergent segregation and loci (polymer-segment) localization 
properties of $12$  different architectures \cite{dna2}.
% Our study focuses solely on the consequences 
% of the change in polymer topology due  to the formation of loops in the bacterial chromosome, 
% but we cannot comment on the precise mechanism or identify the protein(s) complexes which cross-links 
% specific points of the chromosome. }

 A consequence of the the choice of introducing cross-linkers between points on 
the contour of the ring polymers is that we have smaller internal ring-polymer segments within the chromosome
attached to each other. It is already well established by previous studies on semidilute systems 
and melts of ring polymers that ring polymers form compact configurations as compared to linear 
polymers \cite{loop_heerman,halverson2011molecular,rosa2014ring,chubak2021multiscale,halverson2011Statics,pachong2020melts}. Furthermore, rings have effective
repulsive interactions between them and try to exclude each other in space \cite{narros2010influence,narros2014multi}.
 Thus, the different loops of the Arc-2-2 architecture
entropically repel each other and occupy different segments of the cell along the long axis, and thereby 
also localize different loci that are part of the loop as well as speed up the segregation process. An 
appreciation of the underlying physics of this phenomenon, viz., principles of the loci localization at
different points along the long-axis by choosing different sizes of internal loops can be  
achieved by referring to \cite{dna2}. We give a brief review of this mechanism in this paper, before we present our 
results on modeling fast growth.

 We chose two of the $12$ architectures discussed in \cite{dna2}, which we named Arc-2 and Arc-2-2. We 
 used their localization properties to quantitatively match our simulation results with
 the organization of loci as in FISH data for the chromosome of the bacterium for {\em E.coli} in slow 
 growth conditions.  We also matched broad features of domain formation as seen in Hi-C maps 
 of the {\em E.coli} chromosome with the calculated contact map from simulations using the Arc-2-2 architecture and also discussed differences between the two \cite{dna1}.
Furthermore, using our systematic study of  $12$ other polymer-topologies, we showed that we can 
also match our model predictions with experimental Hi-C and FISH data for another bacterium, viz,
{\em C.crescentus} \cite{dna2}, by suitable choice of a different polymer topology. Note 
that we always try to add only $2$ to $4$ CLs in our studies  to minimally modify the topology. 
%In this paper, we now show that  

In fast growth conditions of {\em E.coli} with multifork replication in {\em E.coli}, there exists four (or more)  
chromosomes of different lengths at different stages of the replication process. This makes the segregation 
and faithful division of multiple strands of chromosomes a much more complex task.
 An active machinery that could possibly direct the newly replicated chromosomes
to move in opposite directions to get segregated by directed application of forces, 
might end up in daughter chromosomes remaining spatially overlapped
in this complex life cycle \cite{Youngren2014}. Refer to the schematic figure in SI-2 for a more detailed discussion
of this point.
% This is because there are overlapping rounds of replication 
% that take place and each chromosome  (of each generation) must move in different directions such that 
% they occupy a specific region in the cell, without being mixed with the chromosomes of the
% previous/subsequent generations.
Therefore, an entropic model without any actively-driven segregating
machinery is a worthwhile avenue to pursue to decipher the mechanism of chromosome organization within a cell. 
\begin{figure}[!hbt]
    \centering
    \includegraphics[width =0.9\columnwidth]{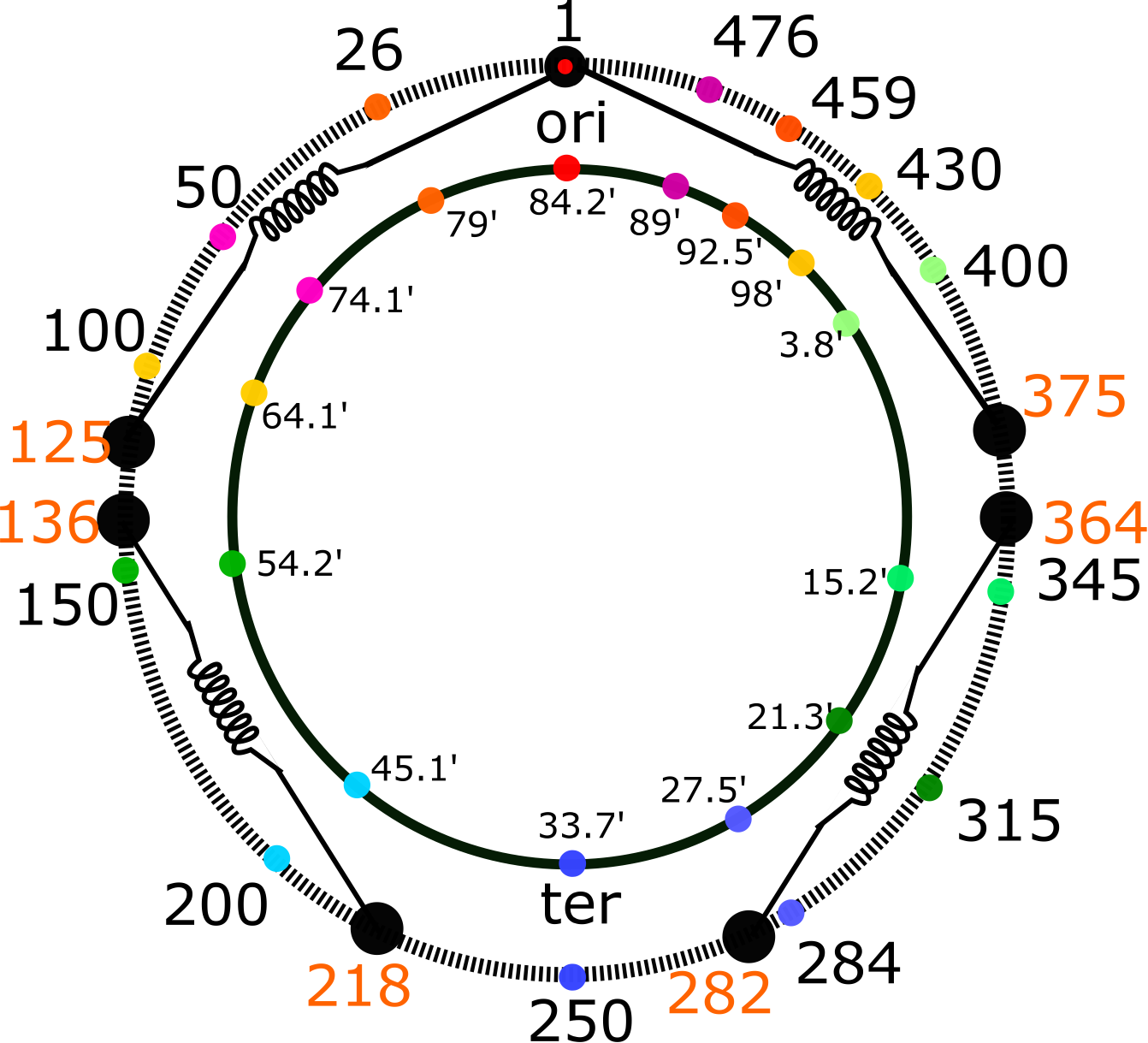}
    \caption{\textbf{Schematic of the tagged loci:} The figure shows a schematic of chromosome loci tagged in 
    experiments along the chain contour and the corresponding monomer indices for a $500$ monomer chain by colored-filled small 
    circles.  Experimentally, the circular chromosome is tagged at different sections,
    where different loci along the chain contour are denoted in terms of minutes and seconds. 
    %The full chromosome is considered as $100$ minutes. 
    The inner circle in the schematic corresponds to the loci fluorescently labeled in the experiment \cite{Youngren2014}. 
    The outer circle denotes the monomer indices corresponding to these labels in our 
    model system of ring polymer, where the {\em oriC} is denoted by monomer index $1$, and {\em dif-ter} by monomer index $250$. 
    % As multifork replication proceeds, there can be more than one loci of a given label within the cylinder (cell) at 
    % a particular stage of the cell cycle. 
    We have also shown the monomers which are cross-linked by springs of equilibrium 
    length $a$ to create the Arc-2-2 architecture, as shown  in Fig.\ref{fig:Arc2_2}.}
    \label{fig:Loci}
\end{figure}

We establish in this article that the same simple model mechanism that was earlier developed to explain the 
organization of chromosomes and the replication fork in slow growth conditions 
also explains the organization of chromosomes in the much more complex 
% scenario of segregation and organization in
fast growth conditions.  We introduce 
appropriate modifications to our previous model to incorporate overlapping rounds of replication.  
We aim to obtain the spontaneous segregation of multiple DNA strands as the different rounds of replication(s) proceed. 
As an emergent phenomenon, the loci of partially replicated chromosomes position themselves at different sections 
of the long axis.  
% We do not model cell division.

%The question however is, does the same entropic mechanism which explains loci-organization in slow growth
%also explain the organization of segregating chromosomes in  fast growth? 

 The organization of genomic loci and that of replication forks self-emerge in our model simulation as 
 a consequence of modified topology of  bead spring model ring polymer(s), which represents
 the bacterial chromosome. The organization we obtain
 is similar to that seen {\em in-vivo}.  We further show that the organization of the replication forks 
 is a direct consequence of the entropic repulsion between the different loops of the chromosome-polymer. 
 In the last section of the article, 
 we suggest a plausible topology based mechanism to try to understand a
 long-standing problem pertaining to the organization of the chromosomal arms with respect to each other. 
 As indicated by the data \cite{woldy,Youngren2014},
 in fast-growth conditions the chromosome arms are arranged in a longitudinal (doughnut-like) fashion. 
 To this end, we further modify Arc-2-2 topology, and establish here that the doughnut-like
 organization can be obtained by introducing smaller loops along the arms of the chromosome-polymer. 
 The introduction of smaller loops along the chain contour leads to entropic segregation of the two 
 arms along the radial axis of the cylinder. 
 
 We emphasize that we introduce a minimal number of topological modifications in a 
 coarse-grained minimal bead-spring polymer to obtain emergent organization of loci and 
 the segregation of chromosomes in fast growth conditions. These modifications are likely
 consequences of active processes within the cell. Thus, we incorporate the 
 internal loops formed by linker proteins (e.g. MukBEF or other protein complexes) 
 in our polymer model, and investigate 
 the consequences of these loops without incorporating the details of the mechanism of formation 
 of the same. For our model, we assume that the loops of Arc-2-2 are permanent, i.e.  
 {\em effectively} long-lived {\em in vivo}.

 We do not explicitly incorporate the effects of many other relevant non-equilibrium biological 
 phenomena, which might contribute to the smooth functioning of the cell. Thus we neglect the presence 
 of organelles in the cells, which are often modelled as crowders \cite{woldringh2024compaction}, and the details of the replication bubble.
 We do not incorporate the effects of supercoiling \cite{JieXiao24} in our model, as models of supercoiling and its 
 consequences are still under investigation in simpler idealized scenarios \cite{junier2023dna}. We also do 
 not explicitly model the process of extrusion, which creates loops of $\sim 100-200$ kilo-BPs. It has been 
 shown by simulations that internal transient loops formed by extrusion indeed speed up  the segregation process 
 of DNA-rings in a cylinder \cite{harju2023loop}, consistent with our previous results \cite{dna2} 
 with permanent CLs. In our minimal model, we do not expect to be able to 
 capture every aspect of the chromosome organization and dynamics.  While we've identified some of the 
 mechanisms that drive  {\em E.coli } chromosome organization at the $100$nm to $\mu$ length scales in 
 fast growth, it is imperative to acknowledge the inherent limitations in our ability to comprehensively 
 capture all the intricacies of experimental observations related to chromosome organization. In our 
 future investigations, we intend to add some of these effects to our current model.
 
We now describe the outline of the rest of the paper. In section II, we describe the model 
 we used to study the system. In section III, we briefly review our previous understandings to introduce 
 the basic principles of entropic repulsion between internal loops to the readers of the paper. This is to
 facilitate our readers to appreciate our new results on loci-localization in fast growth conditions presented 
 in this paper, without referring to our previous papers \cite{dna2}. In section IV we describe in detail 
 the spatio-temporal organization of the different loci along the long axis in fast growth using our simulations. 
 We also point out when we get differences with experimental observations. We not only show results for the 
 different loci but the localization of replication forks as well. We show that the organization of the 
 replication forks arises as a consequence of modifying the topology of the polymer.  After discussing the 
 radial organization of the loci, we summarize our results in section V.

%In this paper, we establish that
%it is indeed the case and demonstrate that the same mechanism can also be used to obtain the organization 
%of loci along the  cell long axis in fast growth conditions with overlapping  replication cycles, 
%and establish quantitative  consistency with previously published experimental data \cite{Youngren2014}.

%Thus, our proposed mechanism is likely  a generic mechanism to obtain chromosome organization in bacterial 
%DNA,

\section{Model}
We use Monte Carlo simulations of the bead spring model of a polymer 
with $500$ monomers  to model a single {\em E.coli} chromosome (with $4.6$ million base pairs) 
within a cylinder, which represents the {\em E.coli} cell. Thus, each coarse grained monomer subsumes $9.2$ kbp of DNA.

The equilibrium distance between two neighboring monomers (beads) along the chain
contour is $a$, and they interact {\em via} the harmonic spring potential with energy 
$V_H = \kappa (r - a)^2$, where $r$ is the distance between the adjacent monomers. The unit of
length in our study is $a$. The 
spring constant $\kappa$ is $100 k_BT/a^2$. 
 %{\color{blue} This choice of $\kappa$
 % allows us to keep the mean extension of individual springs due to thermal effects 
 % to $7\%$. Since different springs along the polymer may be extended or compressed, 
 % the effective variation in the length of the chain will be less than $7\%$, while
 % allowing us Monte Carlo acceptance rates between $10-50\%$.} 
The excluded volume (EV) interactions between monomers are modeled 
by the WCA (Weeks Chandler Anderson) potential \cite{allen2017computer} and the diameter of each monomer is given by $\sigma=0.8a$, unless specified otherwise. This particular choice of parameters,
allows us to keep the mean extension of individual springs due to thermal effects to $7\%$ and prevent 
chain crossing while allowing for reasonable acceptance rates of monomer displacements in the Monte simulations. 
These parameters are appropriate to model a normal inorganic polymer where chains do not cross each other. 

The form of the WCA potential is
\begin{equation}
V_{WCA}=4\epsilon[(\sigma/r)^{12} - (\sigma/r)^6] + \epsilon_0, \forall r<2^{1/6}\sigma.
\end{equation}
The quantity $\epsilon_0$ is needed to ensure that the potential goes smoothly to zero at the cutoff and 
$V_{WCA}=0$ for values of $r$ greater than the cutoff.

We model the chromosome replication and segregation 
over one doubling time $\tau$ inside an elongating cylinder, representing the growing 
{\em E.coli} cell. To model replication at the replication forks(RFs), we add monomers (DNA
segments) at regular intervals to the chain which represents the daughter-DNA-2. 
 A schematic  figure is given in SI-3.   After the RF moves 
to the adjacent site (monomer)  to make a copy of the DNA-segment of the mother (which 
leads to addition of a new monomer to DNA-2 in our model), we rename the monomer of the
mother-DNA such that it now notionally belongs to daughter-DNA-1. Thus the lengths of 
daughter-DNA-1 and DNA-2 keep increasing as the RFs move towards 
{\em ter} along the two arms of the mother DNA.
The simulation starts from the state right after cell division, equivalent to the 
state shown at time $t=0$ in Fig.\ref{fig:cellcycle}. At that stage, two new cells are 
just born from their parent cell. We follow the replication of chromosomes in just 
one cell, and the simulation ends just before the cell is ready to divide into two
daughter cells, i.e. the stage shown at $t=55$ min in Fig.\ref{fig:cellcycle}.  The newly born mother cell (M-cell)  at $ t=0$  
has $80\%$ partially replicated mother M-chromosome, i.e., there are two daughter D-chromosomes, 
schematically marked in green with $400$ monomers each, and 
$20\%$ mother chromosome marked in blue with $100$ monomers; refer Fig.\ref{fig:cellcycle}.
Thereby, there are two {\em oriCs} at the start of simulations.  The way the system is
initialized is described in  SI-4.

Monte Carlo (MC)  simulations is used to update the position of the monomers, where one
Monte Carlo  step (MCS) consists of $N$ attempts to update the position of  the $N$  
monomers, chosen at random. Since we model replication and thereby add monomers at the RFs 
at regular intervals, $N$ keeps increasing as the simulation proceeds. To update monomer 
positions,  a trial move is made to displace the monomer in a random direction, and the 
move is accepted or rejected using the Metropolis criterion. The polymers explore different microstates 
 as the monomers undergo local diffusive motion, as the simulation proceeds. 
 
 The successive Monte Carlo steps cannot and should not be interpreted as the time evolution,
  and we do not mention the elapsed time when we discuss DNA-organization at different stages of the cell cycle. 
  In a Monte Carlo step, the DNA-polymer reaches a different microstate without following the 
  detailed kinetic pathways, and the probability of reaching the new microstates assumes conditions 
  of equilibrium statistical mechanics. But our simulations break detailed balance in multiple ways, which we discuss in some detail at a later stage in this paper. Hence, when we present our simulation results, we avoid 
  the mention of time in units of minutes and instead refer to the stage of cell cycle measured in terms of 
  the progress of replication and position of the RF on the polymer contour.
  However, we do provide a discussion of time scales at the end of this section.

During the course of our simulation, the two RFs reach the {\em dif-ter} loci such that the 
mother is completely replicated to form two (green) D-chromosomes and the cell enters the D-period, refer Fig.\ref{fig:cellcycle}. Simultaneously, a new round 
of replication starts such that each of the two (green) D-{\em oriCs} each divide into two (orange) 
GD-{\em oriC}s. The simultaneous start of the second round and the end of the first round is a consequence of the values of
of $\tau$, $\tau_C$ and $\tau_D$ in experiments of \cite{Youngren2014}, which we choose to model in this paper. 
For modeling replication, we add monomers at a fixed rate of $1$ monomer every $f_{rep}=2\times 10^5$ MCS 
at each RF. We keep $f_{rep}$ identical 
to that used in our study for slow-growth conditions \cite{dna1}.

As the cell cycle proceeds, the cylinder length doubles in small steps over the course of
the simulation while the diameter remains fixed as observed for {\em E.coli} cells {\em in 
vivo}. We increase the length of the cylinder every $f_{rep}$ MCS. The polymer is confined within a cylinder of diameter $7a$ ($\equiv 1\mu$m, the 
typical diameter of the cell), and the cylinder length doubles from $21a$ ($\equiv 3\mu$m) 
to $42a$ as our simulation proceeds.   
We consider the walls of the cylinder to be hard (with infinite potential) such that 
we reject any Monte Carlo trial move in which a monomer-center attempts to occupy a position located outside the
cylinder. The dimensions of the confining cylinder correspond to the volume accessible to the center of the monomers. 

We modify the ring polymer architecture by introducing chromosomal loops by bridging 
specific loci along the chain contour of daughter DNAs after the RFs cross the 
corresponding loci of the mother DNA,  refer Fig.\ref{fig:Arc2_2} and Fig.\ref{fig:Loci}. These loops are created in our simulations by
introducing additional springs that cross-link between specific pairs of monomers
along the chromosome contour \cite{dna1}, using insights from \cite{dna2}. 
Figure \ref{fig:Loci}  also shows the position of tagged loci in fast growth experiments of 
\cite{Youngren2014}, and the corresponding monomer indices in our polymer model with $500$ monomers.
We have explicitly checked that a change in the choice of monomer to be
cross-linked by  $\approx 5$ monomers along the chain of $500$ monomers will not
change the global localization pattern of polymer segments significantly; refer SI-5. Larger
changes will affect the size of loops and, thereby, the relative strength of 
entropic interactions and hence modify the localization and organization patterns
of polymer segments along the cylinder long axis.

Moreover, to mimic the role played by topoisomerase within the 
living cell, we allow topological constraint release (TCR) 
at regular intervals at rates we used previously in \cite{dna1,dna2}, 
i.e.,  every $f_{TCR} = 10^4$ MCS.  We reduce the excluded volume 
interaction by  changing the $\sigma$ to $0.1a$, for the next $900$ MCS. 
This allows the chains to cross through each other. 
We do not model cell division.
% We have also outlined  
% the mechanism of topological constraint release in our simulations in SI-6.
We track the position of all the (available) {\em oriCs}  and other monomers as the simulation 
proceeds. 
 As we have previously shown in \cite{dna2},  topological
constraint release is crucial for successful segregation to occur. If we switch off
the step of TCR by chain crossing, the success rates of segregation 
for a pair of topologically modified polymers is low.
If we decrease $f_{TCR}$ by a factor of $10$, i.e. $f_{TCR}=1000$ (say) and thereby allow chains to cross 
more frequently, then we also reduce the excluded 
volume interaction and therefore the entropic effects. However, once segregated the localization of the loops
and the loci are not affected by TCR as we show later.

Though we use Monte Carlo simulations to investigate chromosome organization as the cell 
goes through its life cycle, the simulation is quintessentially a {\em non-equilibrium} simulation scheme.
In the simulations, we (a) add effects of polymer chains crossing each other to release topological 
constraints, (b) add monomers to the simulation box at regular intervals at different points along the 
contour, i.e., at the position of the  RFs to mimic replication and formation of two 
chains from one, (c)  add cross-links at certain stages of the simulation, and lastly (d) increase the 
length of the cylinder as the simulation proceeds. These are energy-consuming non-equilibrium active
processes inside the cell, and these steps break detailed balance in the simulation. MC is used
primarily to model the diffusion of monomers and explore different conformations of polymers in a 
confined space, assuming local equilibrium \cite{kikuchi1991,sharon}.  As we demonstrate
in SI-6, a polymer ring with $120$ monomers relaxes in $\approx 5 \times 10^4$ MCS, whereas, we add new
monomers and increase the length of the box every $f_{rep} = 2 \times 10^5$ MCS. Just as a reference,
Loop-1 and Loop-2 have $125$ monomers each. A single monomer of the $120$ monomer polymer chain 
takes $\approx 400$ MCS to diffuse its own diameter in confinement  

 As mentioned before, we avoid mapping Monte Carlo Steps (MCS) 
directly to time in terms of minutes but rather mention the progress of the cell 
cycle in terms of the replication stage.  Converting MCS into real time units 
by comparing diffusion rates need not be accurate also because chains can freely cross each other during the TCR step
in our model, but we do not have an estimate 
of how many chain crossing events actually occur per minute.
Incorporating the detailed process of releasing topological constraints by topo-isomerase {\em in-vivo} 
is out of scope in our current coarse grained model. This also involves the unavailability of the time 
scales of the release of topological constraints {\em in-vivo}.
% Moreover, we do not know
% how frequently topo-isomerase to releases topological constraints and how to account for that
% in Monte Carlo steps.
However, we provide an estimate of the number of MCS
for the completion of replication and segregation in units of the number 
of MCS taken for a monomer to diffuse its own diameter to check if the ratio is compatible 
with experimental estimates. However, it must be remembered that a chromosome segment undergoes 
sub-diffusive behavior {\em in-vivo} \cite{amitai2017polymer,AMITAI2017}. While estimating the number of MCS 
taken for a monomer to diffuse its own size, we do not consider the presence of organelles (crowders).
For the comparison, we use $a=150$nm, which is consistent 
with our choice of the cylinder diameter $D=7a=1050$nm. This is shown in SI-7.

\section{ Review: Mechanism of entropic localization.}
We explain the mechanism of the emergence of localization of loci as a consequence of the underlying modified internal
topology of the ring polymer. We show that entropic repulsion between the internal loops of the modified architectures 
lead to the mutual self-avoidance of the loops, and thereby loops occupy different sections of the cylinder along the long axis. 
The monomers belonging to different loops get localized as a consequence.

A systematic and detailed analysis of the mechanism of entropic repulsion between internal loops and their 
consequence on segregation and loci localization of $12$ different architectures was studied in \cite{dna2}. 
These studies started with two overlapping polymers with modified topologies, and then as the simulation proceeds
the polymers segregate to two halves of the cylinder. Thereafter, we get localization of different internal loops 
of each polymer within the half-cylinder that they occupy.  
We recreate a similar simulation and analysis in this section to clarify the role of loops and their organization.

For the results presented in this section, we 
 do not (a) add monomers as the simulation proceeds, (b) change the length 
 of the cylinder and (c) introduce additional CLs during the course of the simulation. Moreover, 
 we present data (i) without topological constraint release as well as (ii) with topological 
 constraint release at rates as used in the next sections. We hope this simpler 
 case study will help the reader appreciate the more complex case of loci-localization with multiple rounds of 
 replication in parallel, as happens for chromosome replication in fast-growth conditions. 

\subsection{Architecture-2: Arc-2}
 % A single polymer with modified 
 % architecture Arc-2 (with two internal loops) or Arc-2-2  (with two additional smaller loops) is confined in a cylinder of
 % length $21a$ and diameter $7a$, also refer Fig.\ref{Arc2_2}. We perform $50$ independent
 % Monte carlo simulations (without modeling replication) starting from random different initial conditions to study the 
 % localization properties of the COMs (center of mass) of the loops and specific monomers of the polymer. 
 Two polymers, each with $500$ monomers, and  with modified architecture Arc-2 are confined in a cylinder of diameter $7a$ 
 and fixed length $42a$, consistent with the dimensions used for the fast-growth model.  Our previous slow growth 
 simulations were performed in a cylinder of length $35a$ \cite{dna1}.
 %We allow for topological constraint release by reducing the diameter $\sigma$ of the monomers and allowing chains to cross each other. We reduce the monomer diameter after every $10^4$ MCS for the next $900$ steps, and this breaks detailed balance. 
 We start our simulations with segregated conformations which were the last configuration of other runs,
 where we had two complete Arc-2 polymers under confinement. Thereafter, we ran the simulations for a further 
 $n_I=4 \times 10^7$ MCS, before we started collecting data 
 over the next $n_P = 2 \times 10^7$ MCS. This data was used to calculate positional distributions of the COMs and other statistical 
 quantities, which we present in this section. The $2$ polymers are connected at monomer $250$, which represents the {\em dif-ter} 
 link that is present in the cell.  From a single production run of $n_P=2 \times 10^7$ MCS, $666$ snapshots  (micro-states)  were collected 
 to calculate statistical averages.  We perform $50$ independent Monte Carlo simulation runs, each with $n_I + n_P=6 \times 10^7$ MCS.

\begin{figure}[hbt!]
    \centering
     \includegraphics[scale = 0.45]{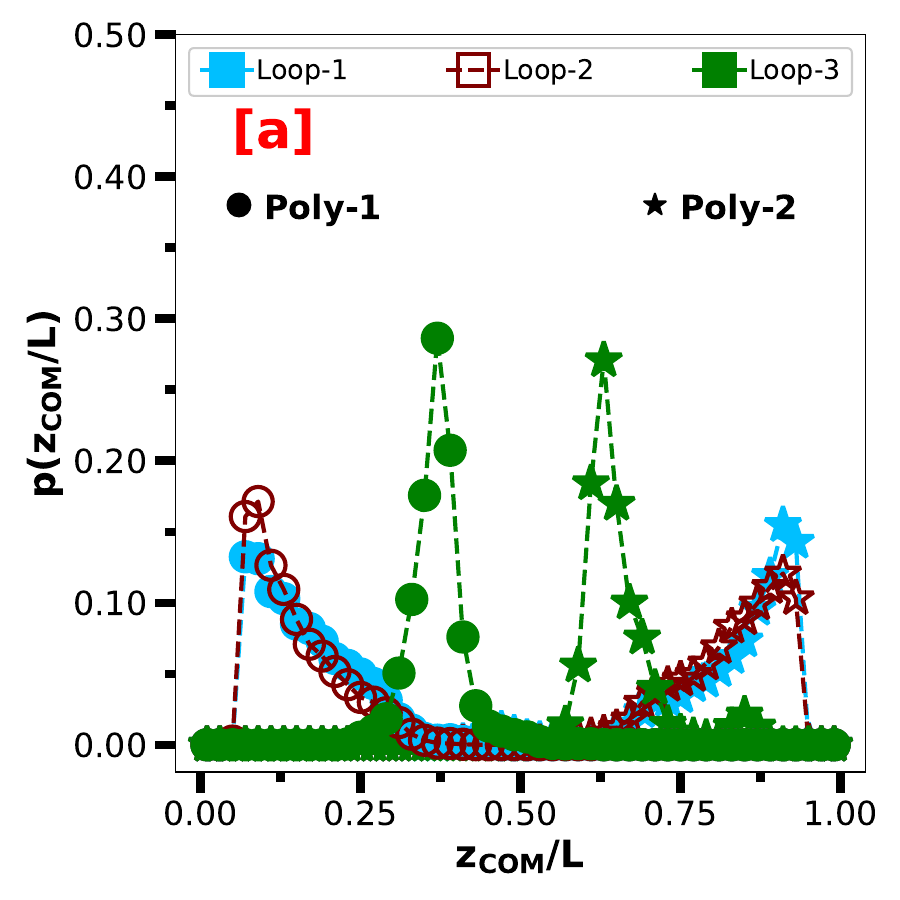}
     \vskip 0.25cm
    \includegraphics[scale = 0.15]{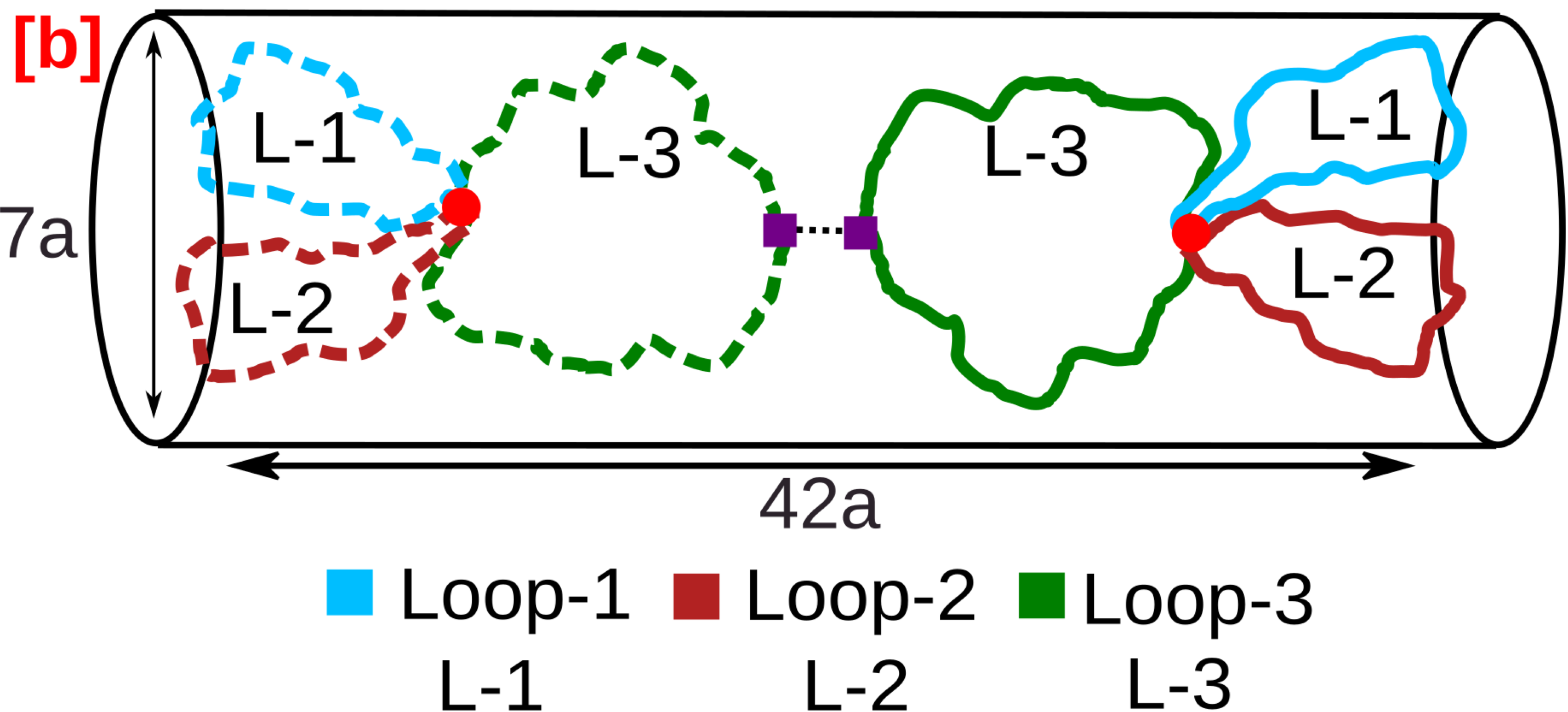}
    \vskip 0.5cm
    \includegraphics[scale = 0.28]{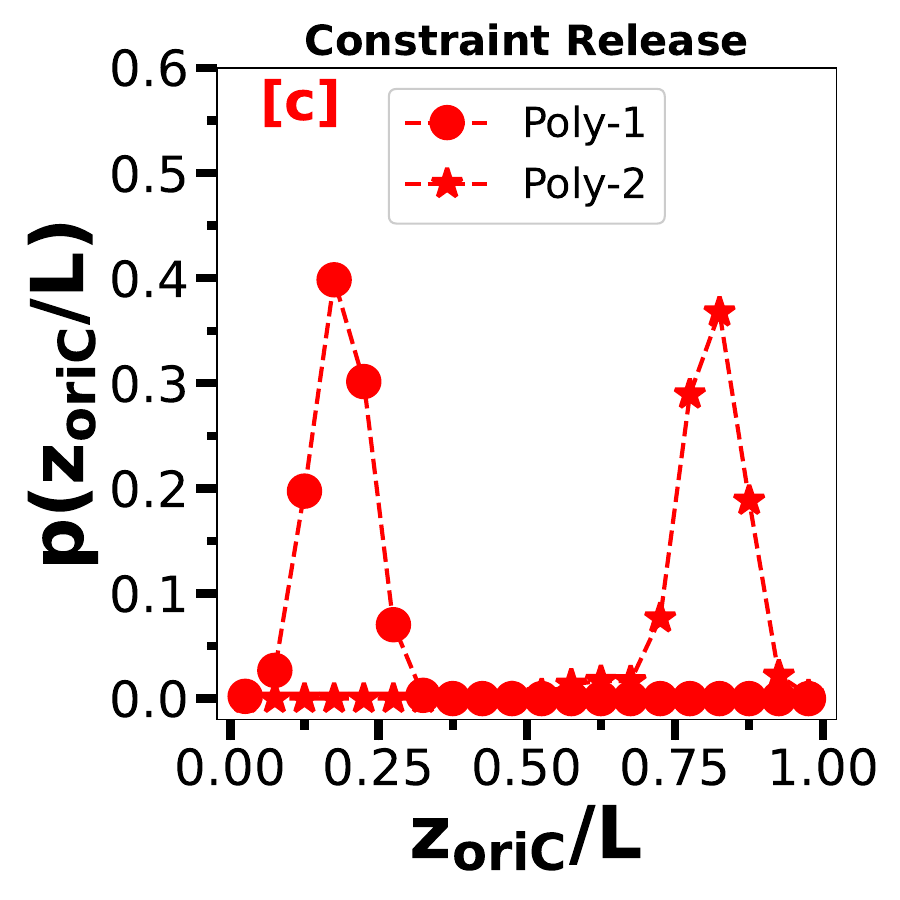}
        \includegraphics[scale = 0.28]{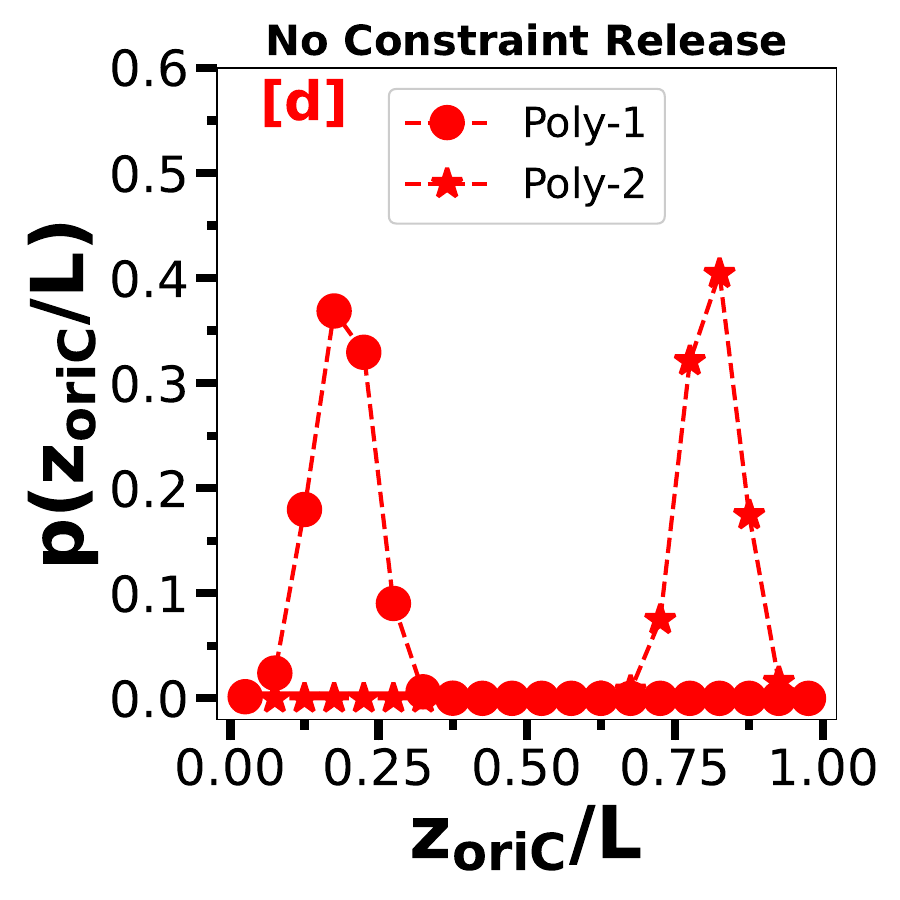}
    \caption{Subfigure (a) shows the probability distribution $p(z_{COM}/L)$ of the centre of mass (COM) of the 
    different loops of a pair of Arc-2 polymers. The position $z$ along the long-axis is normalized by the
    cylinder length $L=42a$. Loop-1 and Loop-2 of each polymer occupy the ends of the cylinder, whereas the 
    COM of Loop-3 occupies distinct regions around the center without overlap. The monomers of different loops
    do partially overlap, and a detailed discussion on the degree of overlap and its relation to different 
    architectures can be found in \cite{dna2}. Subfigure (b) schematically 
    shows the arrangement of the loops as indicated by the previous subfigure. 
    We see that the {\em oriC} is the junction of the loops. In subfigure 
    (c), we see that the {\em oriC} loci ( monomer-1) are also localized close to the quarter positions, as it is at the junction
    between Loops-1 \& 2 and the Loop-3. Subfigure (d) shows {\em oriC} localization in the absence of topological constraint 
    release (TCR).  
    % %For this figure and the next we start out with two   complete polymers (assuming replication is complete).
    }
    \label{fig:Loops_Arranged_Arc2}
\end{figure}

\begin{figure}[hbt!]
    \centering
    \includegraphics[scale = 0.45]{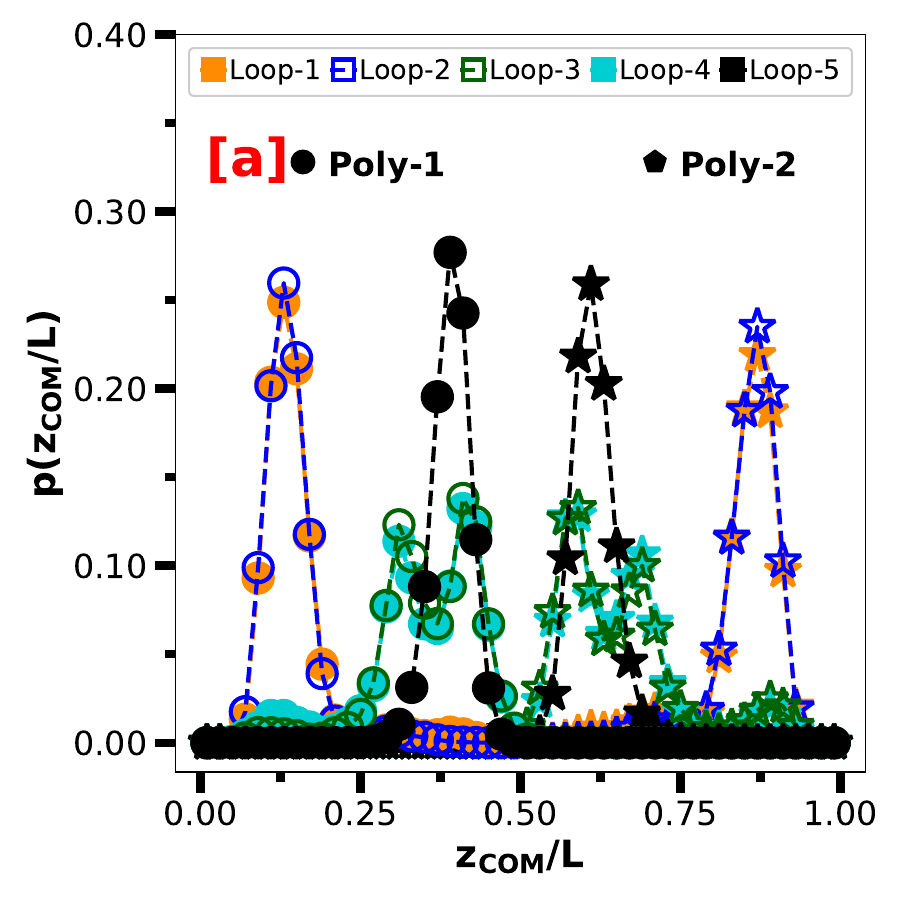}
    \vskip 0.25cm
   \includegraphics[scale = 0.15]{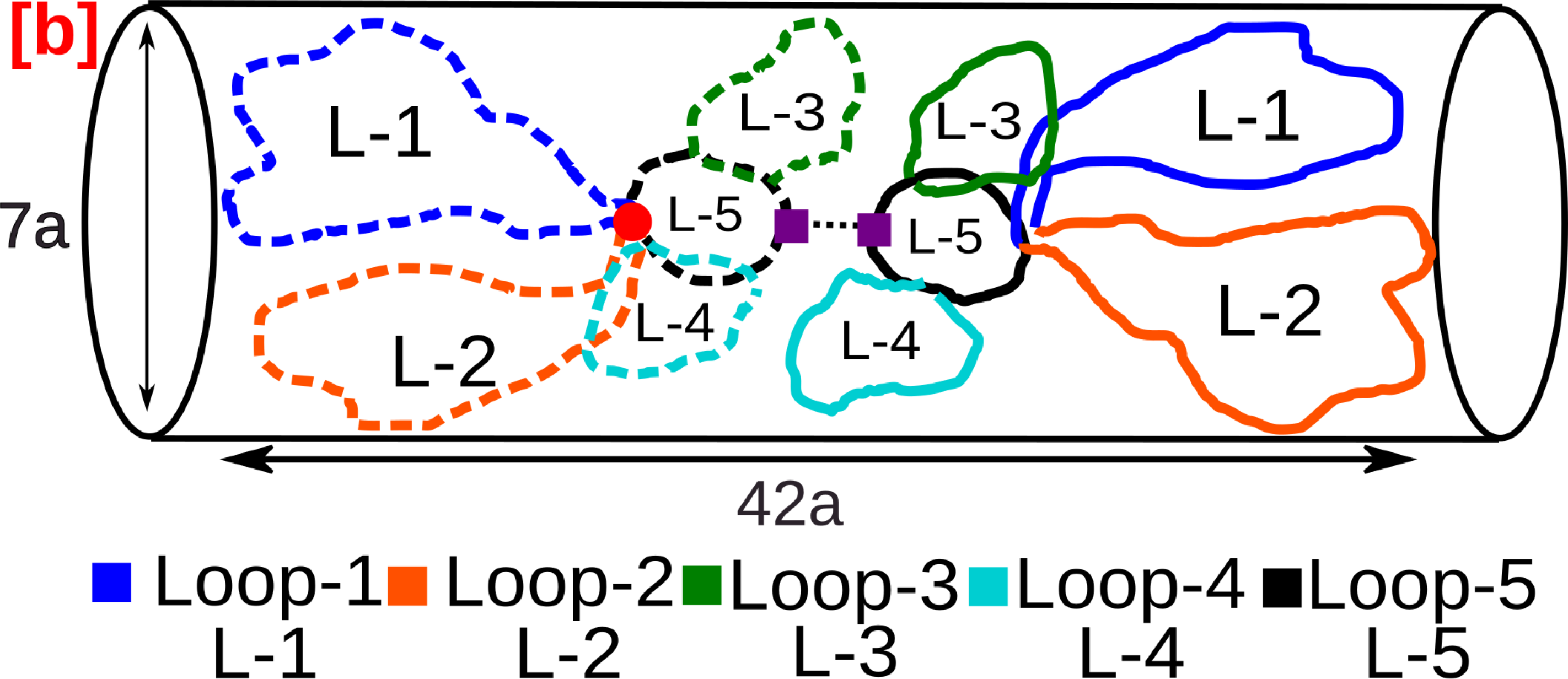}
   \vskip 0.5cm
    \includegraphics[scale = 0.28]{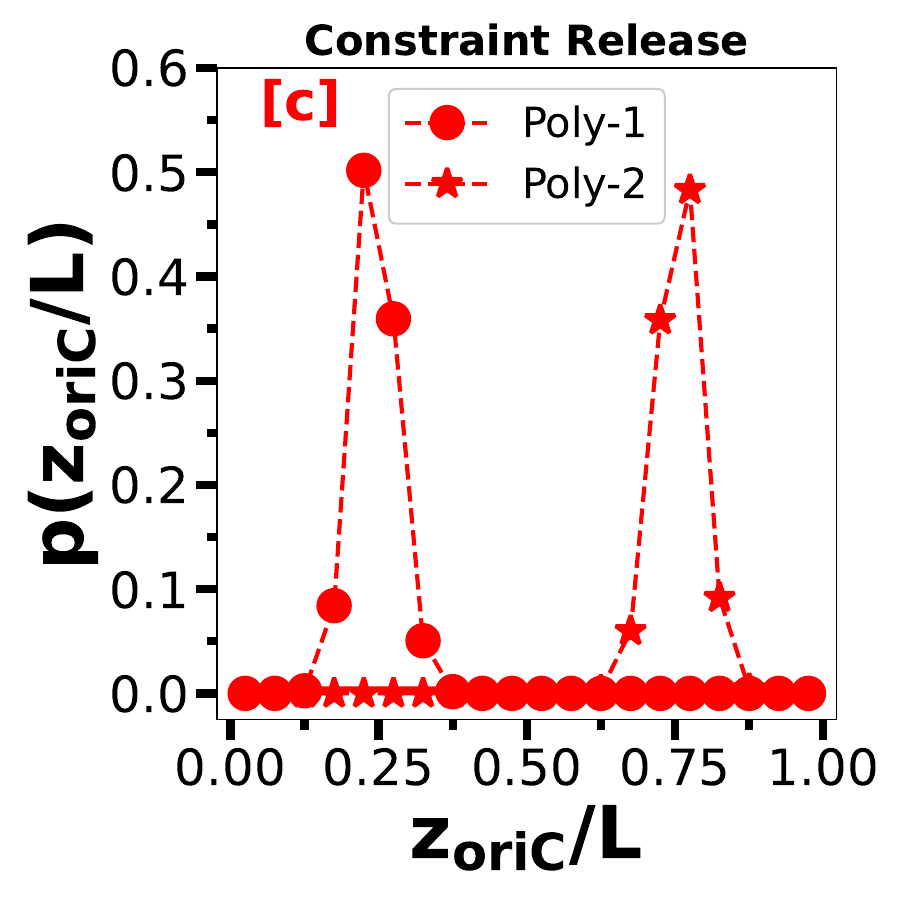}
      \includegraphics[scale = 0.28]{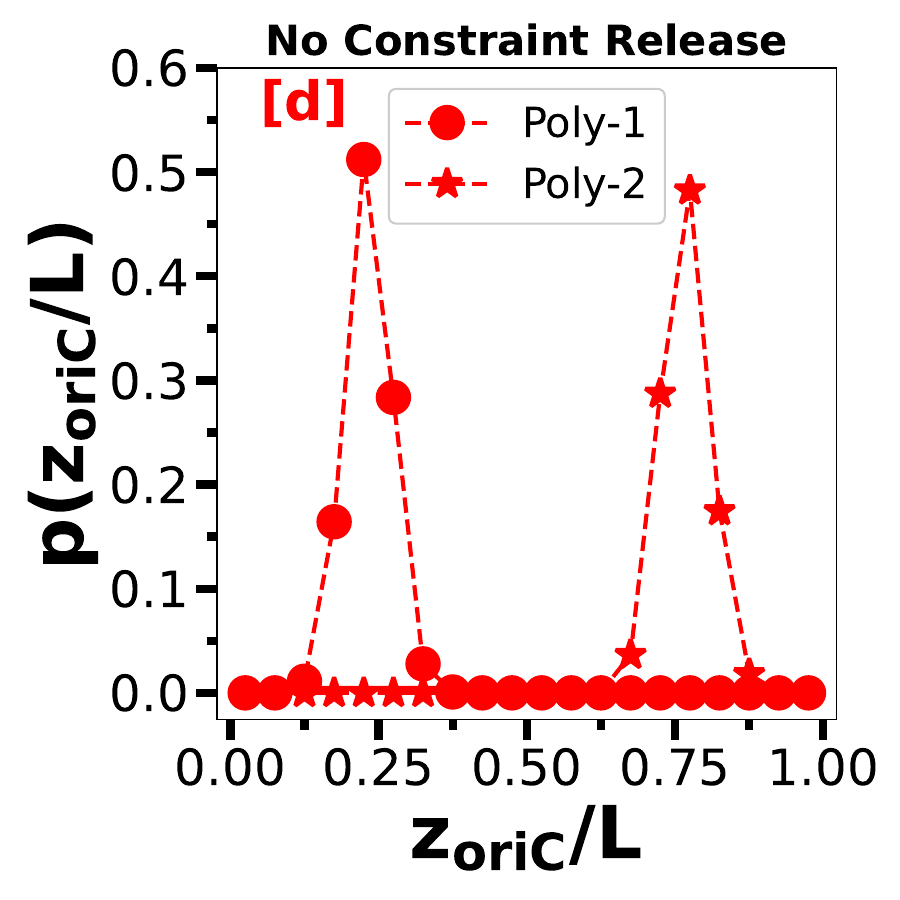}
    
    \caption{Subfigure (a) shows the positional distribution $p(z_{COM}/L)$ of the COMs of different internal loops 
    for a pair of Arc-2-2 polymers along the long-axis $z$. Loops $1,2,3,4$ are defined in Fig.\ref{fig:Arc2_2}, 
    and different colors represent  $p(z_{COM}/L)$ data for COMs of different loops. We label the central
    loop as Loop-5, and its distribution of COM-positions is shown in black.
    Different loops occupy different sections of the cylinder as a consequence of entropic repulsion between loops. 
    The Loops-3 and 4 inter-change positions and remain partially overlapped with loop-5, which shows up the relatively 
    large fluctuations in the peak.  Loop-5 contains the {\em ter}-monomer(s). Thereby Loop-5 is positioned closer 
    to the center of the cylinder due to the presence of ter-crosslink. Subfigure(b) shows a schematic of how the loops are 
    likely arranged in the cylinder. Subfigure(c) and (d)  shows the localisation data of the {\em oriC} when we have TCR and when we switch it 
    off, respectively.}
    \label{fig:Loops_ArrangedArc2-2}
\end{figure}

We refer to the two smaller loops  as Loop-1 (with monomers $1-125$) and Loop-2 (with monomers 
$375-500$). We label the bigger loop as Loop-3 ($125-375$); also refer Fig.\ref{fig:Arc2_2}.
To identify the position of the loops along the cylinder's long axis (z-axis in our simulations), we 
plot the probability distribution of the position of the center of mass (COM).
In Fig.\ref{fig:Loops_Arranged_Arc2}(a), we observe that the peaks of the probability distribution 
$p(z_{COM})$  of the COM of loops belonging to different polymers are well 
separated along the $z$ axis.  Moreover, the two polymers,  referred to as Poly-1 and Poly-2, 
are spatially segregated to two different halves of the cylinder. We also observe 
that the COM of the loops of individual polymers occupy different sections within each half of the 
cylinder. We conclude that the two polymers are arranged in the fashion shown in the schematic
Fig.\ref{fig:Loops_Arranged_Arc2}(b). Probability distribution $p(z_{oriC})$  data for the particular 
monomer $1$, which represents {\em oriC} in our simulations, is presented in Fig.\ref{fig:Loops_Arranged_Arc2}(c). 
As {\em oriC} is the junction of the loops it gets localized to the quarter positions, with distinct peaks 
in  $p(z_{oriC})$.
% We present {\em oriC} localization data when we allow topological constraint release in (c).
In Fig.\ref{fig:Loops_Arranged_Arc2}(c) we show {\em oriC} localization data when we allow topological constraint release. 
We also verified that the position of the peaks in $p(z_{oriC})$ remain unchanged even we 
do not have topological constraint release,  refer Fig.\ref{fig:Loops_Arranged_Arc2}(d). Thus we 
establish that the localization property of loops is also relevant for synthetic polymers, which do not have chain-crossing.

\subsection{Architecture 2-2: Arc-2-2 }
We next present results of organization of loops in cylindrical confinement for a further modified topology,
the Arc-2-2 polymer  with two additional loops near monomer $250$ (the ter-region). For the 
Arc-2-2 polymer, we name the two slightly bigger loops 
as Loop-1 $(1-125)$ and Loop-2 $(375-500)$ similarly the two smaller loops are named Loop-3 $(136-218)$ 
and Loop-4 $(282-364)$, refer Fig.\ref{fig:Arc2_2}. The remaining monomers are clubbed into 
another loop named Loop-5. In Fig.\ref{fig:Loops_ArrangedArc2-2}(a), we plot the probability 
distribution of the position of the COMs of the $5$ loops 
along the long axis of the cylinder. We observe from Fig.\ref{fig:Loops_ArrangedArc2-2}(a)
that the two smaller loops and the two bigger loops occupy different sections
of the cylinder (cell) as each pair of loops repel the other pair entropically. Moreover, 
the monomers of the central loop (in black) occupy the central region of the cylinder. 
Figure \ref{fig:Loops_ArrangedArc2-2}(b) schematically shows how the monomers of 
the different loops could be arranged. As a consequence of this arrangement, the 
{\em oriC} again gets localized in the quarter positions of the cylinder Fig.\ref{fig:Loops_ArrangedArc2-2}(c) 
and Fig.\ref{fig:Loops_ArrangedArc2-2}(d). Another interesting thing to note is that the two smaller 
loops Loop-3 and Loop-4 of both the polymers show two peaks within each half of the cylinder. 
This will be instrumental in understanding how the replication forks also organize in the 
fast-growth simulation in the results section later. 
We have also plotted the distribution of other loci described in Fig.\ref{fig:Loci} in SI-8. In the SI we have explained in detail how the positional distribution of each loci can be explained within the framework discussed above.

In our previous study of DNA-polymer in slow growth conditions \cite{dna1}, DNA-polymers were replicated using 
the same scheme as presented in the model section of this paper, and daughter DNA adopted this architecture as copies 
of the monomer which get cross-linked are created. As a consequence, different loci got localized at different stages of  
replication in our simulation, which corresponds to different stages of the life cycle of the cell.

\subsection{ Other consequences: predictions from Arc-2-2 topology model of the {\em E.coli} chromosome.}
A particular chosen topology (Arc-2-2) not only produced localization of loci as seen as slow growth experiments, but also showed the emergence of domains that are similar to 
macro-domains as seen in Hi-C data for {\em E.coli} \cite{Lioy2018}. Furthermore, we could also resolve other long-standing controversies. 
% consistency with HiC data, and resolved multiple other long standing controversies. 
For example, our simulations agree with the experimental observation that oriCs start moving towards their quarter positions at the stage where 
half the mother DNA has been replicated. This corresponds to a time of roughly 20-25 minutes after replication starts.
Simultaneously the ter loci moves to the mid-cell position: the terminus transition.
Thus, we now understand the mechanism and timing of the ter-transition. 

In addition, our studies using the Arc-2-2 architecture also 
illustrated the mechanism by which the replication forks (RFs) remain localized 
near the center of the cell for most of the replication cycle, despite using the train-track model of replication. 
Thus our model explains the apparent disagreement between the train-track and replication factory model for the 
position of the RFs in the cell. As a consequence of the localization of the replication forks at the center of the cell,
we now understand why the loci split always occurs at the cell center 
for all loci investigated \cite{Cass2016}. Lastly and most importantly, the 
particular {\em position} of these loops along the chain contour helps the polymer significantly increase the rate of 
segregation of the two replicated chromosomes to two different halves of the cell. The importance of internal loops to 
increase the speed of segregation has been independently validated by Molecular dynamics simulations of DNA segregation 
where extrusion is explicitly incorporated in the modeling \cite{harju2023loop}. These  (unexpected) 
predictions and consistencies with other experimental observations, gives us the confidence of adopted particular topology
and our modelling.  All these are discussed in detail in \cite{dna1,dna2}.

As we will see in the later sections of this paper, the same adopted topology provides the mechanism of 
organization of loci for even fast growth experiments, though we did not design the topology keeping the 
fast growth loci-localization data in consideration.  
We do not currently consider the different mechanisms of loop formation e.g. extrusion.
% ( e.g. extrusion which would lead to additional smaller transient loops)  nor do we consider other biological processes which 
% could further aid chromosome organization such as crowders or the formation of the nucleoid.
% NUCLOID OCCUPIES 60\% >>>> WILL LEAD TO ORIC AT QUARTER POSITIONS.
% {\em In vivo}, loops can be formed by  linker proteins, by the phenomenon of extrusion which can create loops with 
% with $10^5$ to $2 \times 10^5$ BPs or when the promoter sequence on the DNA comes close the operator sequence. \cite{amitai2017polymer}.
% These are typically transient loops

% In the last section of this article we further modify the Arc-2-2 topology to add smaller loops to obtain 
% the radial organization of genomic segments as seen {\em in-vivo} for fast-growth conditions. These smaller loops would 
% be a consequence of extrusion, and their positions would vary with time in reality. However, for the sake of simplicity, we do
% not model extrusion thereby, we do not consider loops whose length and position vary with time. 

\section{Results}

Following the principles established in the previous section, we now explain in 
detail how the modified polymer-architecture determines the chromosome organization 
in fast growth. 
We will also plot our simulation data in a manner similar to how the experimental
data was presented in \cite{Youngren2014}.
But FISH experiments have limited resolution, and hence, for certain cases one has difficulty in resolving 
the replicated loci before they are well separated. 
In experiments, it's not possible to identify which foci belongs to which generation, the daughter(D)
or the grand-daughter(GD) chromosome. 
These and other issues (discussed later in this paper)  has the consequences that the plots of the 
experimental  data can be difficult to interpret for a reader.

% \begin{figure}[hbt]
%     \centering
%     \includegraphics[scale = 0.5]{InitPosWithLoopsV2-eps-converted-to.pdf}
%     \caption{ {\color{blue} The figure schematically shows various mother and daughter chromosomes 
%     (bead-spring polymer) with internal loops (topological modifications)   in our simulation at the 
%     beginning of our simulation run. This corresponds to the stage of the cell cycle just after the 
%     birth of the (mother) M-cell (cylinder). Eighty percent of the M-chromosome has been replicated to 
%     two (daughter) D-chromosomes. The two RFs are at the $200$-th and the $300$-th monomers. There are 
%     two copies of the loci $1, 125, 375$ and loci $136, 364$ but only one copy of loci $218, 282$. 
%     These are the loci which create Loop-1, Loop-2, Loop-3 and Loop-4 by CLs, refer Fig.\ref{fig:Arc2_2}.
%     Therefore, in this initial configuration, there are a pair of Loop-1 and Loop-2 on the pair of 
%     daughter chromosomes (polymers), but only one Loop-3 and Loop-4.}
%     }
%     \label{fig:schematic_sim_begins}
% \end{figure}

In simulations, we can identify the number of existing loci and their positions along the long-axis at any
point during the simulation run without ambiguity. Hence, we present the data in a way that
makes it easy to appreciate the underlying mechanism of loci localization. Thereafter, we plot our data 
in the way similar to how experimental data has been presented, incorporating the consequences of the inability to
distinguish two foci, which are spatially close to each other.  This enables a direct
comparison of modelling data in Fig.\ref{long_axis_data}(and other figures) with the experimental data in Fig.\ref{youngren}, which we reproduce from \cite{Youngren2014}
in this paper. Our simulation run models one cell-cycle, and corresponds to the time $t=0$ to $t=\tau$ 
of Fig.\ref{fig:cellcycle}. For the data presented below, we have multiple rounds of replication as 
detailed in the model section.

% \begin{figure*}[hbt]
%    \centering
% \includegraphics[width =0.34\columnwidth,angle=90]{Arc2_2Finalstage.png}\\
%         \includegraphics[width =0.39\columnwidth,angle=90]{Arc0Finalstage.png}
%     \caption{\textbf{Snapshot of simulations for Arc-2-2 and Arc-0(color online)} We show representative simulation 
%     snapshots taken at the last stage of the cell-cycle before cell-division, for Arc-2-2 (top frame) 
%     and Arc-0 (bottom frame), polymer topologies.  The green and red small spheres are the monomers 
%     of the daughter chromosomes. We note that for Arc-2-2 the grand-daughter chromosomes (shown in orange,
%     violet, blue and deep-green) get spatially localized to specific regions along the long axis, while 
%     for Arc-0 the grand-daughter chromosomes show greater spatial overlaps with each other. The four big
%     red spheres represent the four {\em oriCs}, whereas the two big orange spheres are the 
%     two {\em dif-ter} loci, they are connected to each other as is seen {\em in-vivo}. 
%     % Note that they
%     % are of the same size as other monomers in the simulation. The sizes of the monomers of the daughter 
%     % chromosomes(red and light green in colour) have been reduced for aid of visualization. 
%     Note that all the monomers in the simulation have the same size. The sizes have been changed here for aid of visualization. 
%     In the subsequent
%     sections we show that loci show localization patterns for Arc-2-2 but Arc-0 do not show such 
%     localization patterns.}
%     \label{Comparison}
% \end{figure*}

{\em Initialization:} As mentioned earlier, we start the simulations with the birth of a new
M-cell corresponding to $t=0$ of Fig.\ref{fig:cellcycle}. Before we start our simulation , 
we allow the two D-chromosomes to relax over $3 \times 10^7$ MCS, where they can take different 
independent conformations.  In this initialization process, the two D-chromosomes remain connected at the 
two RFs on the M-chromosome assuming completion of $80\%$
replication of both arms. The RFs are located at monomer indices $200$ and $300$, equidistant 
from the {\em dif-ter} loci at the $250$-th monomer.  Refer to
figure in SI-9 for a schematic to follow the inital configuration.
{\em In-vivo}, up to just before cell-division and birth of the M-cell,
the two replicated {\em dif-ter} loci remain linked to each other in the parent cell, and are 
localized in the middle of the parent cell. Thus, in a 
newly born cell one expects the ter-loci to be near the new pole, which is formed in the middle of the 
parent (M-cell). To emulate this, we keep {\em dif-ter}(the $250$-th monomer) tethered to one of the poles 
of the cylinder (corresponding to the new pole of the cell) during initialization, and released the tether 
just at the start of the simulation. We allow chain crossing by TCR. Thus at the beginning of the 
simulation, we have a polymer architecture that has the CLs that create Loop-1 and Loop-2 in each of 
the D-chromosomes, which we name D1 and D2. The mother M-chromosome has the CLs, which results in Loop-3 
and Loop-4; refer Fig.\ref{fig:Arc2_2} and SI-4 for other details of the initialization process.

{
% In the previous section, we study the positions of particular monomers 
% along the chain contour corresponding to particular loci specified in the experimental work of \cite{Youngren2014}. 
% Spatio-temporal localization of monomers from our calculations 
% matched well with FISH results of corresponding loci. REVIEW IS WITH FG LOCI

 {\em Segregation \& chromosome organization}: We present a snapshot from our 
simulations in SI-10 to show that Arc-2-2 DNA-polymers from different rounds of replication 
remain well segregated along the cylinder long axis at the end of the cell cycle. The snapshot 
corresponds to a configuration at the end of the simulation run when $80\%$ of replication 
of each of the D-chromosomes to $2$ GD-chromosome
is complete, and the cylinder has grown from $21a$ to $42a$.
For comparison, we show a snapshot from the end of the simulation run with Arc-0 polymers, i.e. polymers 
with the unmodified ring-polymer topology, and we see that DNA-polymers from different rounds of replication 
remain relatively more mixed as compared to the previous case.

Thereafter, we follow the positions of the same monomers that correspond to loci tagged in 
\cite{Youngren2014} as simulation proceeds, also refer Fig.\ref{fig:Loci}. As the simulation proceeds, 
the RF moves along the two arms of the chromosome. The position of the RF and amount 
of replicated DNA in the simulations is used to mark the evolution of the cell cycle and determine 
the different stages of the life cycle of the cell, even as multiple rounds of replication
are in progress.   We quantify loci organization by plotting their spatial distribution in 
{\em five equally divided intervals} of their life cycle as in experiments, refer 
Fig.\ref{long_axis_data}. 
% The simulation data can be directly compared to experimental data from \cite{Youngren2014},  and the relevant experimental figures has been reproduced
% from the original paper for ease of direct comparison in Fig.\ref{youngren}.
We average our data over $50$ independent runs corresponding to a life cycle in $50$ cells.

\begin{figure*}[ht]
\includegraphics[width=2\columnwidth]{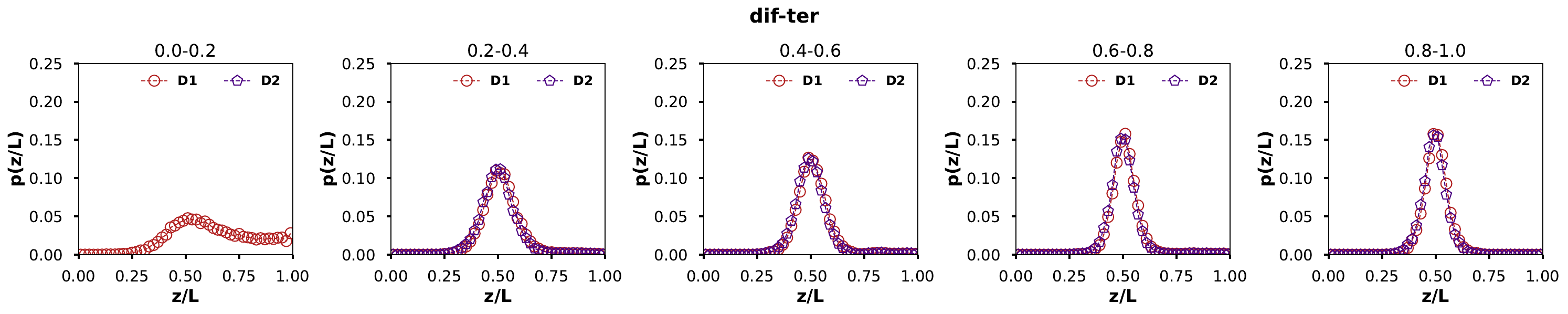}
\includegraphics[width=2\columnwidth]{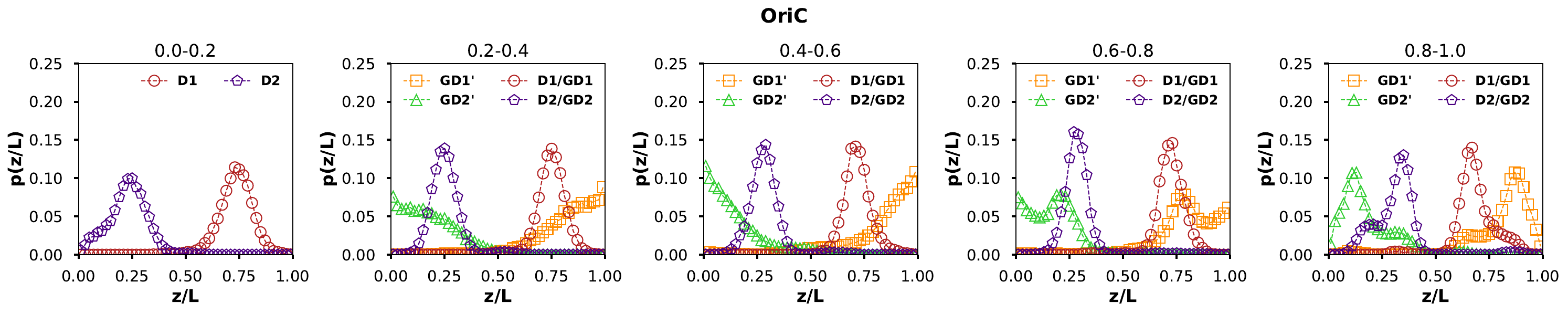}
\includegraphics[width = 2\columnwidth]{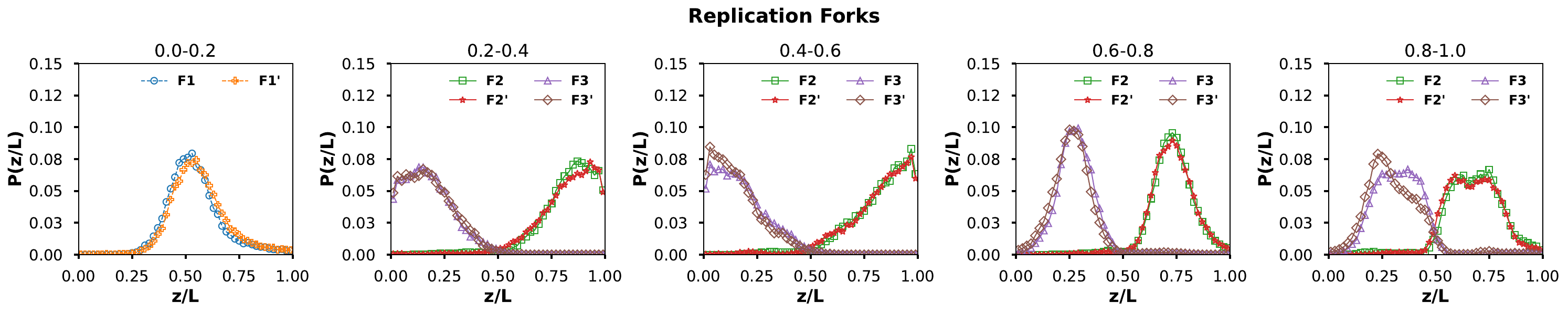}
\caption{\textbf{Long axis distributions for {\em dif-ter}, {\em oriC} and the replication forks:}
We plot the spatial probability distributions, $p(z/L)$ of the position of different loci, 
where $z$ denotes the along the long axis of the cylinder (cell), and $L$ is the length of the cylinder
at that stage of the simulation run. Data is shown for {\em dif-ter} locus (first row), {\em oriC} locus (second row)
and the RFs (third row) for various intervals, as indicated at the top of each subfigure, during the life cycle. 
D1 chromosome on getting replicated creates GD1 and GD1' chromosomes. 
Similarly, one obtains GD2 and GD2' from the replication of D2 chromosome. 
Once a particular locus gets replicated, 
then the corresponding D monomer is renamed as the GD monomer (eg: D1 is renamed to GD1). 
% New monomers 
% are introduced due to the replication protocol (and labelled GD1'). 
% The {\em dif-ter} locus gets replicated once.
% In the stage corresponding to
% $0-0.2$ we have only the  D1 {\em dif-ter} locus. In the next stage, the D1 locus has been replicated and there
% are two overlapping distributions corresponding to D1 and D2 (which remain cross-linked).  The {\em oriC} locus of D1 and D2 
% gets replicated at the start of the $(0.2-0.4)$ interval and hence the plots show four different data sets. 
The localization of the {\em oriC}s and the {\em dif-ter}s can be visualized from our representative simulation video `Vid-3' (refer SI-1).
In the third row we plot the spatial distributions of the RFs as they are assumed to move from one monomer to 
the next, along the chain contour.
In the $(0-0.2)$ interval there only two RFs which move along the contour
of the M-chromosome (shown in orange(plus marker) and blue(circle marker)). Thereafter, one has four RFs branching out from the two 
{\em oriC}s of the D-chromosomes, which get replicated to form the GD chromosomes. Note that GD1 and GD1'
chromosomes occupy one-half of the cell while GD2 and GD2' occupy the other half. The RFs have been named in a specific way to clearly demarcate those which are traversing along different arms of the same chromosome. For instance, F1 and F1' are the two replication forks moving along the two arms of the mother chromosome, while F2 and F2' denote the RFs moving along D1 and corresponding F3 and F3' moving along D2. }
\label{long_axis_data}
\end{figure*}

\begin{figure*}
    \centering
    \includegraphics[width = 1.5\columnwidth]{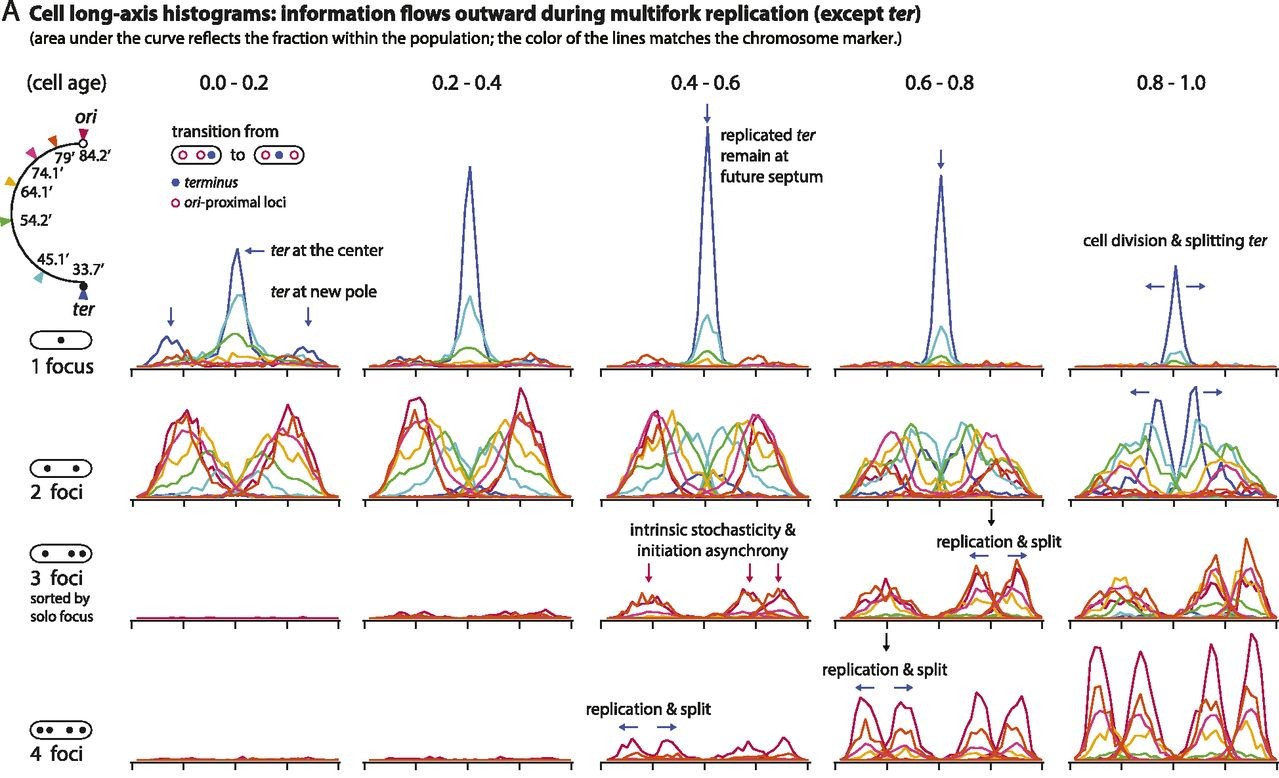}
    \includegraphics[scale = 1.2]{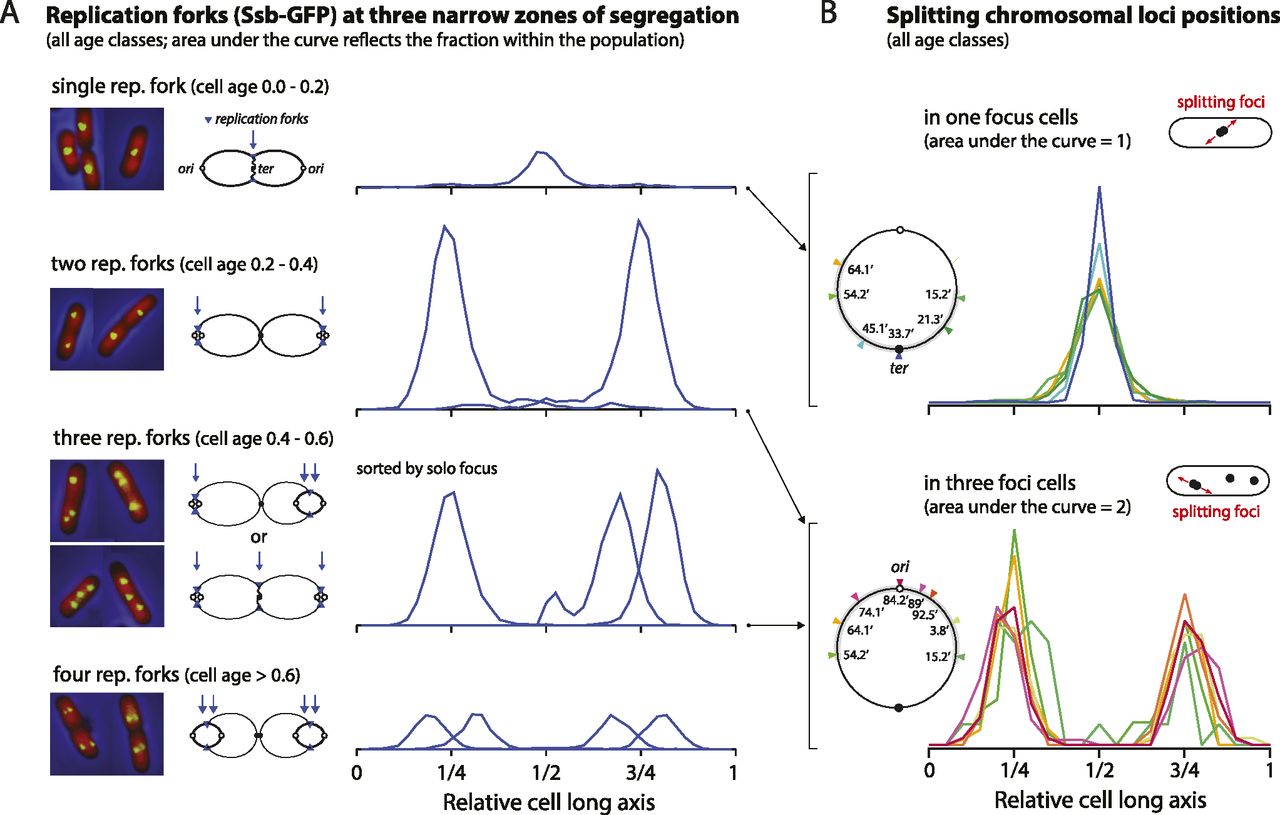}
    \caption{\textbf{Experimental data of loci positions during cell-cycle:} This figure 
    has been reproduced from previously published data in \cite{Youngren2014}.{We reproduce  
    two figures:  
    Fig.2 and Fig.3 respectively} from the paper of \cite{Youngren2014},  (after having
    obtained requisite permissions) for aid of comparison with our modeling results, presented 
    in Fig.\ref{long_axis_data}, Fig.\ref{fig:oric_exp}, Fig.\ref{fig:ExpForks}, Fig.\ref{fig:LociSplit} and Fig.\ref{long_axis_data1}. 
    The top panel with 4 rows shows spatial distributions for different fluorescently tagged loci 
    at different stages of the cell cycle. This data has been extracted from a 
    large number of cells at different ages, each with a different number of foci.
     % This data has been extracted from a large number of cells at different stages of their life cycle, and each with different number of foci.
     The positional distribution of different tagged loci along the long axis is 
     plotted for an ensemble of cells with $1, 2, 3$ or $4$ foci in the four different 
     rows of the figure, respectively.
    The figure further shows how the positional distributions of foci change as the cell ages. 
    This is shown in the 5 columns. Note 
    % As the cell ages, the total number of foci in the cell can change due to replication. 
    % Cells at the beginning of the life cycle do not typically have more than two foci.
    % Other details can be found in the text and in \cite{Youngren2014}. 
    At the bottom 
    panel of the figure, we show the positional distribution of the replication forks 
    (bottom left). The replication forks in different cells show different distributions of its 
    positions, indicating different positions w.r.t to each other at different intervals
    of the cell cycle. Using our simulation and modeling, we establish that this is a consequence
    of the internal topology of the ring polymer in Fig.\ref{long_axis_data} third row. In 
    bottom right of this figure, we show the experimental plot which marks the position 
    along the long-axis at which one foci-spot split into form two distinct foci.  }
    \label{youngren}
\end{figure*}

{\em dif-ter localization:} Soon after the simulation starts, the {\em dif-ter} monomer (i.e., the $250$-th 
monomer) will be near the end of the cylinder where it had been tethered during initialization.
Thus, the probability distribution shows non-zero values near one of the ends of the cylinder in the 
first stage $(0-0.2)$ interval of its life cycle; refer to top row of Fig.\ref{long_axis_data}.
But even as the RF-(s) move from monomer $200$ (and $300$) towards the {\em dif-ter} loci and cross 
the $218$-th monomer (and $282$-th monomer in the other arm), 
we introduce  CLs between the newly introduced $218$-th monomer and the previously replicated 
$136$-th monomer to form a new 
Loop-3. Correspondingly, we introduce CLs between newly replicated $282$-nd and previously replicated
$364$-th monomers to create a new Loop-4 for the  D2-chromosomes. Refer to SI-11 for the 
schematic diagram to better understand how we implement the cross-linking in our simulations.

The entropic repulsion between the loops of the two D-chromosomes ensures the segregation of the two
polymers into two halves of the cylinder. This also relocates the {\em dif-ter} loci to the middle of the 
elongating cylinder. As a consequence, we also observe a peak for the probability distribution of 
{\em dif-ter}  at the middle of the cylinder at all stages of the life-cycle, refer to the first 
row of Fig.\ref{long_axis_data}.
As the {\em dif-ter} monomer is replicated, the two copies are connected by a spring, as the 
{\em dif-ters} are connected {\em in-vivo} upto cell-division. We call this the ter-CL. 
The peak remains unchanged in subsequent intervals as the ter-CL 
maintains the position of the two 
{\em dif-ters}. For reference, we have provided 
experimental plots from \cite{Youngren2014} in Fig.\ref{youngren}. 
In the $(0.8-1)$ interval in Fig.\ref{long_axis_data},
we don't see two peaks in the {\em dif-ter} distribution as we don't model the breaking of 
the ter-CL and consequent separation of {\em dif-ter} loci before cell division; in contrast to the two-foci {\em dif-ter} data in Fig.\ref{youngren}.

% In the $(0.8-1)$ interval,
% we don't see two peaks in the {\em dif-ter} distribution as we don't model the breaking of 
% the ter-CL and consequent separation of {\em dif-ter} loci before cell division;
% compare the data in Fig.\ref{long_axis_data} with the two-foci {\em dif-ter} data in Fig.\ref{youngren}.

{\em oriC localization in initial stages of life cycle:} As the two  D-chromosomes occupy two different halves of the growing cylinder, 
the {\em oriC}-monomer moves to the quarter positions of the cylinder, 
in the  $(0-0.2)$  and $(0.2-0.4)$ intervals of the cell life cycle, as seen {\em in-vivo}. This is due to the 
repulsion between internal loops within each D-chromosome  as outlined in the previous section. 
This organization emerges spontaneously in our simulations as a consequence of the adoption of the 
Arc-2-2 polymer topology, as shown in the second row of Fig.\ref{long_axis_data}.

% {\bf Comparing data from  experiments \& simulations:} Since the two D-chromosomes are connected at the {\em dif-ter}, the {\em dif-ter} continues to remain 
% in the cylinder center. This continues  even when  the next  round of  replication starts at the two
% {\em oriC}s of the D-chromosomes. When comparing our results to that of experiments, there are some 
% caveats related to conventions of presentation of data which need to be accounted for.  Sometimes,
% these bring up apparent differences between the data obtained from experiments 
% and our simulations, the reasons for which  has been explained in subsequent paragraphs.
% For example, we don't see two peaks in the {\em dif-ter} distribution as we don't model cell division:
% compare the data in Fig.\ref{long_axis_data} with the two-foci data in Fig.\ref{youngren} for $(0.8-1)$
% interval.

We have two {\em oriCs} in the $(0-0.2)$ interval in our simulations of the cell life cycle. 
At the beginning of the $(0.2-0.4)$ interval, the D1 and D2 {\em oriC} replicates. According to our
model of replication, the D1 {\em oriC} is re-named GD1
and another monomer is added, which we call GD1'. Similarly, for the D2 oriC, D2 is re-named to GD2,
and another monomer, which we call GD2, is added.

Thus from the $(0.2-0.4)$ interval onwards, our simulations have four {\em oriCs}. We can track 
the four oriCs independently and plot their spatial probability distributions in the last four 
intervals of the cell cycle. However, in experiments, the two just replicated {\em oriCs} cannot
be distinguished in FISH data in the $(0.2-0.4)$ interval,  since the newly replicated  
{\em oriCs} segregate after a certain interval of time, known as the cohesion time. Hence, for 
experimental data shown in Fig.\ref{youngren}, the spatial distribution of four {\em oriC}s do not appear 
till the $(0.4-0.6)$ interval of the life cycle.  Consistent with the observation of cohesion time,  
the distributions of the {\em oriC}s overlap significantly in our simulations in the $(0.2-0.6)$ 
interval of the life-cycle. This overlap decreases as the {\em oriC}s localize to new positions in 
later intervals.

%The third reason for observation of three oriCs is that the two of the oriCs may lie vertically on 
%top of each other along the line of sight through the microscope. 

In the simulation data, we observe that two of the four GD-{\em oriC}s show relatively high values 
in the spatial distribution near the cylinder ends for the three intervals corresponding to $(0.2-0.8)$ 
intervals of the life cycle.
This is not seen in experimental data. This is 
because {\em in-vivo}  the bacterial chromosome is condensed to stay in a region called the nucleoid
within a sphero-cylinder due to crowders or other mechanisms e.g. presence of
spherical end of cell \cite{Wu2019}.
These effects, which could move loci distribution away from the poles,
have not been incorporated in the current study. 
Moreover, in experimental plots data  presented in rows $2$ and $3$ of Fig.\ref{youngren}, we do observe 
rather broad oriC distributions for the $2$ and $3$-foci data in the $(0.4-0.8)$ interval of cell-cycle;
and even wider distributions in $(0.8-1)$ interval for data with two foci, i.e. when the two loci 
cannot be distinguished.  Thus, our results are consistent with those obtained from experiments.

{\em Absence of 3 oriCs in the plots:} In Fig.\ref{long_axis_data}, 
we do not have a scenario with three {\em oriC} foci. But {\em in-vivo}, one can observe
$3$ {\em oriC} foci in the cell because the replication starts need not be
perfectly synchronous (as in simulations). Moreover, segregation of the newly
replicated {\em oriCs} may proceed at slightly different rates due to inherent
stochasticity.
% , which may govern cross-linking by binding proteins. 
This can give rise to different cohesion times for different pairs of
replicated {\em oriC}s, which can result in the observation of three {\em oriC}s. 
 This difference, and the fact that in experiments, one cannot
distinguish between the two to identify whether a locus belongs to D or 
GD-chromosomes also hold true for the data of the spatial distribution
of other loci, as presented later. We observe $3$ peaks of the oriC when we
plot our data in Fig.\ref{fig:oric_exp} and SI-12, similar to how the experimental data is plotted, to enable direct
comparison with experiments. In experiments, one can only track the number of
cells with two, three, or four fluorescent foci and plot the spatial distributions.

In Fig.\ref{fig:oric_exp} (and in figures shown in SI-12), we have plotted the distribution 
of oriC (other loci) using the convention that if the distance along the long axis, between 
two loci is less than a cutoff $a_c$, the two loci will be counted as one focus. We do this to 
have a more direct comparison to the experimental data. This method allows us to easily have 
the positional distribution of {\em oriC} data with two, three, or $4$ foci. As seen in 
experimental data for fast growth conditions, one never has a situation when there is only one {\em oriC}
because the cell is born with two copies of the {\em oriC}. We have also started with such a 
configuration. In Fig.\ref{fig:oric_exp}, we have used the above convention to sort 
the positional distribution data into sets corresponding to the number of foci, as is done in experiments. Although multiple copies of a particular 
locus might exist, they will not be distinguished as separate foci if spatially close. We have chosen $a_c = 2a \approx 0.1L $ for data shown in Fig.\ref{fig:oric_exp}. 
For a different choice of $a_c =a \approx 0.05L$), data is shown in SI-12. 
Although this allows 
for a direct comparison, we lose out on the resolution of our simulation. 

%   In Fig.\ref{fig:oric_exp}, in the interval (0.2-0.4), we observe that the rather quick separation of {\em oriC}s in simulations contrasts with the experimental data, 
% where one can primarily distinguish positions of only $2$ foci in this particular interval, 
% though there is likely to be $4$ {\em oriC}s in the cell since the beginning of the interval. 
In Fig.\ref{fig:oric_exp}, in the interval (0.2-0.4), we observe that the rather quick separation of {\em oriC}s in simulations contrasts with the experimental data.
In the experimental data, one can primarily distinguish positions of only $2$ foci in this particular interval, 
though there is likely to be $4$ {\em oriC}s in the cell since the beginning of the interval. 
As the GD1' monomers get added due to replication, the GD1' chain remains connected at the RF's to 
the D1 chain. This ensures that the two {\em oriC}s of GD1 and GD1' are spatially proximal post replication; 
this contributes to the $2$ foci data in the plot. As the length of the GD1' chromosome increases the 
entropic repulsion between the existing Loop-1 and Loop-2  and the GD1' strand increases. This pushes the 
{\em oriC} of GD1' towards the cell pole, giving rise to the $4$ foci distributions in the interval $(0.2-0.6)$ 
of the life cycle. A reminder to the reader is that the monomers which get cross-linked to form 
Loop-1 and Loop-2 for GD1' chromosome do not replicate until the middle of the $(0.6-0.8)$ interval.
Also, {\em in-vivo} the chromosome occupies only $60\%$ of the volume of the cell. If we incorporate
this aspect in our simulation, one could expect these distributions to be shifted closer to the 
centre. This would also lead to low probabilities of occupation at the cylinder ends, 
as seen from our calculations presented in the third row of Fig.\ref{fig:oric_exp}.
In the $(0.6-0.8)$ interval, we mostly see $4$ segregated foci, but we also see a contribution 
to probability distribution corresponding to data when only $2$ foci are seen. This is  
because the  $125$-th  monomer (and $375$-th monomers) of the D1 and D2 chain get replicated, 
and two additional cross-links between these and the {\em oriC}s of GD1' and GD2' are created.
This brings the pair of {\em oriC}s close to the quarter positions in each half of the cylinder. 
In the last interval of the cell cycle $(0.8-1)$, there are mostly only $4$ peaks, as seen in 
the plots above, which match well with experimental data. The mutual repulsion between the 
Loop-1 and Loop-2 of GD1 and GD1', and similarly the GD2 and GD2', keep all {\em oriC}s well separated.

\begin{figure*}[ht]
    \centering
    \includegraphics[scale = 0.4]{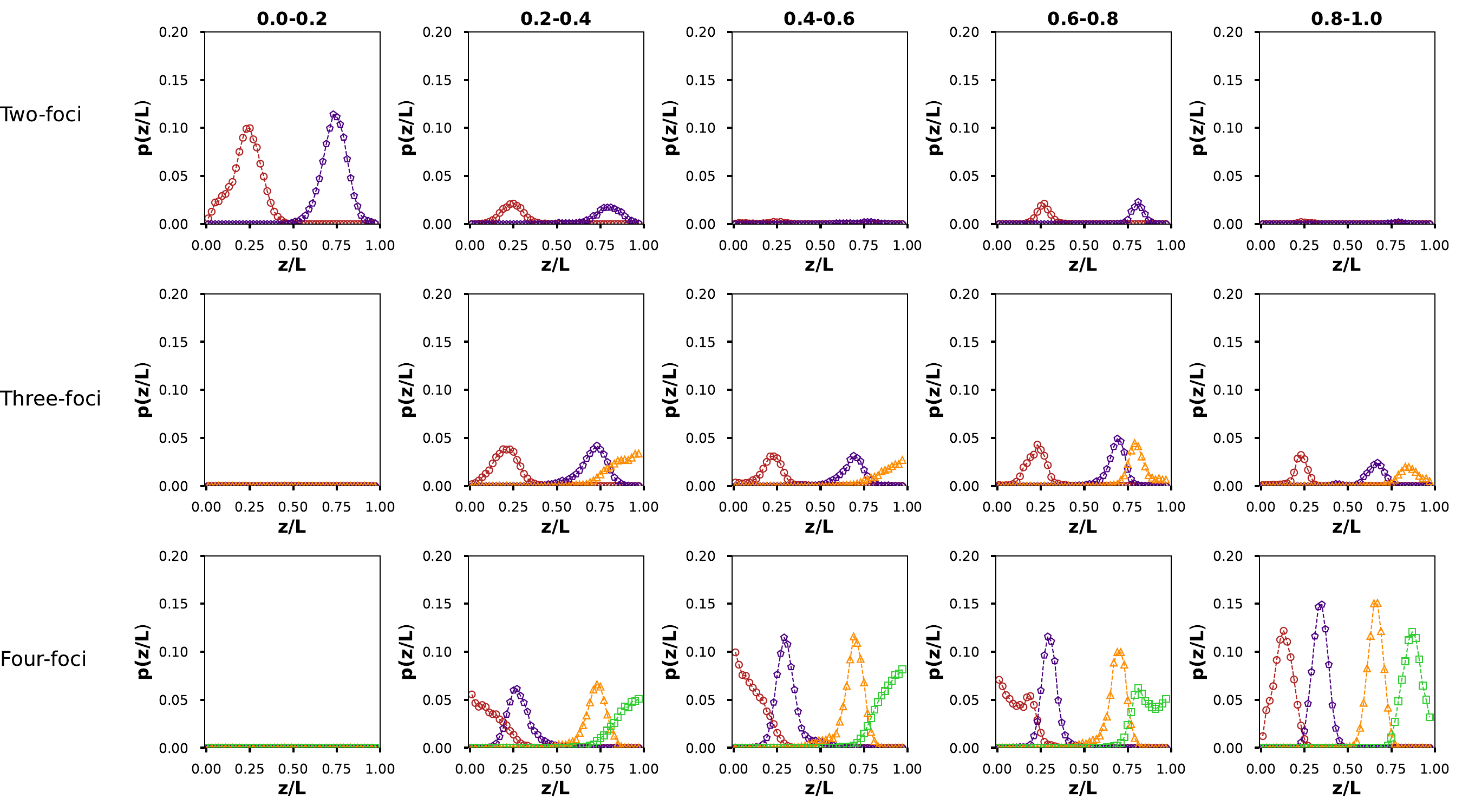}
    \caption{In the figure, we have plotted the positional
    distribution of {\em oriC} loci with a convention similar to that used for
    plotting data from experiments. If two loci cannot be
    distinguished due to spatial proximity, the two loci appear as one
    focus under the microscope. In steady state fast growth, one does  
    not have a cell with a single focus for {\em oriC}, and hence we show 
    spatial distribution data for only $2$ foci, $3$ foci and $4$ foci
    cases in the different rows of the figure. The different columns refer 
    to the different intervals in the cell's life cycle.
    We see that in
    the interval $(0-0.2)$, the two {\em oriC} loci are localized to the
    quarter positions, as can be deduced from the peaks in the probability distributions. 
    In the second interval i.e., $(0.2-0.4)$,
    there are $4$ oriCs, and hence, in simulations, there are four distinct
    peaks indicating $4$ distinct foci as seen in the third row. We also 
    plot two foci peaks corresponding to the set of configurations, 
    where the distance between {\em oriC}s is less than $2a$. The sum of probability distribution in individual subfigures is not $1$, but the sum of probabilities 
summed over subfigures in a column will be $1$. }
    \label{fig:oric_exp}
\end{figure*}

%Extra text from the above figure
% In simulations two of the
% four newly formed GD chromosomes, one on each side of the cylinder, do 
% not have the internal Loop-1 and Loop-2 till the middle of the interval $0.6-0.8$. We refer to them as GD1' and GD2'. This absence of internal loops leads to relatively 
%  slow separation of oriCs due to lack of sufficient entropic repulsion. Thus for interval $0.2-0.4$ we also get probability distribution for 3-foci and 2 foci case. 
% As more
% monomers of the GD's are formed, the two GD1' and GD2' oriC's get pushed towards cell ends to
% avoid overlap with Loop-1 and Loop-2 of GD1 and GD2. This is seen in the time period $0.4-0.6$. 

\begin{figure*}[ht]
    \includegraphics[scale = 0.4]{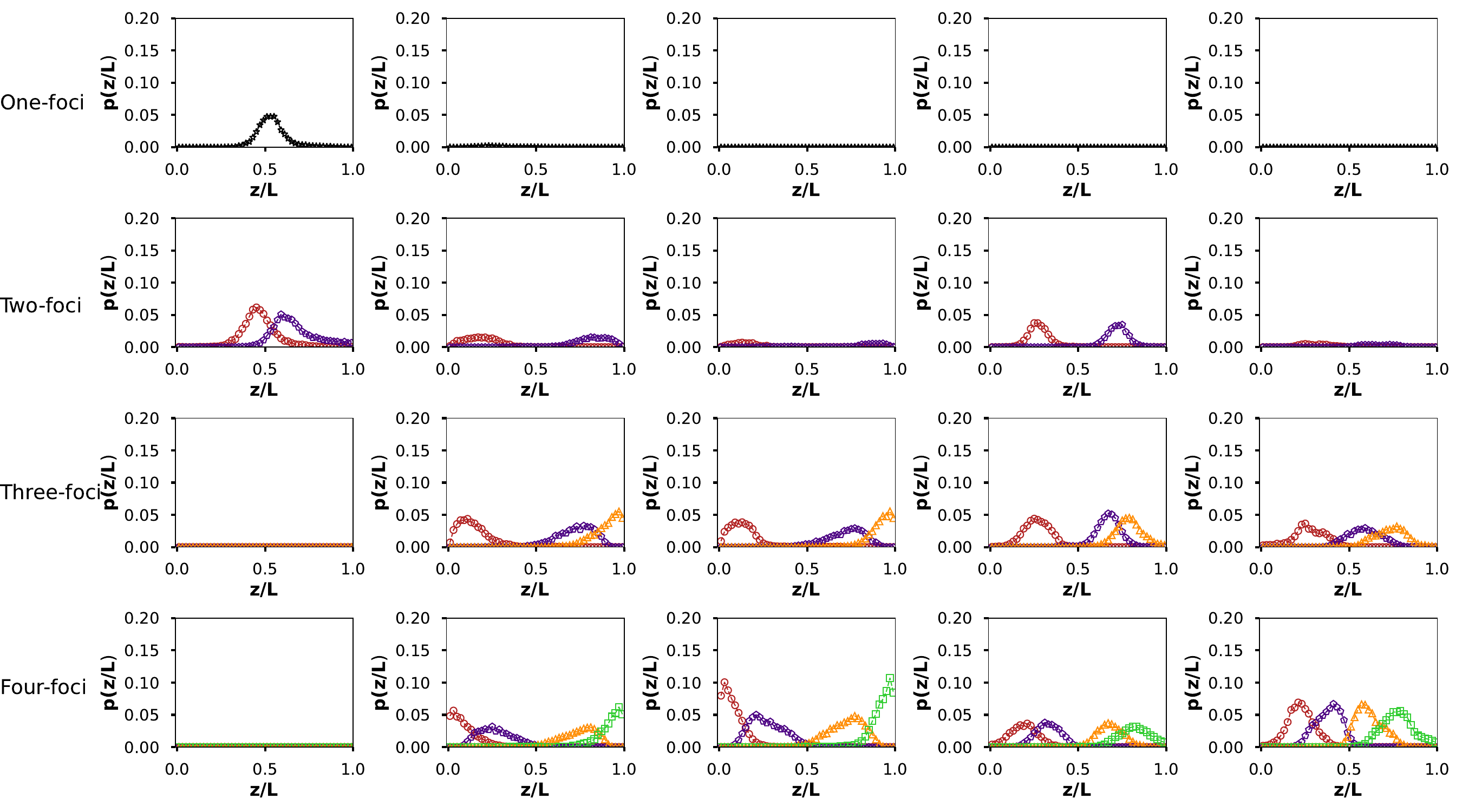}
\caption{  The probability distribution of the spatial position of the replication forks (RFs) at 
different intervals of the life cycle. The plot follows the conventions of Fig.\ref{fig:oric_exp}
which enables direct comparison with the RF positions observed experimentally and presented in Fig.\ref{youngren}. In interval $(0-0.2)$, the two RFs of the D1 and D2 chromosomes can be distinguished, and hence their spatial distributions is shown in the two-foci data: second row.
However, towards the end of the interval as the two RFs reach close to the {\em dif-ter} loci,
the spatial separation is often less than $2a$, and for those cases, spatial distribution of 
the two RFs is shown in the one-foci row as a single peak near the cylinder center, where 
the {\em dif-ter} is located. In the intervals $(0.2-0.6)$, the pair of RFs in each half of the 
cylinder, can be distinguished at times and hence RFs show up their distributions in the 
two-foci, three-foci and four foci rows. In the middle of the $(0.6-0.8)$ interval, the RFs are 
spatially close to each other near the quarter positions, as they replicate the $125$-th and 
the $375$-th monomer and start moving towards the {\em dif-ter}. Depending on the distance between
the RFs in different microstates, the spatial distribution can contribute to the two-foci,
three-foci and four-foci rows. In the last $(0.8-1)$ interval, the RFs can mostly be distinguished
due to the repulsion between internal loops in the GD chromosomes, and thus one sees contribution 
in from $4$ foci for the $4$ loci in the last row. The data is suitably normalized depending on the 
number of configurations for the four different foci-cases in a particular interval.}
    \label{fig:ExpForks}
\end{figure*}

{\em oriC localization in the later stages of life cycle:}
In simulations, as the RFs move from monomer $1$ ({\em oriC}) towards
monomers $125$ and $375$, the monomers belonging to D1 and D2 chains get re-named to GD1 and GD2 monomers.
In addition, monomers are added to create the GD1' and GD2' chains, whose lengths keep increasing as RFs keep moving 
away from oriCs. The CLs between the $125$-th and $375$-th monomer and the {\em oriC} of D1 (and also for D2)
remain present, as  D1 gets converted to GD1 chromosomes (and similarly from D2 to GD2 chromosome) in the $(0.2-0.4)$ and $(0.4-0.6)$ intervals.
We refer to them as GD1 and GD2 only for ease of communication; otherwise, there is nothing to 
distinguish between GD1 and GD1' in our simulations.    
Additional CLs are introduced to create Loop-1 and Loop-2 in the GD1' and GD2' chains 
in the middle of $(0.6-0.8)$ interval. It is only then that the additional $125'$ and $375'$ monomers 
get introduced in the cylinder, i.e. after the RF's pass these points on the chain contour of 
the D-chromosmes.  
Thereafter, one has four Loop-1 and four Loop-2, one pair on each of the four GD-chains. 
Thus from the middle of $(0.6-0.8)$ interval of the life cycle,  the two {\em oriCs} belonging to GD1 and GD1'
start to occupy the $1/8$  and $3/8$-th positions in one half of the cell. 
Correspondingly, the other two {\em oriC}s (belong to GD2 and GD2', respectively)  
occupy $5/8$-th and $7/8$-th  positions in the other half of the cell, leading 
to the appearance of four peaks at these position distribution plots for $(0.8-1)$ interval in 
Fig.\ref{long_axis_data} and Fig.\ref{fig:oric_exp}. This is seen  {\em in-vivo} and in our simulations, and 
the localization occurs due to entropic repulsion between GD-loops. The peaks 
are enhanced in the last interval of the life cycle. 

% In addition, consistent to our claim about the effect of transient loops ,  
% once Loops1,2 are formed we observe that 
% the {\em oriC} avoids the poles of the cylinder. 

{\em Position of Replications Forks (RFs)} 
The confidence in our model is further strengthened by the reconciliation of the spatial distribution of the RFs
from our model and {\em in-vivo} results. 
We have access to the spatial coordinates of the monomers on which the RFs are located
throughout our simulation, corresponding to a cell's life cycle. 
% In our simulation, we have access to the spatial coordinates of the monomers on which the RFs are located
% at different stages of the simulation corresponding to the cell's life cycle; 
% We have all the positional information of the position of the 
% model RFs as we know the position of the RFs on the chain contour at different stages of the 
% cell's life cycle; 
refer $3-$rd row of Fig.\ref{long_axis_data}. In the $(0-0.2)$ interval, the two RFs are situated on the M-chromosome arms, 
and move from the $200$-th (and the $300$-th) monomer to the {\em dif-ter} position. 
The cell center gets occupied also by the {\em dif-ter} loci,
soon after cell division due to reasons already explained previously. 
Thereby, there is a peak 
in the spatial distribution $p(z/L)$ of RFs at the center of the cell. 
At the end of the $(0-0.2)$ interval, the replication of 
the M-chromosome is complete and one has two complete D-chromosomes connected at the {\em dif-ter} in the Arc-2-2 
architecture. 
The Arc-2-2 architecture ensures that the {\em oriC} are in the quarter positions at this stage of the life cycle. 
At the start of $(0.2-0.4)$ interval, replication of the two D-chromosomes begins from the two {\em oriC}s. 
Thus, four new RFs start at the position of the {\em oriC}s, i.e. at quarter positions, and start creating the GD-polymer segments. 
Thus, the RFs are found at the quarter positions in our simulations in the $(0.2-0.4)$ interval, and are consistent 
with the experimental data reported. Thereafter, the RFs start moving along the 
two arms of each daughter chromosome assuming the train track model of RF-movement. 
% Their positions are maintained around the quarter position as the cell elongates, though they also get pushed out towards the 
% cylinder ends by the unreplicated sections of D-chromosomes. 
The RFs move towards the poles because the loci that are being replicated, are also 
closer to the poles due to Arc-2-2 architecture, as explained before.
As an example, at the end of $(0.2-0.4)$ interval the RF reached locus $50$ and $450$ on the left arm, which is located near the poles. Also refer the review section and SI-8,
where we show loci localization of monomer $50$ when replication is switched off.
Thereby RFs show a higher propensity to be in cylinder-ends in the 
$(0.2-0.6)$ stage, though it peaks near the quarter positions as in experiments. Experimentally,
the presence of the nucleoid ensures that the probability of finding the RFs at the cell poles is zero.

The peaks in the spatial distribution of RFs near the quarter position can be observed for the 
$(0.6-0.8)$ interval in Fig.\ref{long_axis_data}, and can be understood as follows. 
In this stage, the  RF's traverse the D chromosomes 
from the monomer $100$ ($400$ on the other arm) to $150$ ($350$ on the other arm),
and reach $125$ ($375$) at $0.7 \tau$. We have cross-links 
at $125$ and $375$ on D1 and D2. These monomers are connected to {\em oriC}. As shown before {\em oriC} 
is at the quarter positions. Any monomer connected to {\em oriC} will also be at the quarter
positions. This implyies that $125(375)$ are at quarter positions. Since the RF's are traversing around 
these monomers, the RF's also localize at the quarter positions.
% At the first part of this interval, the RFs move along  the contour towards the site of the 
% CLs, i.e. monomer $125$ and $375$ which create Loop-1 and Loop-2, and are close
% to the quarter positions. This is because they are linked with {\em oriC}. 
% After new CLs get introduced to create Loop-1 and Loop-2 of the GD1' and GD2' polymers
% at time $0.7\tau$,  the RFs follow the the oriC and the CLs to their new positions. 
% After the Loop-1 and Loop-2 of  GD1'  (and correspondingly GD2') are formed 
% at $0.7 \tau$,  the {\em oriCs} of the GD-chromosomes move to the $3/8$-th and $5/8$-th positions
% due to enhanced entropic repulsion between internal loops. 
% Therefore, the RFs will also be found at this position towards the end of the interval $(0.6-0.8)$.
The same raw data has been analyzed and presented in the convention, which allows for a more direct 
comparison with experimental data in Fig.\ref{fig:ExpForks}.

\begin{figure}
  \centering
  \begin{tikzpicture}
    % Include your images
    \node[anchor=south west,inner sep=0] at (0,0) {\includegraphics[scale=0.25]{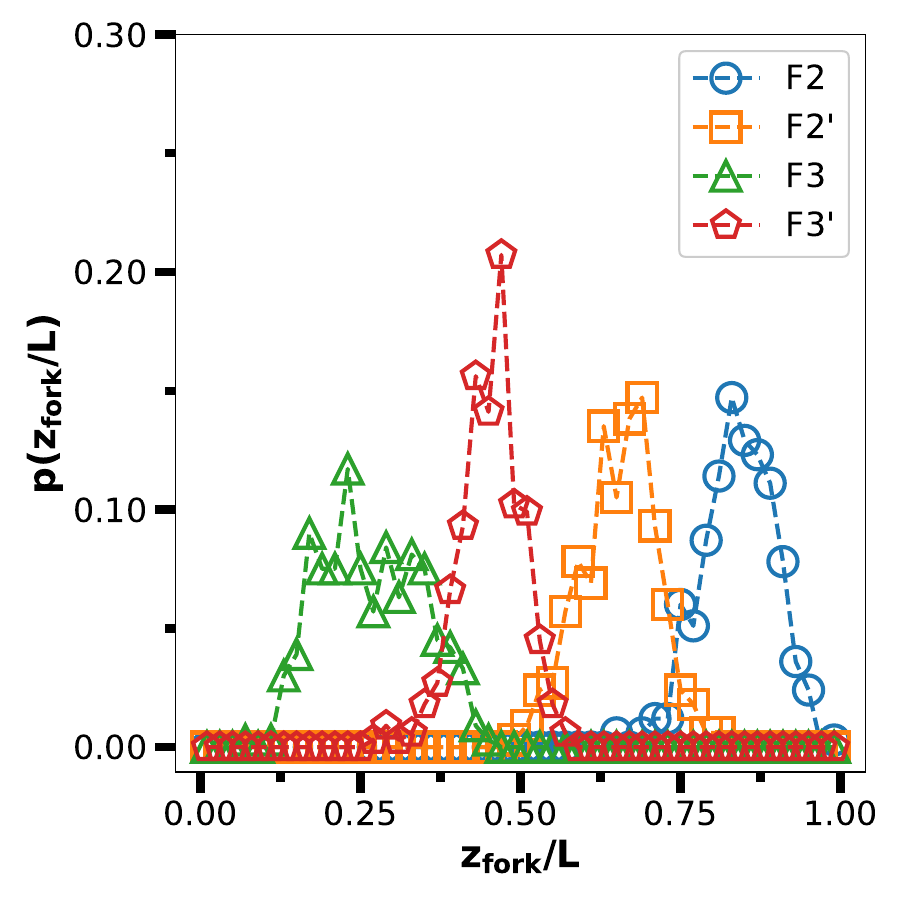}};
    \node[anchor=south west,inner sep=0] at (4,0) {\includegraphics[scale=0.25]{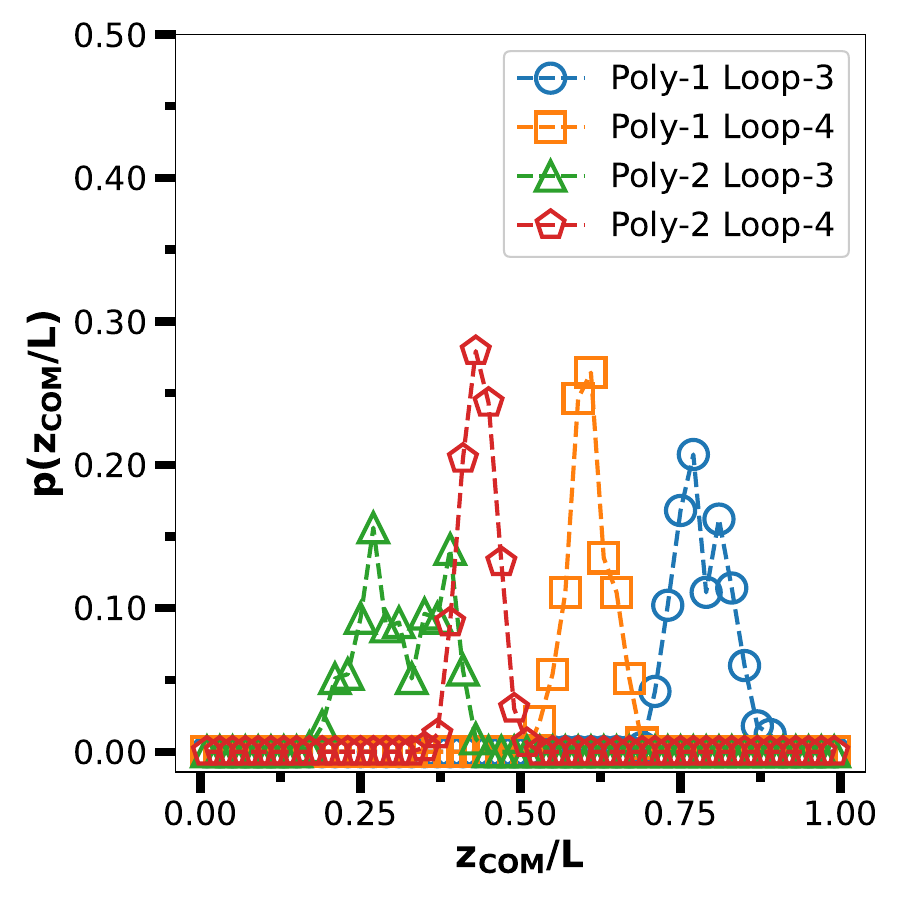}};
    \node[anchor=south west,inner sep=0] at (0,-1.6) {\includegraphics[scale = 0.35]{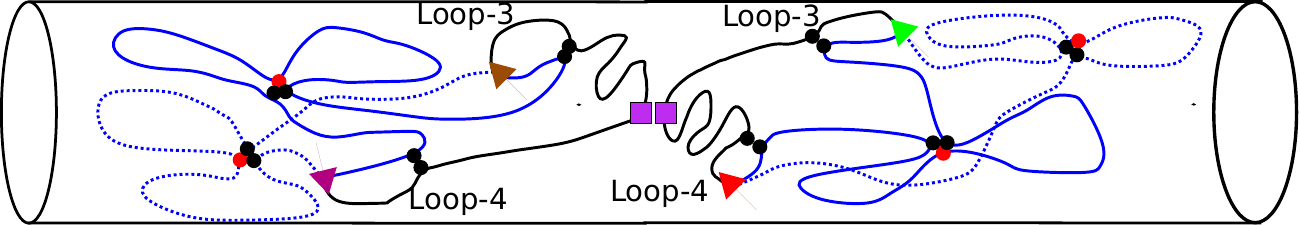}};
    
    % Add text on the figures
    \node at (1.2,3.3) {\color{red}\textbf{[a]}};
    \node at (5.2,3.3) {\color{red}\textbf{[b]}};
    \node at (0.2,-0.1) {\color{red}\textbf{[c]}};
    
    % Your caption
  \end{tikzpicture}
      \caption{ The plots show data spatial probability distribution data for 
    (a) the RFs and  (b) the center of mass (COM) of loops Loop-3 and Loop-4 
    from a \textbf{single} independent run for the interval $(0.8-1)$. We draw attention
    to the similarity in position of peaks in the spatial distribution of the quantities
    plotted in the two graphs. The replication forks have higher probability
    of occupying the same positions occupied by the COM of Loop-3 and Loop-4. The F2 and F2' denote
    the RFs of D1 which create GD1 and GD1', whereas F3 and F3' are the RFs of D2.  Below we show a schematic on how the loops could be
    arranged in this complex scenario.}
    \label{fig:ForksLoops}
\end{figure}

% \begin{figure}[ht]
%     \centering
%     \includegraphics[scale = 0.25]{ForkHistRun40-eps-converted-to.pdf}
%     \includegraphics[scale = 0.25]{LoopsHistRun40-eps-converted-to.pdf}
%     \includegraphics[scale = 0.35]{RFOrganisation-eps-converted-to.pdf}
%     \caption{ {\color{blue} The plots show data spatial probability distribution data for 
%     (a) the RFs and  (b) the center of mass (COM) of loops Loop-3 and Loop-4 
%     from a \textbf{single} independent run for the interval $(0.8-1)$. We draw attention
%     to the similarity in position of peaks in the spatial distribution of the quantities
%     plotted in the two graphs. The replication forks have higher probability
%     of occupying the same positions occupied by the COM of Loop-3 and Loop-4. The F2 and F2' denote
%     the RFs of D1 which create GD1 and GD1', whereas F3 and F3' are the RFs of D2.  Below we show a schematic on how the loops could be
%     arranged in this complex scenario.}}
%     \label{fig:ForksLoops}
% \end{figure}

{\em Role of Loop-3 and Loop-4 in RF positioning:} In the last $(0.8-1)$ interval of the cell cycle,
the RFs move towards {\em dif-ters} of the two D-chromosomes, starting out from the CL sites, as mentioned
in the previous paragraph. Thus, there is a higher propensity for them to be near the cell-center, 
but there are indications of some spatial separation along the long axis in the position of the peaks 
of the two RFs in one-half of the cell in the data from our simulations as seen in
Fig.\ref{long_axis_data} (3rd row). In contrast, the experimental data for cell-age 
$>0.6$ in Fig.\ref{youngren}  clearly shows a prominent separation of peaks in the 
spatial distribution of RFs. 

To check the reason for this discrepancy of our data (averaged over $50$ runs) from
experimental data, we plot the spatial distribution of the RFs from individual runs in 
the $(0.8-1)$ interval of the life cycle: refer to SI-13 for data on spatial positions of 
RFs from $50$ individual runs. Many of the data from individual runs clearly show 
$3$ to $4$ peaks, consistent with experiments. Note that when 
we plot the averaged data shown in Fig.\ref{long_axis_data} (3rd row), there is no clear 
separation in the positions of different peaks of the distribution.  This because the
RFs can be positioned differently relative to each other, across different simulation
runs. In a particular run, the RF from the left arm (of say D1)  can be closer to the 
middle of the cylinder, whereas the RF from the right arm of D1 might be closer to center 
in another run. We compared the spatial distribution of RF data obtained with Arc-2-2
with that obtained using the Arc-2 architecture  in SI-13 and SI-14, respectively. 
In comparison, the data obtained using Arc-2 architecture shows lesser distinct
separation between the peaks, which indicates that Loops-3 and Loops-4 (absent in Arc-2 architecture) play 
a role in the separation of peaks, seen in the spatial distribution of RFs.

\begin{figure}[h]
    \centering
  \begin{tikzpicture}
    % Include your images
    \node[anchor=south west,inner sep=0] at (0,0) {\includegraphics[scale=0.35]{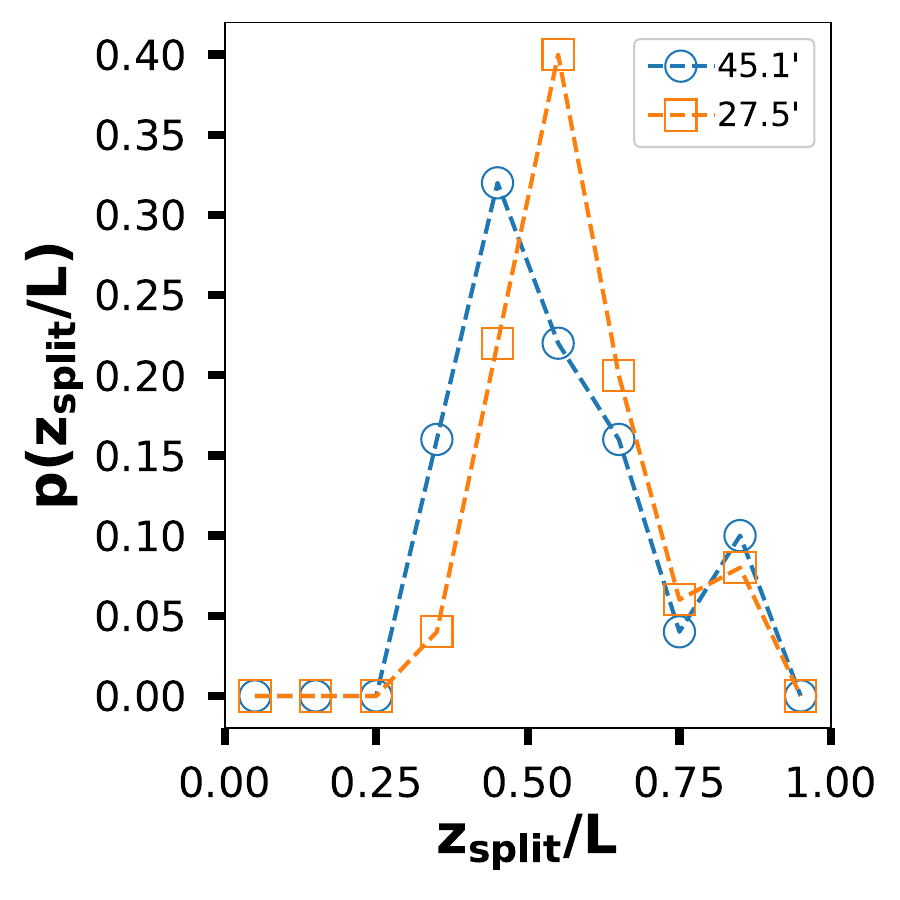}};
    \node[anchor=south west,inner sep=0] at (0,-4.5) {\includegraphics[scale=0.35]{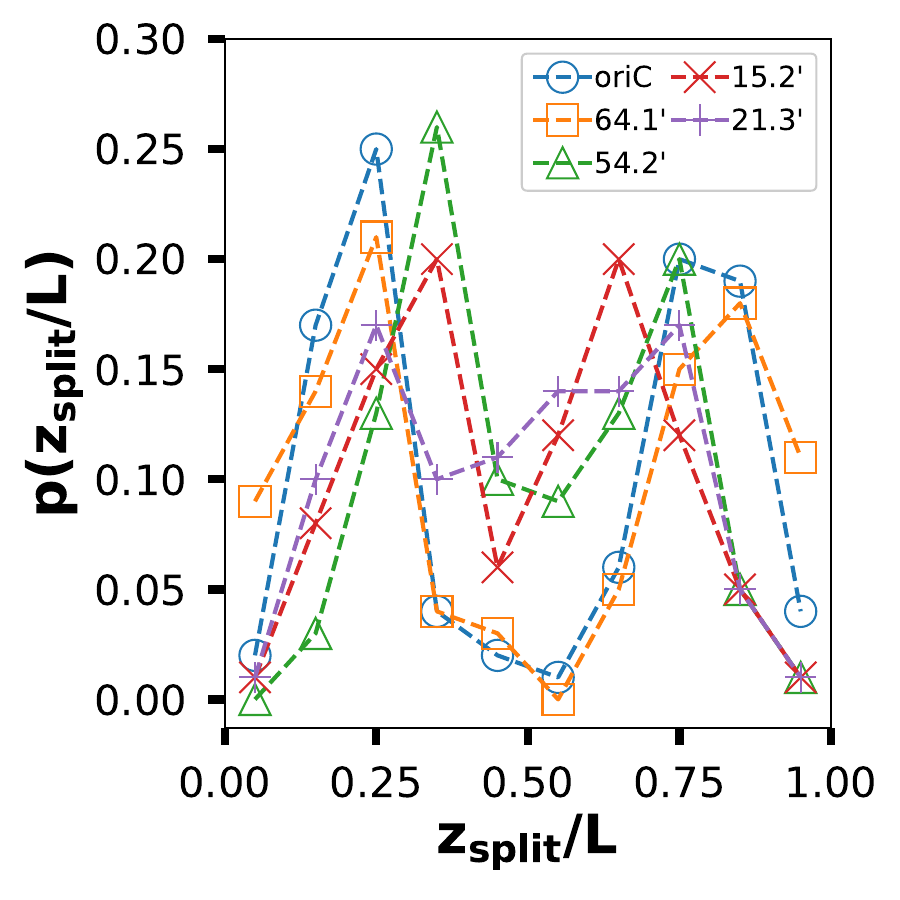}};
    \node[anchor=south west,inner sep=0] at (0,-9) {\includegraphics[scale = 0.35]{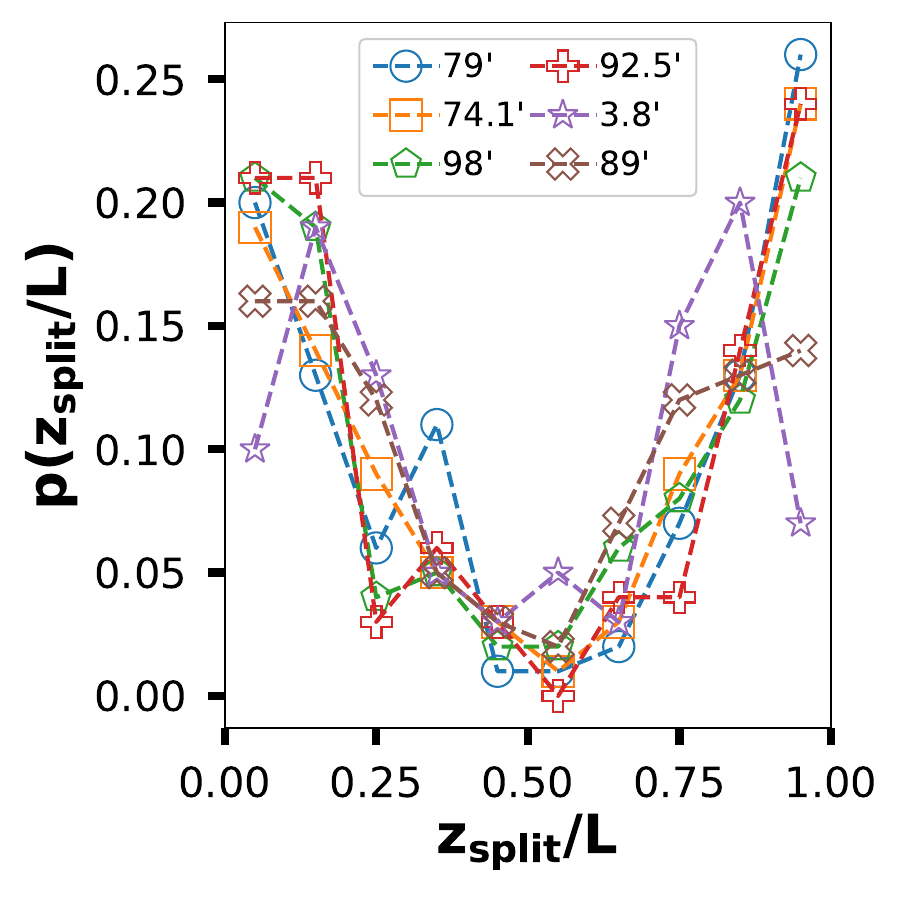}};
    
    % Add text on the figures
    \node at (1.8,4.7) {\color{red}\textbf{[a]}};
    \node at (1.8,0.2) {\color{red}\textbf{[b]}};
    \node at (1.8,-4.3) {\color{red}\textbf{[c]}};
    
    % Your caption
  \end{tikzpicture}
    % \centering
    % \includegraphics[width = 0.5\columnwidth]{MidSplitV2-eps-converted-to.pdf} 
    % \includegraphics[width = 0.5\columnwidth]{QuarterSplitV2-eps-converted-to.pdf} 
    % \includegraphics[width = 0.5\columnwidth]{PoleSplitV2-eps-converted-to.pdf}
    \caption{The probability distribution of the position of split for pair of replicated loci along the long axis. There is a distinctive two peak distribution for the loci initially having
two copies, i.e. for D1 (\& D2) loci replicating to GD1-GD1' (to GD2-GD2') loci and splitting
thereafter. There is a single peak distribution for the loci that only have a single copy initially,
i.e. for M-loci replicating to D1 and D2. We specify that the loci are spatially segregated if 
their axial distance is greater than $2a$. When the axial distance becomes $>2a$,
we calculate the midpoint of the two loci and identify the point as the
point of split. Since we have $50$ independent runs, there are only $100$ replication events (data points) to calculate the position of ‘split’ for the
two-peaked distributions. Moreover, we have $50$
points for single peaked distribution.}
    \label{fig:LociSplit}
\end{figure}

For further investigation of the above observations, we plot the spatial distributions
of the center of mass (COM) of  Loop-3 and Loop-4 in interval $(0.8-1)$ for each of 
the individual 50 independent runs, refer to data 
provided in SI-15.  Furthermore, we compare this distribution with the distribution of RFs for 
Arc-2-2 architecture for each independent runs. We do this because
at this stage of the life cycle, the RFs are traversing along the contours of Loop-3 and Loop-4. 
We find reasonable one-to-one correspondence in the position of the peaks of the distribution for
the COM of loops and the RFs. 
The representative data for one particular run is shown in  Fig.\ref{fig:ForksLoops}(a) and Fig.\ref{fig:ForksLoops}(b).
% and a comparison between the two sets of plots for all $50$ runs can be found in SI-13 and SI-15.
Refer to SI-16
for one-to-one comparison for $5$ different independent runs out of $50$, which are shown individually in SI-13 and SI-15.
This implies that the RFs remain 
separated along the cell's long axis in the region between the quarter position 
and the center of the cylinder. The Loops-3 and 4 can exchange positions as they occupy  
positions along the long axis with only partial overlap. This is a consequence of 
mutual entropic repulsion between Loops-3 and 4. Entropic interactions with Loops-1 
and 2 of both the replicated chromosomes position the Loops-3 and Loop3 away from the cylinder poles. 
All loops jostle for space to avoid each other and often end up interchanging positions along the long axis.

\begin{figure*}[ht]
\centering
\includegraphics[width=1.7\columnwidth]{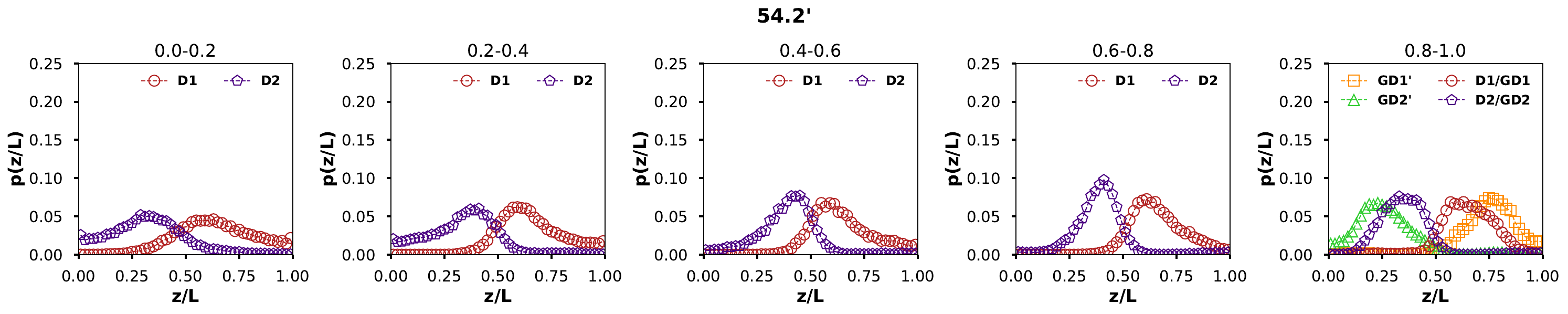}\\
\includegraphics[width=1.7\columnwidth]{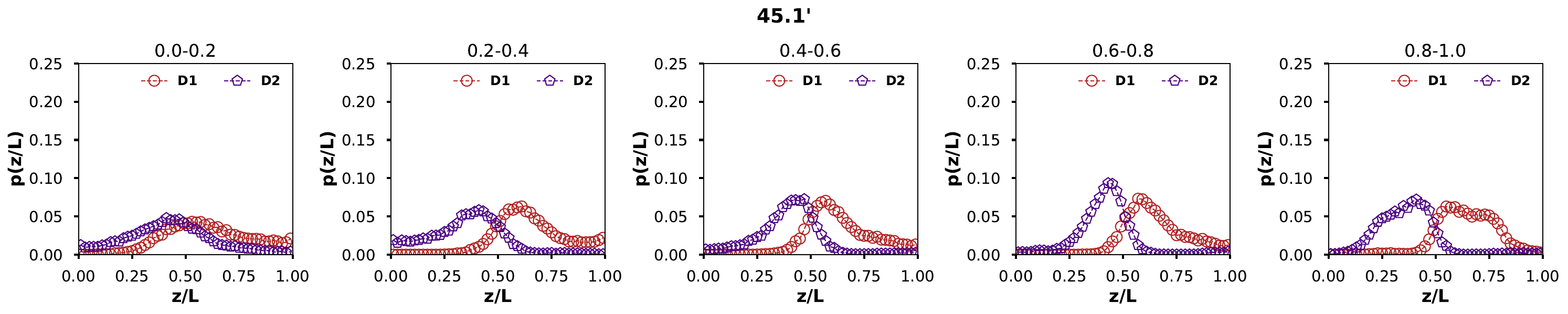} \\
\includegraphics[width=1.7\columnwidth]{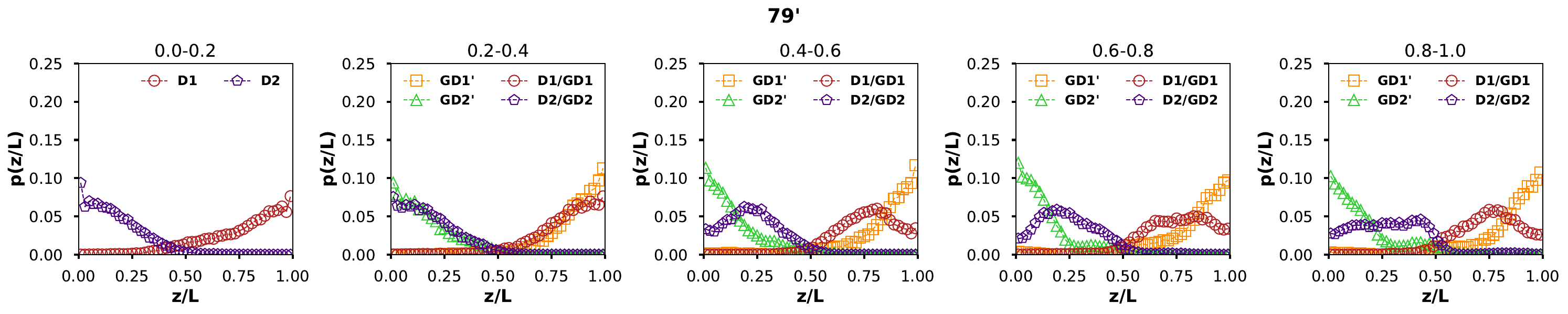} \\
\includegraphics[width=1.7\columnwidth]{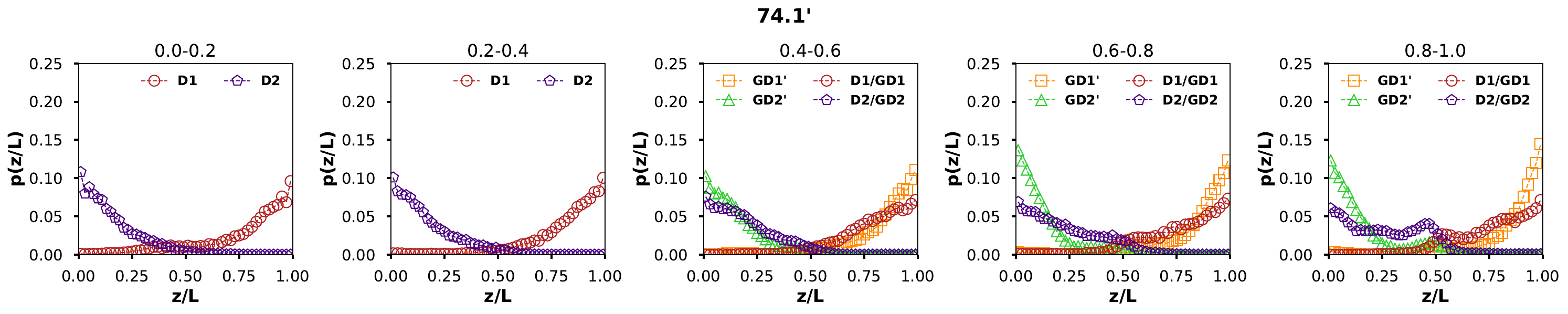} \\
\includegraphics[width=1.7\columnwidth]{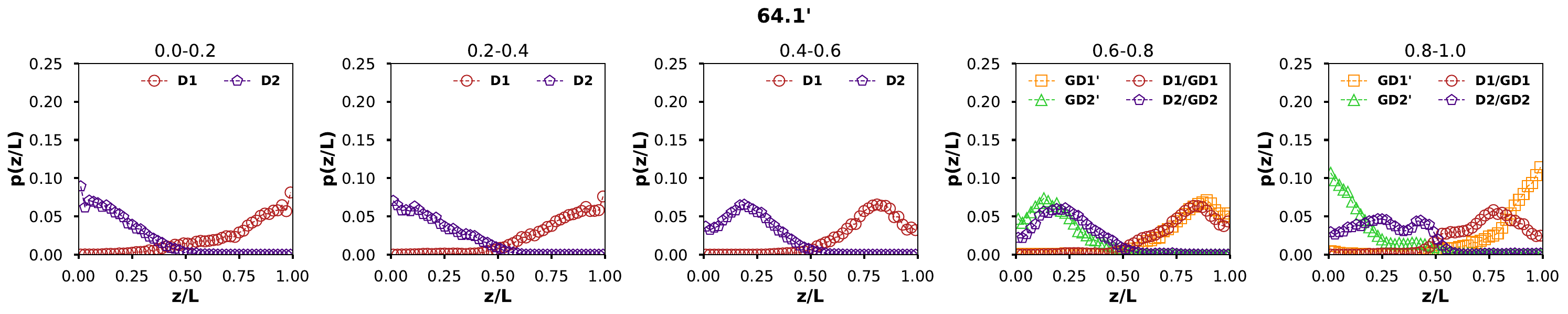} 
\caption{\textbf{Long axis distribution for other tagged loci: } We plot the spatial 
probability distributions $p(z/L)$ of the position of different loci, where 
$z$ denotes the position along the long axis of the cylinder (cell), and $L$ 
is the length of the cylinder at that stage of the simulation run. Data is shown for 
$54.2'$ locus corresponding to monomer $150$ in our simulations (first row),  for 
$45.1'$ locus corresponding to monomer $200$ (second row),
$79'$ locus corresponding to monomer $26$ (third row),
$74.1'$ locus corresponding to monomer $50$ (fourth row) and 
$64.1'$ locus corresponding 
to monomer $100$ (fifth row),
during the life cycle.  The corresponding monomer indices are at the top of each row.
The other plotting conventions are same as in Fig.\ref{long_axis_data}.}
\label{long_axis_data1}
\end{figure*}

{\em Difference in the behavior of Loop-1 and Loop-2 vs. Loop-3 and Loop-4:} Why do the 
spatial distributions of the RFs in the $(0.2-0.4)$ and $(0.4-0.6)$ intervals not show 
four peaks in contrast to the distribution in the $(0.8-1)$ interval?
This is because in the earlier stages of the life cycle, the RFs traverse along the
chain contours of Loop-1 and Loop-2 of DNA-1 and DNA-2.  They reach Loops 3 and 4 only 
in the last stage of the life cycle. The Loops-3 and Loop-4 behave differently from 
Loop-1 and Loop-2, as
they are closer to the {\em dif-ter} CL, and thereby try to avoid overlaps not only 
with each other but also Loop-3,4 of the other polymer and the ter-segments (Loop-5) 
from both polymers. As a consequence Loop-3 and Loop-4 have a greater propensity to interchange positions  
along the cell long axis,  as compared to that of loops 1 and 2. The interchanging of 
loops are better visualised in Fig.\ref{fig:ForksLoops}. To have a 
improved understanding of how different polymer architectures affect the organization of loops
with respect to each other, refer to our article \cite{dna2}.

{\em Position of loci-split:} The location of each locus as they get 
separated from its copy (post-replication) was studied in experiments 
\cite{Youngren2014}. The distribution of these positions along  the long-axis
for each locus is reproduced in the bottom right panel of Fig.\ref{youngren} 
for the aid of the reader. 
From our simulations, we can also obtain this distribution of ``position of split"  
of the relevant  replicated monomers corresponding to the tagged loci in experiments;
and we show our data from simulations  in 
Fig.\ref{fig:LociSplit}. We have more resolution than in experiments,
but while comparing our modelling results with the experimental distributions one also 
needs to account for the fact that we do not have nucleoid or the spherical ends of the cell.
The experimental data divides the set of loci into two sets, viz., the ones which 
split at the cell center and those who primarily split at the quarter positions. 
When we present data from our simulations, we provide  the distribution of the position of split 
in three subfigures, i.e. (a) the ones which are located near the  {\em dif-ter} loci along 
the chain contour, and hence spatially close to the position of the {\em dif-ter}, i.e.
near the cell center  (b) the loci which are located closer to the {\em oriC} loci (along the 
chain contour) or the  monomers ($125$ and $375$) cross-linked to it, which are spatially 
located near the quarter positions, and (c)  the monomers which are located on the 
Loop-1 and Loop-2 but away from CLs and these get replicated nearer the cell poles. As a 
consequence the position of split is also spatially closer to these positions. 
% In SI-17  we further 
% discuss why the `split positions'  of some loci as obtained by us differ from that seen in experiments.
% We have also plotted probability distributions of the splitting chromosome loci positions in SI-11. 

% \textbf{THIS PARAGRAPH IS SIMILAR TO THE ONE BEFORE. WE CAN CLUB THEM}
{\em Data normalization: Experiments vs. Simulations} There are some other caveats 
that the reader must take into account while comparing data from simulations with that of experiments. 
In simulations, we never see a pair of 
distinctly separated peaks of the {\em ter} loci distributions, as we do not model
cell division.  However, the experimental data shows finite probability for two 
{\em dif-ters}, even when the cell is in its $(0.6-0.8)$ interval of life cycle 
(as deduced from the length of the cell) as well as for the interval $(0.8-1)$.  
In experiments, the cell lengths are used as a proxy for the age of the cell. 
These could give rise to discrepancies when analyzing data by image processing software.
Moreover, there are differences in the methods of collecting data and normalizing the 
spatial distributions. In the given experimental data, the cells were first 
differentiated by their cell size to categorize the age of the cell. For each 
such category, the number of distinguishable foci in each cell is observed and 
spatial distribution data corresponding to the number of foci observed,  
they are normalized with respect to the number of cells in each sub-category.
The experimental images cannot discern if the foci  belong to the D-chromosome 
or GD-chromosome.   Furthermore, if the foci cannot be spatially resolved, 
a $4$-locus cell may be erroneously categorized as a three-foci or a two-foci cell. 
The three foci scenario may also arise due to asynchrony in the replication 
initiation process or stochasticity in the cohesion times. In the simulations, 
we do not have asynchrony in the replication initiation process,
however, there maybe stochasticity in the cohesion times of the loci. But, we 
have access to the positions of each monomer at all stages of the simulation run. 
Thus, our normalization protocol is different from that adopted in experiments.

In simulations, we have $50$ independent runs to collect data over the entire life cycle.
We precisely know the stage of the life-cycle of the model cell, and the stage when a loci is replicated
to two loci of the next generation. The RF moves to the next monomer on the contour every $2 \times 10^5$ MCS.
We thereby normalize by the precise number of micro-states 
relevant for a particular monomer, depending on when that monomer has been introduced within that
interval of the life cycle. For each independent run having a total of $5 \times 10^7$MCS, 
we store data to calculate distributions every $3 \times 10^4$ MCS. Thereby, we know the number 
of contributing microstates in each stage of the life cycle for each loci.

{\em Spatial distribution of other loci:} We now calculate the spatial distributions 
of the other loci that were tagged in experiments and compare the distributions obtained 
by our simulations to those obtained by \cite{Youngren2014}, refer Fig.\ref{youngren} for comparison. 
Here, we provide data for five such loci from the left arm in Fig.\ref{long_axis_data1}. 
The data for the loci on the right arm are given in SI-17. 

For the locus marked as $54.2 ^\prime$ (monomer $150$), we see only two distributions for $(0-0.8)$ intervals of 
the life cycle, as the monomers of the daughter chromosomes get replicated only at the end of the fourth interval.
Our modeling data is in fair agreement with the experimental data. We find that there are four peaks 
towards the end of the life cycle. 
Experimentally, it is not possible to distinguish between the two GD loci near
the end of the life cycle, as they cannot be spatially resolved if are spatially proximal. 
In our simulations, we can uniquely identify the loci of each GD chromosome,
and thereby, we obtain four distinct spatial distributions, albeit they overlap. In experiments, 
the loci distributions (that we obtain) will appear as broad distributions having only two
distinct peaks. In this sense, our spatial distribution of loci is in agreement with the 
experimentally obtained distributions \cite{Youngren2014}.

\begin{figure*}[ht]
    \centering
          \includegraphics[width =0.6\columnwidth]{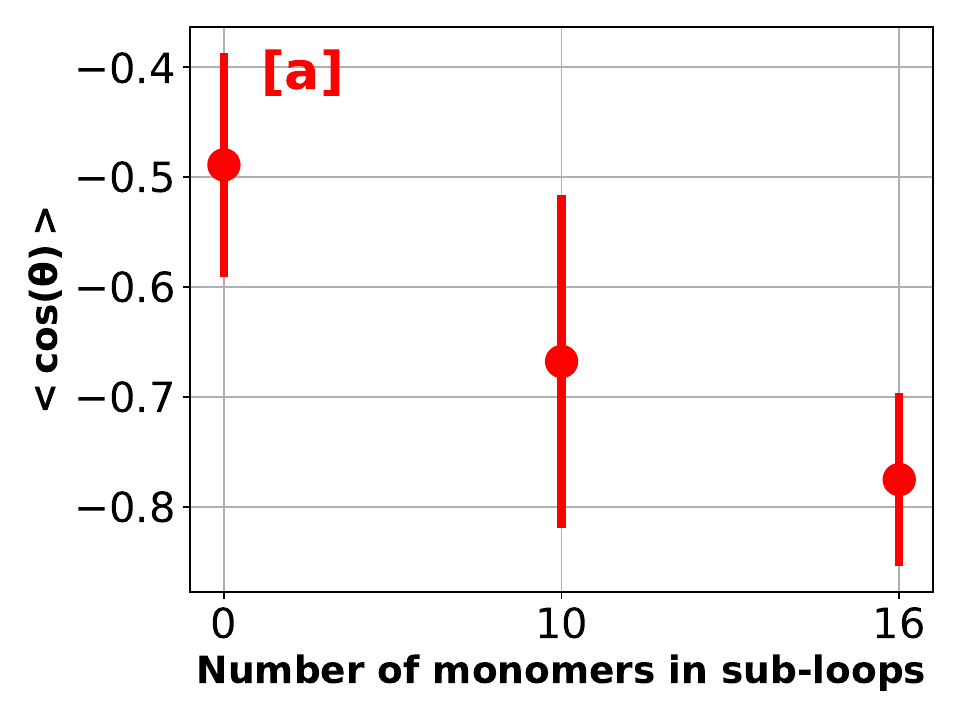} \\
    \includegraphics[width = 0.5\columnwidth]{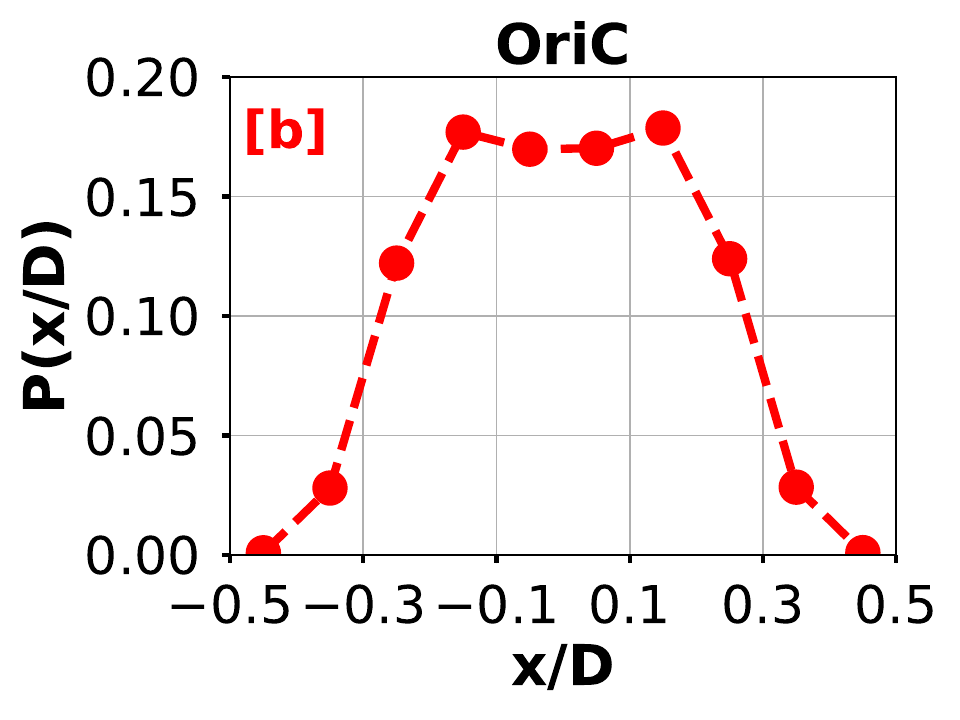}
    \includegraphics[width = 0.5\columnwidth]{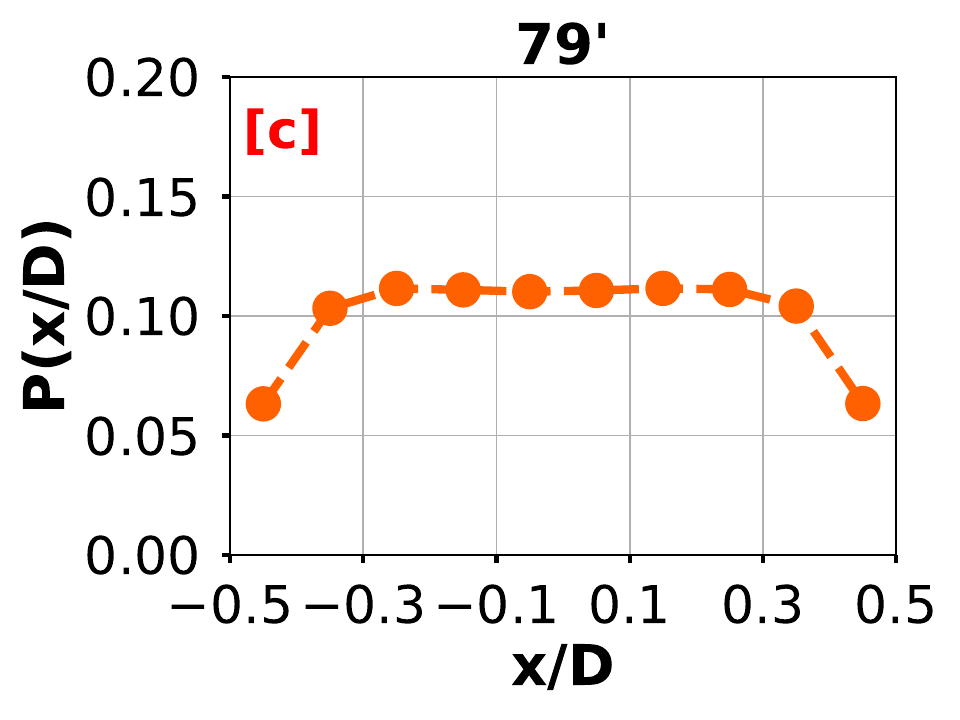}
    \includegraphics[width = 0.5\columnwidth]{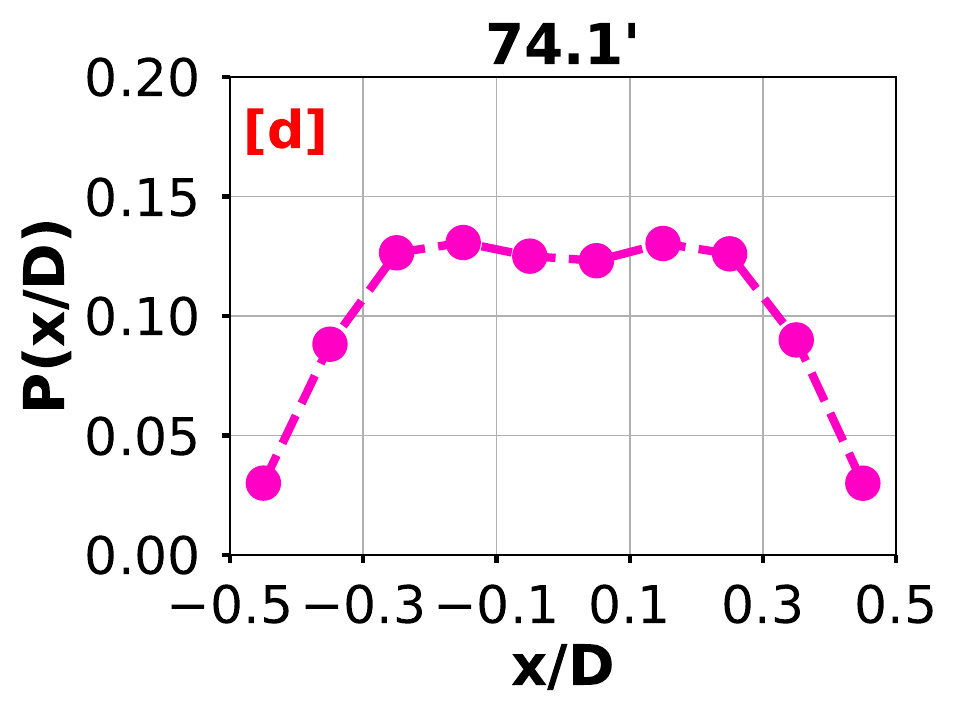} \\
    \includegraphics[width = 0.5\columnwidth]{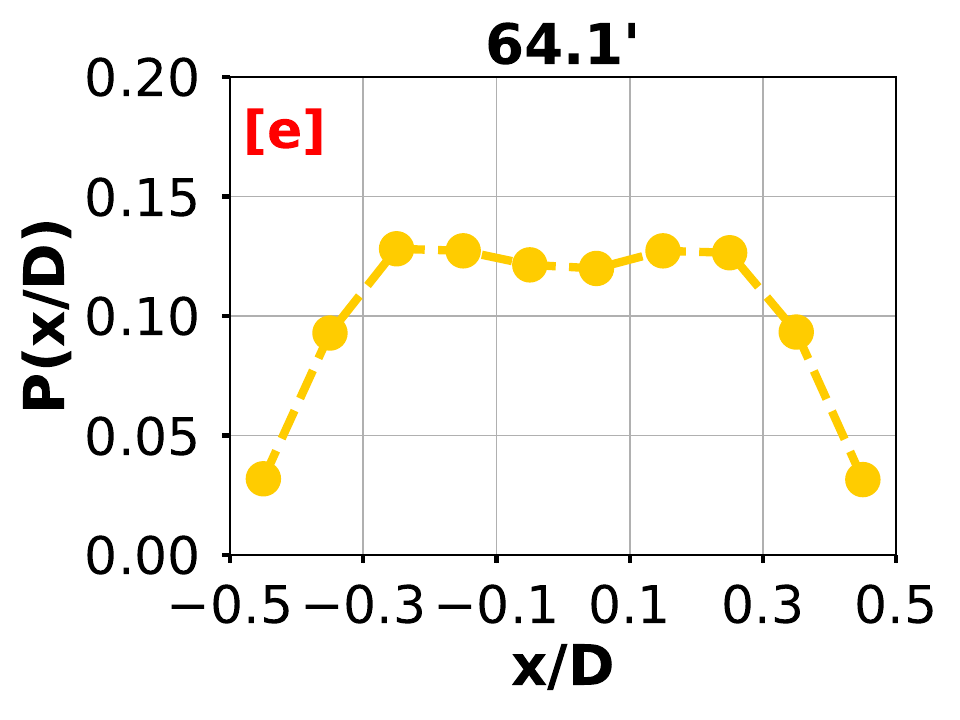}
    \includegraphics[width = 0.5\columnwidth]{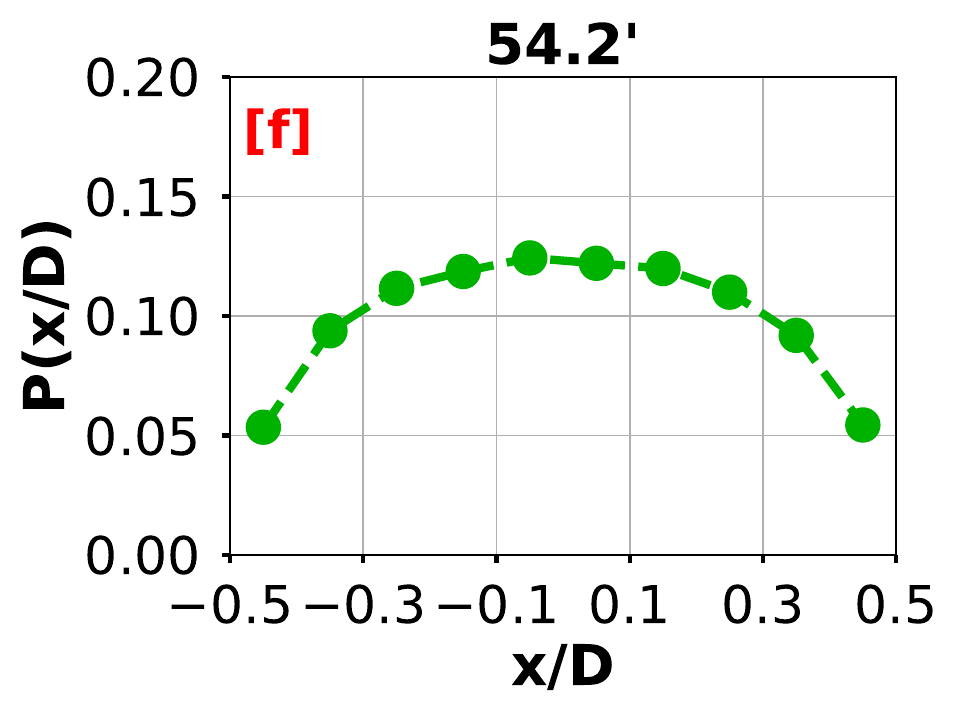}
    \includegraphics[width = 0.5\columnwidth]{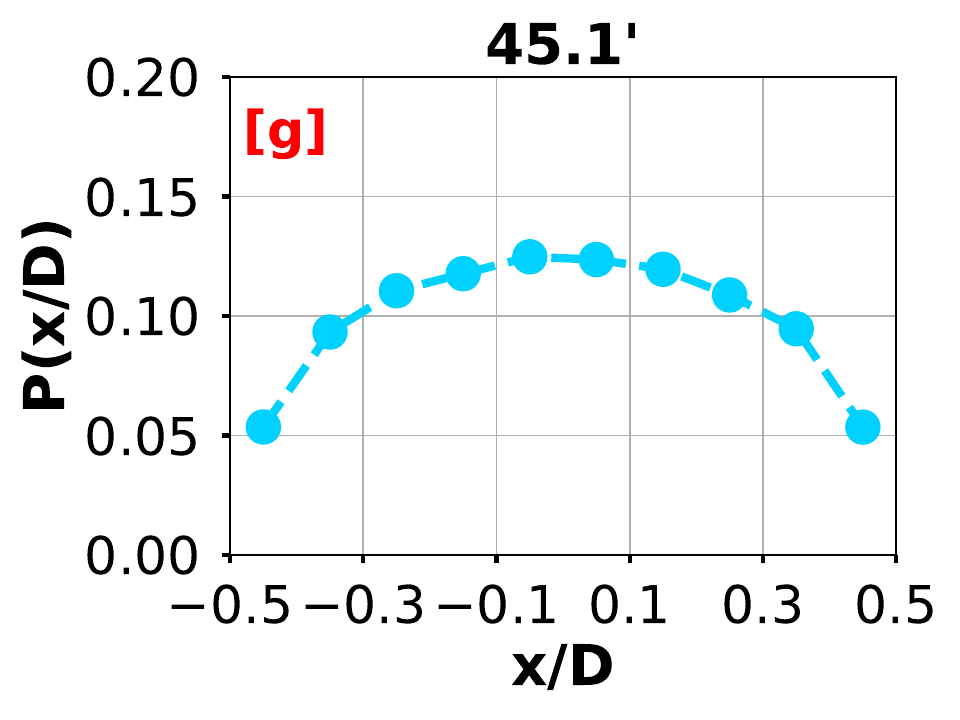} \\
    \includegraphics[width = 0.6\columnwidth]{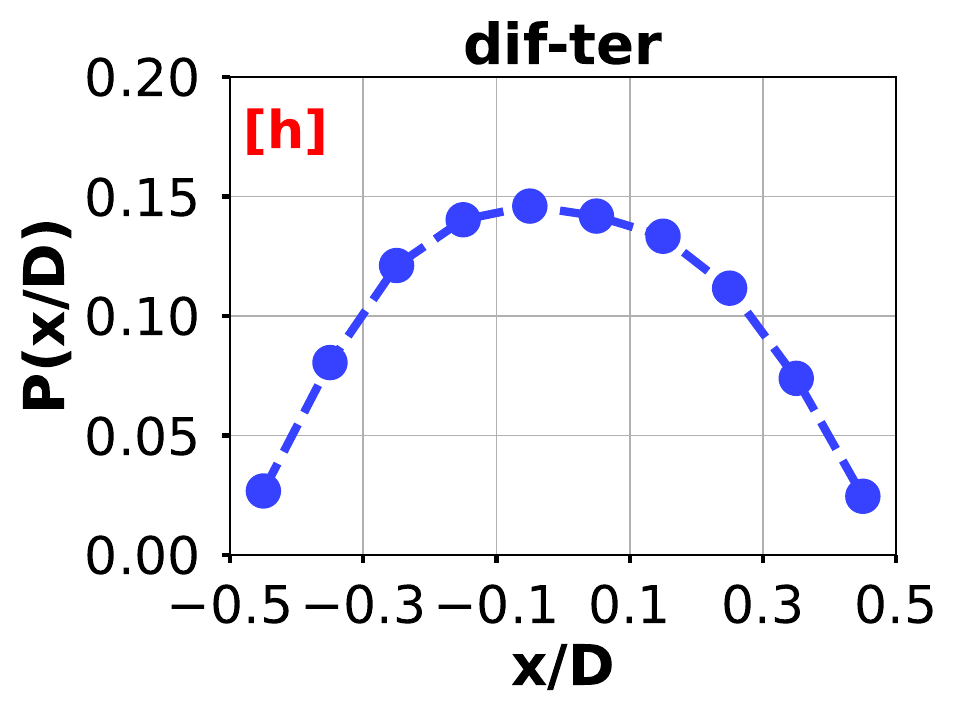}
    \caption{ \textbf{Organization of chromosomal arms and radial distribution of loci: }Subfigure (a) shows $\langle cos(\theta)\rangle$, where $\theta$ denotes the angle between vectors $\Vec{l}_1$ and $\Vec{l}_2$(refer text). 
    % Vectors $\Vec{l}_1$ and $\Vec{l}_2$ have been constructed by joining the centers of the cylinder to the COMs of loop-1 and loop-2 respectively.
    % We only consider the $\hat{x}$ and $\hat{y}$ components of the vectors.
    A high negative value of $\cos(\theta)$ indicates that the two loops (belonging to the two arms of the chromosome) lie on different cell halves along the radial axis.
    % We have computed the average value of $cos(\theta)$ by considering $1000$ microstates. 
    We observe that the average $cos\theta$ value is more negative in the cases with smaller loops (within Loop-1 and Loop-2) as compared to the case without smaller loops.   
    % We note that if we do not incorporate additional loops then $\langle cos(\theta)\rangle\approx0$ which indicates that the two loops may overlap significantly along the radial axis. We also show that if smaller loops are introduced (of sizes $10$ monomers and $16$ monomers) we obtain a greater separation of the arms along the radial axis.
     Subfigures(b-h) show the radial probability distributions of the monomers for several loci in the presence of smaller loops. These smaller loops are of size $10$ monomers each and are placed along loop-1 and loop-2. We show here the data for the loci of the left arm, while the corresponding data for the right arm loci have been provided in the Supplementary(SI-20). We note that we obtain a bimodal distribution for some loci by introducing these smaller loops. The distributions we obtain match those found in the experiments for the loci {\em oriC},79', 74', 45.1' {\em dif-ter} along the left arm only. The experimentally obtained radial distributions for the loci have also been reproduced in SI-21 for aid of comparison. We do not obtain a match for the locus 54.2' and data for some other loci on the right arm (presented in SI-20). We also notice that we obtain a double peaked distribution of the {\em oriC} while the experimentally obtained distribution has a single peaked distribution. 
     % However, we have explicitly checked that one can obtain an exact match with experiments by tuning the size and location of these loops along the chain contour. But this is outside the scope of this study. Here we have just established the mechanism by which one may obtain the bimodal peaks in the radial distribution plots of genomic loci. 
     }
    \label{radial_loci}
\end{figure*}

The data obtained from simulations  for  the locus marked as $45.1'$  (monomer $200$) 
also shows good agreement with the experimental data for $2$ foci. We do not have four 
peaks for this locus in the $(0.8-1)$ interval (as in experiments) since the simulations 
( cell cycle) are stopped just as this specific locus gets replicated. This is because cell 
division takes place at this stage of the cell cycle. Correspondingly, the experimental data 
for $4$ foci also doesn't have any contribution from this locus in the $(0.8-1)$ interval.
Furthermore, comparing our data in other intervals with the $2$ foci data in 
Fig.\ref{youngren}, the distribution of the locus $45.1'$ is peaked near the center 
as this locus is close to {\em dif-ter} at initial intervals. As the cell ages, the loci 
move away from each other, hence delocalizing from the center of the cell due to the 
presence of mutual repulsion between Loop-3 and Loop-4.  But once the Loop-1 and Loop-2 
of the GD1' and GD2' chains are formed, they are pushed away from the poles in the 
$(0.6-0.8)$ interval. Thus, our data are consistent with the experimentally obtained 2-foci data 
shown in Fig.\ref{youngren}.

In the third row of Fig.\ref{long_axis_data1}, we show data for the locus $79'$ (monomer $26$). 
We note that the distributions from our simulations peak around the cell  poles at the 
$(0-0.2)$ interval. This differs
from two foci data from experiments in the $(0-0.4)$ intervals, where they find 
that this locus is localized around the quarter positions. However, if one looks carefully 
at the 2-foci data of Fig.\ref{youngren}, one notices that the distribution has more 
contribution from regions between the poles and the quarter position. In simulations, 
we have not incorporated the effects of the nucleoid, the hemi-spherical poles at the 
end of the cylinder in our current work. If 
we introduce these effects in our simulations, we presume that this locus will stay away from the poles. 
In later parts ($0.6-1$ interval) of the cell cycle and for data with two or more foci, 
the experimental distribution is either broad or four peaks are observed for the 4-foci case. 
In simulations, we do have distributions that are spread out over the length of the cell. 
We observe the spatial distribution of the $4$ loci, where two of the four distributions are closer to the center,
though we do not observe distinct peaks in the distribution.

%To compare directly with experiments, one has some consider the sum of the GD1 and GD1' distributions
%(and correspondingly GD2 and GD2'), as experimentally they cannot be distinguished.

%A similar mechanism would be valid while comparing the spatial distributions 
%(as obtained from simulations) 
%of the $74'$ locus (monomer $50$) with experimental data. 
We may reason along similar lines as to why the distributions obtained for the $74'$ 
($50$-th monomer) locus appear quantitatively different to those seen in experiments.
This locus again lies within the Loop-1, and would remain away from the cell-center 
which is the position of the {\em dif-ter}. While there are broad similarities between 
the experimental and modelling data, we still have high probability to obtain the loci right at the poles. 
This differs from the experimental data which has $4$ distinct peaks for the 4-foci data 
at the intervals $0.6-1$ interval. In our case Loop-1, which caries monomer $50$, from 
GD1 and GD1' ( and correspondingly GD2 and GD2') may keep interchanging positions 
along the long axis. In our ongoing studies without replication
(unpublished data), but in the presence of a rosette of small loops in the region $1-125$ and 
$1-375$, we observe that the loops repel each other and rarely interchange positions, which 
would lead to the sharp peaks observed in experiments.

The $64.1'$ loci ($100$-th monomer) is again on Loop-1 but closer to the position of CLs
along the chain contour. Hence we expect this loci to primarily occupy the region between the 
quarter positions and the poles in the $0-0.6$ interval in the 2-foci data.  This foci gets 
replicated at the end of the $0.4-0.6$ interval, and hence 4-foci do not appear till after 
this stage. This is exactly what we see in Fig.\ref{youngren}. After the middle of the 
$0.6-0.8$ intervals, the Loop-1 and Loop-2 of all four GD chromosomes are formed. Thereby,
this loci will be near the $1/8$ and $3/8$-th position but relatively away from the cell center.
The Loop-1 from GD1 and GD1' may keep interchanging positions along the long axis. Consequently,
we see broad distributions in simulations and in experiments. Since the loci $64.1$ and $79$
are equidistant from the position of the CLs which create Loop-1 and Loop-2, the reader can 
observe that the spatial distribution of loci is nearly identical in the $0.8-1$ interval.
However, $79'$ is replicated at the middle of the $0.2-0.4$ interval, whereas $64.1'$ is replicated
at the end of $0.4-6$ interval. So the $4$-foci data appear at a later stage for the $64.1'$ loci.

{\em Organization of chromosomal arms:} 
The organization of the arms of the {\em E.coli} chromosome has been a topic of considerable interest
in the literature. It has been suggested that in slow-growth conditions, the arms occupy different 
sections of the cylinder along the cell long axis, while in fast growth conditions, the arms occupy different halves along the short axis \cite{woldy,Youngren2014} i.e., if the arms would 
occupy different halves if the cylinder were sliced into two halves parallel to the
long axis. In this context, the work of \cite{Youngren2014} also provides experimental data for the radial distributions of loci along the short axes. They find that many loci show double-peaked distributions, indicating that the two arms lie preferentially on opposite sides 
along the short axis.

We plot spatial distributions using the loci's $x$ or $y$ coordinate, as both can be used to measure ``radial" distance from
the central axis. Data from simulations using the Arc-2-2 polymer 
topology has been plotted in Figures shown in SI-18 and SI-19 to estimate the preferential 
radial distance of loci. We find that these distributions show, in general, broad distributions
that peaked at the middle of the cylinder and thus do not agree with what is observed experimentally.
Experiments show doubled peaked radial distribution for some of the loci.

% This indicates that the monomers are preferentially localized near the cell periphery while in our simulations the monomers are found towards the middle of the cylinder along the radial axis. 

Thus, we propose modifications to a simpler version of our model with replication 
switched off and discuss the loop-based mechanism to
obtain the longitudinal organization of chromosomal arms. The longitudinal organization of
the arms can be inferred from the bimodal radial distributions of loci.
%We propose an outline of the mechanism by which one can obtain such bimodal  distribution as seen in experiments.
We introduce $6$ smaller subloops in each of Loop-1 and Loop-2 of size $10$ monomers each. 
These subloops are equally spaced with $10$ monomers in between. Introducing these 
subloops enhances the entropic repulsion between Loop-1 and Loop-2 along the short axis.
Consequently, the monomers belonging to those loops show a bimodal distribution. 
We establish this emergence of a bimodal distribution through simulations of two (modified) Arc-2-2 
polymers that have segregated along the long axis, with additional subloops of size $10$ monomers 
in Loop-1 and Loop-2. These subloops are created by introducing extra CLs between monomers separated 
by $10$ monomers, i.e., between $11$ and $20$, $31$ and $40$ and so on. 
We do not incorporate replication since we only outline the mechanism by which one 
may obtain such bimodal distributions.
% These might be realised by transient loops which we have not modeled explicitly.
%The existence of macro-domains in Hi-C maps indicates the existence of smaller loops
%in the DNA-polymer
%\cite{loop_heerman,Lioy2018,Le2014,Le2013,jobdekker2011,Maji2020,tejal2018,tejal2019,tejal2019_2,Mondal2020,sbsmodel,jost,D2SM00612J}. 
The entropic repulsion will likely act through transient loops {\em in-vivo} unlike the long-lived permanent loops we consider for our model.

 To establish that Loop-1 and Loop-2 lie in different halves of the cylinder, along the radial axis, we carry 
 out the following calculation. We construct a vector $\Vec{l}_1$  joining the mid-point of the cylinder and the 
 COM (center of mass) of Loop-1. Similarly, we construct another vector $\Vec{l}_2$ joining the center of the cylinder 
 and the COM of Loop-2. Note that we only consider the $\hat{x}$ and $\hat{y}$ components of the vectors. The angle 
 between the two vectors is denoted by $\theta$. Then, if $cos\theta\approx-1$, the two vectors are anti-parallel 
 to each other. This implies that the two loops, Loop-1 and Loop-2 (belonging to different arms of the chromosome) 
 lie in different cell halves along the short axis. As can be inferred from Fig.\ref{radial_loci}(a),
 introducing smaller additional loops leads to the separation of arms along the short axis.
 We further note in Fig.\ref{radial_loci}(b-h) that the introduction of subloops also leads to bimodal  
 distributions of some loci along the short axis, similar to what is seen in \cite{Youngren2014}.
 Other loci show broader radial distributions as compared
 to radial distributions for Arc-2-2 polymers without smaller loops (refer SI-18 \& SI-19), even if they do 
 not show bimodal distributions. 
 To check for the robustness of our conclusions, we conduct similar simulations with five smaller subloops within Loop-1 
 (and Loop-2) with $16$ monomers in each of the subloops. 
 This also shows the separation of these arms as can be inferred from values of $\langle cos(\theta) \rangle$,
 refer Fig.\ref{radial_loci}(a). This however fails to show bimodal radial distribution. 
%\begin{figure}[!]
%    \centering
%    \includegraphics[width =0.8\columnwidth]{ComLoopsCos-eps-converted-to.pdf}
 %   \caption{ The figure shows $\langle cos(\theta)\rangle$, where $\theta$ denotes the angle between vectors $\Vec{l1}$ and $\Vec{l2}$. Vectors $\Vec{l1}$ and $\Vec{l2}$ have been constructed by joining the centers of the cylinder to the COMs of loop-1 and loop-2 respectively.  We only consider the $\hat{x}$ and $\hat{y}$ components of the vectors. A high negative value of $\cos(\theta)$ indicates that the two loops (belonging to the two arms of the chromosome) lie on different cell halves along the radial axis. We have computed the average value of $cos(\theta)$ by considering $1000$ microstates. We note that if we do not incorporate additional loops then $\langle cos(\theta)\rangle\approx0$ which indicates that the two loops may overlap significantly along the radial axis. We also show that if smaller loops are introduced (of sizes $10$ monomers and $16$ monomers) we obtain a greater separation of the arms along the radial axis. }
%    \label{fig:COM of loops 1 and 2}
%\end{figure}

 Furthermore, one may also introduce smaller loops along the rest of the Arc-2-2 polymer (outside
 loop-1 and loop-2) to obtain the bimodal distribution of other monomers. In the experimental data of
 \cite{Youngren2014}, some loci show single peaked distributions while others show bimodal distribution.
 Even in our simulations, some monomers show single peaked distributions while others show bimodal 
 distributions, although an exact match is not obtained with experiments. To obtain an exact match, 
 one needs to optimize the size and location of loops along the contour, which is outside the scope
 of the current manuscript.We have also shown in SI-22 that the  localization of
 the tagged loci along the long axis which was discussed previously, remains unaffected  with 
 the introduction of smaller loops

\section{Discussion}
We establish that entropic repulsion between internal loops is a viable mechanism through which the chromosomes 
segregate and organize themselves. We show that the organization of {\em oriC}, {\em dif-ter}, and other loci 
emerge spontaneously in our model, both along the longitudinal and the radial axis. Moreover, we outline the key mechanisms governing the localization of loci both along the radial and long axis of the {\em E.coli} cell.  Our model also successfully reconciles other 
experimental observations, such as the spatial organization of replication forks. Though we have implemented the 
replication process following the train-track model,  our simulations show the localization of the RFs, 
which supports (in spirit) the replication-factory model. Thus, the hypothesis proposed by the authors of \cite{Youngren2014}: 
``The position and dynamics of the replication forks are likely the consequence of the spatial organization of the 
chromosomes rather than vice-versa" is supported by our simulations. With this work and our past paper on the organization 
of {\em E.coli} chromosomes in slow growth conditions using the Arc-2-2 architecture \cite{dna1}, we propose that this
model of the  {\em E.coli} chromosome provides a viable mechanistic understanding of chromosome organization in all growth conditions. To the best of our knowledge, 
we are the first to explicitly model the replication and evolution of the organization of the chromosomes 
in the complex case of overlapping rounds of replication.

% With this work and our past paper on the organization of {\em E.coli} chromosomes in slow growth conditions using the Arc-2-2 architecture \cite{dna1}, we present a unified model of the {\em E.coli} chromosome organization for all growth conditions. 

Despite the many successes of the model in reconciling the broad features of chromosome organization in fast-growth conditions, future studies may improve upon this model in some areas. We were not able to match all the details presented in \cite{Youngren2014}. For instance, we note in the data presented in \cite{Youngren2014} that the probability distributions of loci keep shifting towards the quarter positions along the long axis. The time taken by each locus to move to the quarter positions is a function of distance from {\em oriC} along the contour of the polymer. Our simulations fail to capture this aspect. Moreover, some loci in our simulations have a higher probability of being at the cell poles, which is absent 
in the experimental data. We attribute this to not incorporating the presence of crowders which are known to condense the chromosome through depletion interactions \cite{Joyeux2015,Joyeux2018,Jun2012,woldringh2024compaction,kim2015polymer,odijk1998osmotic,yang2020effects}.

 We remind the reader that 
we constructed a minimalist model with only 4 additional Cls to obtain a mechanistic understanding of the localization patterns of genomic segments as seen  {\em in-vivo}, from FISH experiments. The aim was to provide 
an underlying mechanistic explanation for the experimental observations seen in both slow \cite{dna1} and fast growth conditions. We are fully aware that this minimal model cannot be a complete 
and accurate description of all the phenomena associated with the {\em E.coli} chromosome. The  {\em E.coli} chromosome is significantly more complex than our description of it with a minimalist model, with a variety of different sub-cellular phenomena affecting its properties. Thus, the model 
has ample scope for extension to incorporate relevant biological processes more accurately and make more
improved predictions thereafter.

% One may also incorporate the effects of other transient cross-links in addition to the permanent (long-lived) cross-links 
% that we have employed. These transient CLs will further fine-tune the emergent organization of the chromosome at different stages of the cell-cycle. 

% Binding and loop extrusion proteins, which associate and disassociate from the chromosome at different stages of the cell-cycle, are likely to result in transient cross-links and further fine-tune the evolution of the organization. 

The CLs in our simulations are likely mediated by linker proteins, such as MukBEF or H-NS, {\em in-vivo} \cite{Badrinarayanan2015,Sherratt2020,wang2011chromosome,norris2023roles}. 
% Positional variations of up to approximately $5$ monomers have been established without affecting the localization of loci. 
MukBEF complexes, dissociating every $60$ seconds in vivo, are observed in clusters at ori proximal regions. Despite the dissociation of 
individual complexes, replacements within the cluster may create new cross-links. Continuous loading and unloading of MukBEF complexes may lead to slight 
variations in cross-link positions. But we have shown than altering positions of CLs in our simulations by $5$ monomers 
(i.e. $\approx 45$Kbp) on either side does not impact {\em oriC} localization.

Extrusion due to MukBEF can also form smaller transient loops 
 %with an observed average size of about $52$Kbps 
 {\em in vivo} \cite{Sherratt2020}. A series of smaller loops 
 %formed by proximal cross-link clusters near in ori-proximal regions
 can bring distant  DNA segments closer spatially. This can result in a scenario that resembles  ``effective" CLs
 between distant monomers along the polymer chain, such as those we have considered in this paper. %between monomer $1$ and $125$.
 %A systematic investigation of the consequences of smaller loops 
 %on ori-proximal regions 
 %will be 
 %reported in upcoming research.
 Other proteins, like H-NS, potentially mediate long-range interactions\cite{wang2011chromosome,norris2023roles}
 and may act as cross-links.

Simulations by other groups have shown that entropic effects persist even with transient extruded loops.  
Moreover, internal loops within a DNA ring polymer aid the segregation of daughter chromosomes 
\cite{harju2023loop}. However, the authors have not shown the organization or localization of loci of the 
bacterial chromosome with transient loops at random positions.  In contrast, we show that introducing 
smaller internal permanent loops within Loop-1 and Loop-2 in our Arc-2-2 topology keeps the organization of 
the chromosome along the long-axis relatively unchanged. Although in reality, there are likely multiple smaller,
transient extruded loops organized in an hierarchical fashion within the cell, we currently have a 
simplified description of the
genome using just four permanent  (likely to be long lived {\em in vivo}) cross-links.
Future experiments can try to identify the protein complexes which mediate the long-lived links
at the positions of CLs of the Arc-2-2 model, proposed by us. These connect the oriC with points that are  nearly 
half-way along the contour of the  left arm and right-arm of the DNA.
Future work aims to incorporate transient small  loops by introducing additional cross-links  
within Loop-1 and Loop-2, and systematically investigate its consequences, using appropriate coarse 
graining techniques \cite{kadam} to reproduce the same results we obtain in the manuscript.
Note that we keep longer loops {\em viz.} Loop-3 and Loop-4, of length $\sim 800$kBP, consistent with what is 
reported in \cite{thiel2012long,Boccard2021}}

 Many energy-consuming active processes, such as topological constraint release, motion of Replication fork (RF), 
 and formation of loops due to various proteins occur 
 within the cell. The consequences of some of these processes
 are incorporated in an effective manner in the current model. 
Though the energy consuming cellular processes drive the system out of equilibrium, we use Monte Carlo  simulations 
to realize local diffusion of polymer segments. 
 We do not have an estimate of the time scales involved in our simulation in real units.
An estimate of time scales would require using Langevin dynamics simulations.
At our current stage of understanding, incorporating a) replication, b) topological constraint
release due to the presence of topo-isomerase, c) addition of cross-links 
at specific stages in the cell cycle, especially between monomers that might be spatially far 
apart and d) changing the  size of the cylinder systematically is significantly more difficult 
in Langevin dynamics simulations than 
in Monte Carlo simulations.

We have outlined principles by which one may obtain the experimental data of 
\cite{Youngren2014} through our model of the DNA-polymer with a modified topology. 
The experiments were conducted on a specific growth medium of the bacterial cells, which determines the 
value of the doubling time. We have adapted our model similarly to establish a correspondence 
to the experiments of \cite{Youngren2014}. In future manuscripts, we shall communicate our 
results for a different choice of $\tau_C$, $\tau_D$ and the doubling time as realized experimentally using different strains and growth conditions.
 We hope our theoretical predictions can be validated by experiments 
conducted using different growth media. 
% Future studies may also explore other strategies of topological
% constraint release and the resultant organization of chromosome-polymers.
\\

% {\color{blue} {\em In  vivo} one expects transient loops formed by transient links by linker proteins. However, it is known that 
% MukBEF forms clusters near oriC  \cite{Badrinarayanan2015,Sherratt2020}
% , so it is plausible that 
%  another MukBEF present in the cluster replaces a broken link, and thus, one obtains effectively long-lived loops.}

\section{Author Contributions}

SP implemented the model, performed calculations and analysis. The research plan was designed and 
discussed by SP, DM, AC. SP, DM and AC wrote the paper.\\

\section{Acknowledgements}
Authors acknowledge useful discussions with Arieh Zaritsky, Conrad Woldringh, Tejal Agarwal and Suckjoon Jun. A.C., 
with DST-SERB identification SQUID-1973-AC-4067, acknowledges funding by DST-India, project MTR/2019/000078 and CRG/2021/007824. 
A.C also acknowledges discussions in meetings organized by ICTS, Bangalore and use of the computing facilities by PARAM-BRAHMA. 
We acknowledge the support from women in STEM fellowship from DST and IUSSTF which enabled Tejal Agarwal to visit Suckjoon Jun group.
The authors thank the anonymous referees for their detailed comments, which helped in the significant improvement of the paper.
% \bibliography{pnas-sample}
 
\bibliographystyle{unsrt}
\bibliography{ref.bib}

\begin{thebibliography}{10}

\bibitem{Phillips2012}
Rob Phillips, Jane Kondev, Julie Theriot, Hernan~G. Garcia, and Nigel Orme.
\newblock {\em Physical Biology of the Cell}.
\newblock Garland Science, October 2012.

\bibitem{Kuzminov2013}
Andrei Kuzminov.
\newblock The chromosome cycle of prokaryotes.
\newblock {\em Molecular Microbiology}, 90(2):214--227, 2013.

\bibitem{KLECKNER2013}
Jay~K. Fisher, Aude Bourniquel, Guillaume Witz, Beth Weiner, Mara Prentiss, and
  Nancy Kleckner.
\newblock Four-dimensional imaging of {E. coli} nucleoid organization and
  dynamics in living cells.
\newblock {\em Cell}, 153(4):882--895, 2013.

\bibitem{Dewachter2018}
Liselot Dewachter, Natalie Verstraeten, Maarten Fauvart, and Jan Michiels.
\newblock An integrative view of cell cycle control in {Escherichia} coli.
\newblock {\em {FEMS} Microbiology Reviews}, 42(2):116--136, January 2018.

\bibitem{Japaridze2020}
Aleksandre Japaridze, Christos Gogou, Jacob W.~J. Kerssemakers, Huyen~My
  Nguyen, and Cees Dekker.
\newblock Direct observation of independently moving replisomes in
  {Escherichia} coli.
\newblock {\em Nature Communications}, 11(1), June 2020.

\bibitem{Sherratt2020}
Jarno M\"{a}kel\"{a} and David~J. Sherratt.
\newblock Organization of the {{Escherichia}} coli chromosome by a {MukBEF}
  axial core.
\newblock {\em Molecular Cell}, 78(2):250--260.e5, April 2020.

\bibitem{Wiggins2018}
Sarah~M. Mangiameli, Julie~A. Cass, Houra Merrikh, and Paul~A. Wiggins.
\newblock The bacterial replisome has factory-like localization.
\newblock {\em Current Genetics}, 64(5):1029--1036, April 2018.

\bibitem{Badrinarayanan2015}
Anjana Badrinarayanan, Tung~B.K. Le, and Michael~T. Laub.
\newblock Bacterial chromosome organization and segregation.
\newblock {\em Annual Review of Cell and Developmental Biology},
  31(1):171--199, November 2015.

\bibitem{WOLDRINGH2006273}
Conrad~L. Woldringh and Nanne Nanninga.
\newblock Structural and physical aspects of bacterial chromosome segregation.
\newblock {\em Journal of Structural Biology}, 156(2):273--283, 2006.

\bibitem{BenYehuda2003}
S.~Ben-Yehuda.
\newblock {RacA}, a bacterial protein that anchors chromosomes to the cell
  poles.
\newblock {\em Science}, 299(5606):532--536, December 2002.

\bibitem{gogou2021mechanisms}
Christos Gogou, Aleksandre Japaridze, and Cees Dekker.
\newblock Mechanisms for chromosome segregation in bacteria.
\newblock {\em Frontiers in Microbiology}, 12:1533, 2021.

\bibitem{Jun2010}
Suckjoon Jun and Andrew Wright.
\newblock Entropy as the driver of chromosome segregation.
\newblock {\em Nature Reviews Microbiology}, 8(8):600--607, August 2010.

\bibitem{crescentus_nature}
Joris J.~B. Messelink, Muriel C.~F. van Teeseling, Jacqueline Janssen, Martin
  Thanbichler, and Chase~P. Broedersz.
\newblock Learning the distribution of single-cell chromosome conformations in
  bacteria reveals emergent order across genomic scales.
\newblock {\em Nature Communications}, 12(1), March 2021.

\bibitem{caul_loci}
P.~H. Viollier, M.~Thanbichler, P.~T. McGrath, L.~West, M.~Meewan, H.~H.
  McAdams, and L.~Shapiro.
\newblock Rapid and sequential movement of individual chromosomal loci to
  specific subcellular locations during bacterial {DNA} replication.
\newblock {\em Proceedings of the National Academy of Sciences},
  101(25):9257--9262, June 2004.

\bibitem{wigginsrf}
Sarah~M. Mangiameli, Brian~T. Veit, Houra Merrikh, and Paul~A. Wiggins.
\newblock The replisomes remain spatially proximal throughout the cell cycle in
  bacteria.
\newblock {\em {PLOS} Genetics}, 13(1):e1006582, January 2017.

\bibitem{nielsen2006progressive}
Henrik~J Nielsen, Yongfang Li, Brenda Youngren, Flemming~G Hansen, and Stuart
  Austin.
\newblock Progressive segregation of the {Escherichia} coli chromosome.
\newblock {\em Molecular microbiology}, 61(2):383--393, 2006.

\bibitem{marko2009linking}
John~F Marko.
\newblock Linking topology of tethered polymer rings with applications to
  chromosome segregation and estimation of the knotting length.
\newblock {\em Physical Review E}, 79(5):051905, 2009.

\bibitem{Helmstetter1968}
C.~Helmstetter, S.~Cooper, O.~Pierucci, and E.~Revelas.
\newblock On the bacterial life sequence.
\newblock {\em Cold Spring Harbor Symposia on Quantitative Biology},
  33(0):809--822, January 1968.

\bibitem{zaritsky2007changes}
Arieh Zaritsky, Norbert Vischer, and Avinoam Rabinovitch.
\newblock Changes of initiation mass and cell dimensions by the ‘eclipse’.
\newblock {\em Molecular microbiology}, 63(1):15--21, 2007.

\bibitem{zaritsky2011instructive}
Arieh Zaritsky, Ping Wang, and Norbert~OE Vischer.
\newblock Instructive simulation of the bacterial cell division cycle.
\newblock {\em Microbiology}, 157(7):1876--1885, 2011.

\bibitem{Zaritsky2019}
Arieh Zaritsky, Waldemar Vollmer, Jaan M\"{a}nnik, and Chenli Liu.
\newblock Does the nucleoid determine cell dimensions in {{Escherichia} coli?}
\newblock {\em Frontiers in Microbiology}, 10, August 2019.

\bibitem{bremer1977examination}
Hans Bremer and Gordon Churchward.
\newblock An examination of the cooper-helmstetter theory of dna replication in
  bacteria and its underlying assumptions.
\newblock {\em Journal of theoretical biology}, 69(4):645--654, 1977.

\bibitem{skarstad1985escherichia}
KIRSTEN Skarstad, HARALD~B Steen, and ERIK Boye.
\newblock {Escherichia} coli dna distributions measured by flow cytometry and
  compared with theoretical computer simulations.
\newblock {\em Journal of bacteriology}, 163(2):661--668, 1985.

\bibitem{Youngren2014}
B.~Youngren, H.~J. Nielsen, S.~Jun, and S.~Austin.
\newblock The multifork {{Escherichia}} coli chromosome is a self-duplicating
  and self-segregating thermodynamic ring polymer.
\newblock {\em Genes {\&} Development}, 28(1):71--84, January 2014.

\bibitem{Suppli}
See supplementary material at [link] for more information.

\bibitem{Jun2006}
S.~Jun and B.~Mulder.
\newblock Entropy-driven spatial organization of highly confined polymers:
  Lessons for the bacterial chromosome.
\newblock {\em Proceedings of the National Academy of Sciences},
  103(33):12388--12393, August 2006.

\bibitem{Jun2007}
Suckjoon Jun, Axel Arnold, and Bae-Yeun Ha.
\newblock Confined space and effective interactions of multiple self-avoiding
  chains.
\newblock {\em Physical Review Letters}, 98(12), March 2007.

\bibitem{Jun2012}
J.~Pelletier, K.~Halvorsen, B.-Y. Ha, R.~Paparcone, S.~J. Sandler, C.~L.
  Woldringh, W.~P. Wong, and S.~Jun.
\newblock Physical manipulation of the {Escherichia} coli chromosome reveals
  its soft nature.
\newblock {\em Proceedings of the National Academy of Sciences},
  109(40):E2649--E2656, September 2012.

\bibitem{ha1}
Youngkyun Jung and Bae-Yeun Ha.
\newblock Overlapping two self-avoiding polymers in a closed cylindrical pore:
  Implications for chromosome segregation in a bacterial cell.
\newblock {\em Phys. Rev. E}, 82:051926, Nov 2010.

\bibitem{hamain}
Youngkyun Jung, Chanil Jeon, Juin Kim, Hawoong Jeong, Suckjoon Jun, and
  Bae-Yeun Ha.
\newblock Ring polymers as model bacterial chromosomes: confinement{,} chain
  topology{,} single chain statistics{,} and how they interact.
\newblock {\em Soft Matter}, 8:2095--2102, 2012.

\bibitem{hamain2}
Youngkyun Jung, Juin Kim, Suckjoon Jun, and Bae-Yeun Ha.
\newblock Intrachain ordering and segregation of polymers under confinement.
\newblock {\em Macromolecules}, 45(7):3256--3262, March 2012.

\bibitem{hareview}
Bae-Yeun Ha and Youngkyun Jung.
\newblock Polymers under confinement: single polymers, how they interact, and
  as model chromosomes.
\newblock {\em Soft Matter}, 11(12):2333--2352, 2015.

\bibitem{Boccard2021}
Virginia~S. Lioy, Ivan Junier, and Fr{\'{e}}d{\'{e}}ric Boccard.
\newblock Multiscale dynamic structuring of bacterial chromosomes.
\newblock {\em Annual Review of Microbiology}, 75(1), August 2021.

\bibitem{Cass2016}
Julie A. Cass, Nathan J. Kuwada, Beth Traxler, and Paul A. Wiggins.
\newblock {Escherichia} coli chromosomal loci segregate from midcell with
  universal dynamics.
\newblock {\em Biophysical Journal}, 110(12):2597--2609, 2016.

\bibitem{woldy}
Conrad~L. Woldringh, Flemming~G. Hansen, Norbert O.~E. Vischer, and Tove
  Atlung.
\newblock Segregation of chromosome arms in growing and non-growing
  {Escherichia} coli cells.
\newblock {\em Frontiers in Microbiology}, 6, May 2015.

\bibitem{dna1}
Debarshi Mitra, Shreerang Pande, and Apratim Chatterji.
\newblock Polymer architecture orchestrates the segregation and spatial
  organization of replicating {E. coli} chromosomes in slow growth.
\newblock {\em Soft Matter}, 18:5615--5631, 2022.

\bibitem{dna2}
Debarshi Mitra, Shreerang Pande, and Apratim Chatterji.
\newblock Topology-driven spatial organization of ring polymers under
  confinement.
\newblock {\em Phys. Rev. E}, 106:054502, Nov 2022.

\bibitem{loop_heerman}
Andreas Hofmann and Dieter~W. Heermann.
\newblock The role of loops on the order of eukaryotes and prokaryotes.
\newblock {\em {FEBS} Letters}, 589(20PartA):2958--2965, April 2015.

\bibitem{halverson2011molecular}
Jonathan~D Halverson, Won~Bo Lee, Gary~S Grest, Alexander~Y Grosberg, and Kurt
  Kremer.
\newblock Molecular dynamics simulation study of nonconcatenated ring polymers
  in a melt. ii. dynamics.
\newblock {\em The Journal of chemical physics}, 134(20), 2011.

\bibitem{rosa2014ring}
Angelo Rosa and Ralf Everaers.
\newblock Ring polymers in the melt state: the physics of crumpling.
\newblock {\em Physical review letters}, 112(11):118302, 2014.

\bibitem{chubak2021multiscale}
Iurii Chubak, Christos~N Likos, and Sergei~A Egorov.
\newblock Multiscale approaches for confined ring polymer solutions.
\newblock {\em The Journal of Physical Chemistry B}, 125(18):4910--4923, 2021.

\bibitem{halverson2011Statics}
Jonathan~D Halverson, Won~Bo Lee, Gary~S Grest, Alexander~Y Grosberg, and Kurt
  Kremer.
\newblock Molecular dynamics simulation study of nonconcatenated ring polymers
  in a melt. i. statics.
\newblock {\em The Journal of chemical physics}, 134(20), 2011.

\bibitem{pachong2020melts}
Stanard~Mebwe Pachong, Iurii Chubak, Kurt Kremer, and Jan Smrek.
\newblock Melts of nonconcatenated rings in spherical confinement.
\newblock {\em The Journal of Chemical Physics}, 153(6), 2020.

\bibitem{narros2010influence}
Arturo Narros, Angel~J Moreno, and Christos~N Likos.
\newblock Influence of topology on effective potentials: coarse-graining ring
  polymers.
\newblock {\em Soft Matter}, 6(11):2435--2441, 2010.

\bibitem{narros2014multi}
Arturo Narros, Christos~N Likos, Angel~J Moreno, and Barbara Capone.
\newblock Multi-blob coarse graining for ring polymer solutions.
\newblock {\em Soft Matter}, 10(48):9601--9614, 2014.

\bibitem{woldringh2024compaction}
Conrad~Louis Woldringh.
\newblock Compaction and segregation of {DNA} in {Escherichia} coli.
\newblock 2024.

\bibitem{JieXiao24}
Ziqi Fu, Monica~S Guo, Weiqiang Zhou, and Jie Xiao.
\newblock {Differential roles of positive and negative supercoiling in
  organizing the {E. coli} genome}.
\newblock {\em Nucleic Acids Research}, 52(2):724--737, 12 2023.

\bibitem{junier2023dna}
Ivan Junier, Elham Ghobadpour, Olivier Espeli, and Ralf Everaers.
\newblock Dna supercoiling in bacteria: state of play and challenges from a
  viewpoint of physics based modeling.
\newblock {\em Frontiers in Microbiology}, 14, 2023.

\bibitem{harju2023loop}
Janni Harju, Muriel~CF van Teeseling, and Chase~P Broedersz.
\newblock Loop-extruders alter bacterial chromosome topology to direct entropic
  forces for segregation.
\newblock {\em bioRxiv}, pages 2023--06, 2023.

\bibitem{allen2017computer}
Michael~P Allen and Dominic~J Tildesley.
\newblock {\em Computer simulation of liquids}.
\newblock Oxford university press, 2017.

\bibitem{kikuchi1991}
K.~Kikuchi, M.~Yoshida, T.~Maekawa, and H.~Watanabe.
\newblock Metropolis monte carlo method as a numerical technique to solve the
  fokker{\textemdash}planck equation.
\newblock {\em Chemical Physics Letters}, 185(3-4):335--338, October 1991.

\bibitem{sharon}
Sharon~C. Glotzer, Dietrich Stauffer, and Naeem Jan.
\newblock Monte carlo simulations of phase separation in chemically reactive
  binary mixtures.
\newblock {\em Physical Review Letters}, 72:4109--4112, Jun 1994.

\bibitem{amitai2017polymer}
Assaf Amitai and David Holcman.
\newblock Polymer physics of nuclear organization and function.
\newblock {\em Physics Reports}, 678:1--83, 2017.

\bibitem{AMITAI2017}
Assaf Amitai, Andrew Seeber, Susan~M. Gasser, and David Holcman.
\newblock Visualization of chromatin decompaction and break site extrusion as
  predicted by statistical polymer modeling of single-locus trajectories.
\newblock {\em Cell Reports}, 18(5):1200--1214, 2017.

\bibitem{Lioy2018}
Virginia~S. Lioy, Axel Cournac, Martial Marbouty, St{\'{e}}phane Duigou, Julien
  Mozziconacci, Olivier Esp{\'{e}}li, Fr{\'{e}}d{\'{e}}ric Boccard, and Romain
  Koszul.
\newblock Multiscale structuring of the e.~coli chromosome by
  nucleoid-associated and condensin proteins.
\newblock {\em Cell}, 172(4):771--783.e18, February 2018.

\bibitem{Wu2019}
Fabai Wu, Pinaki Swain, Louis Kuijpers, Xuan Zheng, Kevin Felter, Margot
  Guurink, Jacopo Solari, Suckjoon Jun, Thomas~S. Shimizu, Debasish Chaudhuri,
  Bela Mulder, and Cees Dekker.
\newblock Cell boundary confinement sets the size and position of the e.~coli
  chromosome.
\newblock {\em Current Biology}, 29(13):2131--2144.e4, July 2019.

\bibitem{Joyeux2015}
Marc Joyeux.
\newblock Compaction of bacterial genomic {DNA}: clarifying the concepts.
\newblock {\em Journal of Physics: Condensed Matter}, 27(38):383001, September
  2015.

\bibitem{Joyeux2018}
Marc Joyeux.
\newblock A segregative phase separation scenario of the formation of the
  bacterial nucleoid.
\newblock {\em Soft Matter}, 14(36):7368--7381, 2018.

\bibitem{kim2015polymer}
Juin Kim, Chanil Jeon, Hawoong Jeong, Youngkyun Jung, and Bae-Yeun Ha.
\newblock A polymer in a crowded and confined space: effects of crowder size
  and poly-dispersity.
\newblock {\em Soft Matter}, 11(10):1877--1888, 2015.

\bibitem{odijk1998osmotic}
Theo Odijk.
\newblock Osmotic compaction of supercoiled {DNA} into a bacterial nucleoid.
\newblock {\em Biophysical chemistry}, 73(1-2):23--29, 1998.

\bibitem{yang2020effects}
Da~Yang, Jaana M{\"a}nnik, Scott~T Retterer, and Jaan M{\"a}nnik.
\newblock The effects of polydisperse crowders on the compaction of the
  {Escherichia} coli nucleoid.
\newblock {\em Molecular microbiology}, 113(5):1022--1037, 2020.

\bibitem{wang2011chromosome}
Wenqin Wang, Gene-Wei Li, Chongyi Chen, X~Sunney Xie, and Xiaowei Zhuang.
\newblock Chromosome organization by a nucleoid-associated protein in live
  bacteria.
\newblock {\em Science}, 333(6048):1445--1449, 2011.

\bibitem{norris2023roles}
Vic Norris, Clara Kayser, Georgi Muskhelishvili, and Yoan Konto-Ghiorghi.
\newblock The roles of nucleoid-associated proteins and topoisomerases in
  chromosome structure, strand segregation, and the generation of phenotypic
  heterogeneity in bacteria.
\newblock {\em FEMS Microbiology Reviews}, 47(6):fuac049, 2023.

\bibitem{kadam}
Sangram Kadam, Kiran Kumari, Vinoth Manivannan, Shuvadip Dutta, Mithun~K Mitra,
  and Ranjith Padinhateeri.
\newblock Predicting scale-dependent chromatin polymer properties from
  systematic coarse-graining.
\newblock {\em Nature Communications}, 14(1):4108, 2023.

\bibitem{thiel2012long}
Axel Thiel, Mich{\`e}le Valens, Isabelle Vallet-Gely, Olivier Esp{\'e}li, and
  Fr{\'e}d{\'e}ric Boccard.
\newblock Long-range chromosome organization in {E. coli}: a site-specific
  system isolates the ter macrodomain.
\newblock {\em PLoS genetics}, 8(4):e1002672, 2012.

\end{thebibliography}
\end{document}